\documentclass[aps,prd,showpacs,nofootinbib,superscriptaddress]{revtex4}
\usepackage{cancel}
\usepackage{amssymb,amsfonts,amsmath}
\usepackage{graphicx,epsfig,pdfpages}
\usepackage{caption}
\usepackage{subcaption}
\newcommand{\bea}{\begin{eqnarray}}
\newcommand{\eea}{\end{eqnarray}}

\newcommand{\met}{$\cancel E_T$}

\begin{document}
\def\gsim{\mathrel{
   \rlap{\raise 0.511ex \hbox{$>$}}{\lower 0.511ex \hbox{$\sim$}}}}
\def\lsim{\mathrel{
   \rlap{\raise 0.511ex \hbox{$<$}}{\lower 0.511ex \hbox{$\sim$}}}}

\title{\boldmath Light top squarks in $U(1)_{R}$ lepton number model with a
right handed neutrino and the LHC}
\preprint{HIP-2015-23/TH, HRI-P-15-07-001}

\author{Sabyasachi Chakraborty}
\email{tpsc3@iacs.res.in}
\affiliation{Department of Theoretical Physics, Indian Association for
the Cultivation of Science, 2A $\&$ 2B Raja S.C.Mullick Road, Jadavpur,
Kolkata 700 032, INDIA}

\author{AseshKrishna Datta}
\email{asesh@hri.res.in}
\affiliation{Harish-Chandra Research Institute, Chhatnag Road, Jhunsi, 
Allahabad 211019, INDIA}

\author{Katri Huitu}
\email{katri.huitu@helsinki.fi}
\affiliation{Department of Physics and Helsinki Institute of Physics, 
P. O. Box 64, FIN-00014 University of Helsinki, Finland}

\author{Sourov Roy}
\email{tpsr@iacs.res.in}
\affiliation{Department of Theoretical Physics, Indian Association for
the Cultivation of Science, 2A $\&$ 2B Raja S.C.Mullick Road, Jadavpur,
Kolkata 700 032, INDIA}

\author{Harri Waltari}
\email{harri.waltari@helsinki.fi}
\affiliation{Department of Physics and Helsinki Institute of Physics, 
P. O. Box 64, FIN-00014 University of Helsinki, Finland}

%%%%%%%%%%%%%%%%%%%%%%%%%%%%%%%%%%%%%%%%%%%%%%%%%%%%%%%%%%%%%%%%%%%%%%%%%%%%%%%
\begin{abstract}
We investigate the phenomenology of top squarks at the Large Hadron Collider (LHC) 
in a supersymmetric model where lepton number is identified with an approximate 
$U(1)_R$ symmetry in such a way that one of the left chiral sneutrinos can acquire 
a large vacuum expectation value ($vev$) and can play the role of the down-type Higgs. 
This $R$-symmetry allows a subset of trilinear $R$-parity violating interactions, 
which determine the collider phenomenology of this model in a significant way. 
The gauginos are Dirac particles and gluinos are relatively heavy in this class of 
models. The model contains a right handed neutrino superfield, which gives a tree 
level mass to one of the active neutrinos. An order one neutrino Yukawa coupling also 
helps enhance the Higgs boson mass at the tree level and results in a very light 
bino-like neutralino ($\widetilde \chi_2^0$) with mass around a few hundred MeV, 
which is a carrier of missing (transverse) energy (\met). The model can accommodate 
two rather light top squarks, compatible with the observed mass of the
Higgs boson. The lighter top squark (${\widetilde t}_1$) can decay into 
$t\widetilde\chi_2^0$, and thus the signal would be similar to the signal of top 
quark pair production at the LHC. In addition, fully visible decays such as 
${\widetilde t}_2 \rightarrow b e^+$ can give rise to interesting final states. Such 
signals at the LHC combined with other features like a heavy gluino could provide a 
strong evidence for this kind of a model. Our analysis shows that $m_{\widetilde t_1}\lsim 
575~(750)$ GeV and $m_{\widetilde t_2}\lsim 1.2~(1.4)$ TeV can be probed with 
5$\sigma$ statistical significance at the 13 TeV LHC with 300~(3000) fb$^{-1}$ of 
integrated luminosity. Finally, we observe that in the presence of super-light 
carriers of $\cancel{E}_T$, the so-called `stealth' top squark scenario may naturally 
appear in our model.
\end{abstract}
%%%%%%%%%%%%%%%%%%%%%%%%%%%%%%%%%%%%%%%%%%%%%%%%%%%%%%%%%%%%%%%%%%%%%%%%%%%%%%%
\maketitle
%%%%%%%%%%%%%%%%%%%%%%%%%%%%%%%%%%%%%%%%%%%%%%%%%%%%%%%%%%%%%%%%%%%%%%%%%%%%%%%%%%%%
\section{Introduction}
\label{section}
%%%%%%%%%%%%%%%%%%%%%%%%%%%%%%%%%%%%%%%%%%%%%%%%%%%%%%%%%%
The discovery of a Higgs boson at the Large Hadron Collider (LHC) with a mass 
around 125 GeV \cite{ATLAS-higgs, CMS-higgs} is of immense importance in high 
energy physics and, in particular, in the context of electroweak symmetry 
breaking. However, in spite of its enormous success over the years, the Standard 
Model (SM) of particle physics suffers from several drawbacks. From a theoretical 
perspective, the naturalness problem remains a troublesome issue in the framework 
of SM. Supersymmetry (SUSY) renders an elegant solution to this problem and has 
become the most popular choice for physics beyond the standard model (BSM) to date. 
Nevertheless, searches for superpartners by the LHC collaborations (ATLAS and CMS)
in $pp$ collisions at the centre-of-mass energies of $\sqrt s=7$ TeV and 8 TeV have 
shown no significant excess \cite{CMS-bound, ATLAS-bound} over the SM background. 
This has put stringent lower limits on the superpartner masses in many different 
SUSY scenarios. Recent experimental analyses within the framework of a simplified 
phenomenological minimal supersymmetric standard model (pMSSM) have set a lower bound 
of 1.7 TeV~\cite{ATLAS-bound-1} for comparable masses of the gluino and 
the first two generation squarks.

On top of that, finding a Higgs boson with a mass $\sim$ 125 GeV and the non-observation 
of any signals of physics beyond the SM have severely constrained many supersymmetric 
scenarios which are otherwise very well motivated. In view of this, models with 
Dirac gauginos and $U(1)_R$ symmetry have become popular as they can 
significantly lower the current exclusion bounds on the first and second generation 
squarks and at the same time can address the 125 GeV Higgs boson even in the 
presence of lighter top squarks. Other virtues of such scenarios include 
significant suppression of flavor- as well as CP-violating effects. Assorted 
versions of models with Dirac gauginos and $U(1)_R$ symmetry can be found in 
the literature~\cite{Fayet,Polchinski,Hall-1,Hall-2,Jack,Nelson,Fox,Chacko,Antoniadis,
Antoniadis1,Antoniadis-1,kribs,Choi,Amigo,Blechman,Benakli,Belanger,Benakli-m,Kumar,
Fox-1,Benakli-m2,Choi-1,Carpenter,Kribs-1,Abel,Benakli-2,Kalinowski,Benakli-1,
Gregoire,Katz,Rehermann,Davies,ItoyamaMaru,IM1,IM2,Bertuzzo,Davies-1,Argurio,
Goodsell:2012fm,Fok,Argurio-1,Kumar-1,Claudia,Goodsell,Riva,Agrawal,SC-1,Csaki,Dudas,
Beauchesne,Bertuzzo:2014bwa,Benakli:2014cia,SC-2,Goodsell:2014dia,Ipek:2014moa,
Busbridge:2014sha,Diessner:2014ksa,SC-3,Tuhin,Martin:2015eca,Berger:2015qra,
Goodsell:2015ura}.

On the other hand, experiments in the neutrino sector have firmly established the 
fact that neutrinos have tiny masses and non-trivial mixings. Non-vanishing neutrino 
masses and mixings~\cite{Nu-1,Nu-2,Nu-3,Nu-4} are very important indications of new 
physics. An interesting question to investigate is whether models with Dirac gauginos 
could also provide explanation for observed neutrino masses and mixings. As we shall 
describe later on, the introduction of a right-handed neutrino superfield with an 
appropriate $R$-charge and Yukawa coupling `$f$' can give rise to a small neutrino 
mass of the right order at the tree level. At the same time, an order one `$f$' 
generates an additional tree level contribution to the Higgs boson mass. Thus, physics 
in the Higgs sector and the physics in the neutrino sector become intimately related 
in this model. This gives an opportunity to look into the Higgs sector through the 
neutrino-window and vice-versa. The additional tree level contribution to the Higgs 
boson mass also opens up the possibility of having rather light top squarks in the 
spectrum. This can also ameliorate the situation with so called `naturalness' that 
is somewhat compromised in popular SUSY frameworks like the MSSM. There, the top squark 
masses are pushed to higher values ($\sim$ {$\cal O$}(TeV)) to ensure a Higgs boson 
as heavy as observed at the LHC. Another very interesting outcome of this scenario is 
to have a very light bino-like neutralino, also identified as the lightest supersymmetric 
particle (LSP), with a mass in the range of a few hundred MeVs. The scenario violates 
$R$-parity.  Thus, such a light neutralino LSP could decay to SM fermions. However, 
the decay length turns out to be much larger than the collider dimension~\cite{Romao}. 
Hence it would essentially contribute to missing transverse energy (MET; \met). 

In such a backdrop, we study the phenomenology of light top squarks at the LHC in a 
$U(1)_R$ symmetric model, first introduced in references~\cite{,Kumar-1,Claudia} and 
later on augmented by a right handed neutrino superfield in references~\cite{SC-1,SC-2,SC-3}. 
The $R$-charges are identified with the lepton numbers in such a way that the left 
chiral sneutrino $vevs$ can be large, and are not constrained by the Majorana mass of 
the neutrinos. Thus, the sneutrino can play the role of a down type Higgs field. There 
also exists a subset of $R$-parity violating operators, mixings between the neutrinos 
and the neutralinos, as well as, between the charged leptons and the charginos. Once 
$U(1)_R$ symmetry is invoked, the gauginos cease to have Majorana masses. However, they 
can acquire Dirac masses which requires additional chiral superfields living in the 
adjoint representation of the SM gauge group. It is somewhat crucial in the context of 
the present work to note that $R$-symmetric models also prohibit the traditional trilinear 
scalar couplings ($`A'$ terms) and the Higgsino mass parameter ($\mu$ term). To generate
a $\mu$ term, one needs to incorporate two more chiral superfields $R_u$ and $R_d$ with 
appropriate $R$-charges.

The main motivation for having a right handed neutrino superfield is to have a tree level 
neutrino mass~\cite{SC-1}. However, such a simple extension has enormous implication for 
the lightest CP even Higgs boson mass~\cite{SC-1,SC-3} and for the dark matter 
sector~\cite{SC-2}.

The squarks in the present context carry a non-zero $R$-charge ($R = 1$) and hence 
a non-zero lepton number since lepton number is identified with the $R$-charges. 
Top squarks can naturally be light in this model and can have novel signatures at the LHC.
In the present work, we take the obviously natural direction of connecting to the
top squark sector which is very much in the focus of the current LHC programme and
thus could be put to test in a straight-forward way. This work presents for
the first time the collider implications of the very characteristic top squark
sector of the scenario under consideration. Various possibilities in the decays
of both top squarks are discussed in detail. A novel final state in the form of
$b\bar b e^+ e^-$ is highlighted where the final state objects can, in principle,
be reconstructed to the mass of the heavier top squark. The decay of the top squark
to $b e^+$ differs from typical $R$-parity violating MSSM decay modes, and is typical
for this model. This also provides us with an interesting handle, using which the reach 
of $\widetilde t_2$ can be enhanced significantly, so much so, that the enhanced rate 
could also lead to its discovery even before its lighter cousin.

As for $\widetilde t_1$, pair production of top squarks and their subsequent decays 
might lead to signal similar to top quark pair production and provide important information 
on the model and, in particular, on the scenario with an order one neutrino Yukawa coupling 
`$f$'. A characteristic difference in the signal is in the form of a somewhat softer 
$\cancel{E}_T$, when compared to similar $m_{\widetilde t_1}$ values in the MSSM. This 
is because of the presence of a MeV neutralino LSP, which is again a salient feature of 
our scenario. It is also demonstrated how various decay modes of $\widetilde t_1$ remain 
simultaneously open thus necessitating a thorough analysis of the experimental data from 
the 13 TeV run of the LHC.

%For example, pair production of 
%top squarks and their subsequent decays via $R$-parity conserving/violating channels, 
%or a combination of both, might lead to signals similar to top quark pair production 
%and provide important information on the model and, in particular, on the scenario 
%with an order one neutrino Yukawa coupling `$f$'. Note that in the latter case, the 
%presence of the light bino-like neutralino gives rise to signals with MET.
%Furthermore, characteristic signatures for these light top squark states
%at the LHC even have the potential to discriminate between competing scenarios that
%may give rise to such a light pair of top squarks.

The paper is organised as follows. A brief description of the model is presented in 
section~\ref{model}. The neutral scalar sector of the model and its characteristic 
features are discussed in section~\ref{scalar}. Section~\ref{gauginos} describes the 
electroweak gauginos in general. We start with a generic discussion of the neutralino 
sector and its role in generating the tree level neutrino mass both in $U(1)_R$ conserving 
case as well as taking into account mild violation of this $R$-symmetry. We discuss the 
possibility of mixings among the neutralinos and the neutrinos. Later on, we discuss the 
chargino sector of this model and the corresponding mixing between the charginos and the 
electron. The focus area of this work, that is to say, the top squark sector is described 
in section~\ref{top}. Expressions for the decay rates in various relevant modes are 
presented. The latest bounds on the the masses of the top squarks as reported by the LHC 
collaborations are also discussed. The model is incorporated in {\tt SARAH (v4.4.1)}
~\cite{Staub,Staub1,Staub2}. In section~\ref{benchmarks} we present a few benchmark points 
that reflect the characteristic decay patterns of the two top squarks and are found to 
be instrumental in shaping the interesting signatures at the LHC. Section~\ref{collider} 
is devoted to the actual simulation study of the signals and the most relevant backgrounds 
using event generators. Estimations of the reaches in the masses of the  the top squarks 
are also presented. In section~\ref{stealth} we briefly analyse the issue of the `stealth' 
top squark which arises naturally in our scenario. We summarise with some
concluding remarks in section~\ref{conclusion}.

%%%%%%%%%%%%%%%%%%%%%%%%%%%%%%%%%%%%%%%%%%%%%%%%%%%%%%%%%%
\section{The $U(1)_{R}$-lepton number model}
\label{model}
%%%%%%%%%%%%%%%%%%%%%%%%%%%%%%%%%%%%%%%%%%%%%%%%%%%%%%%%%%
We minimally extend an $R$-symmetric model, first discussed in \cite{Kumar-1,Claudia}, by 
a single right handed neutrino superfield $\hat N^c$~\cite{SC-1}. Along with the chiral 
superfields of the minimal supersymmetric standard model (MSSM) superfields, $\hat H_{u}$, 
$\hat H_{d}$, $\hat Q_i$, $\hat U_{i}^c$, $\hat D_{i}^c$, $\hat L_{i}$, $\hat E_{i}^c$, 
the model contains two `inert' doublet superfields $\hat R_{u}$ and $\hat R_{d}$ with 
opposite hypercharges. To prohibit spontaneous $R$-breaking and hence the emergence of 
$R$-axions, the scalar components of $\hat R_{u}$ and $\hat R_{d}$ are barred from receiving 
any nonzero $vev$. This is why $\hat R_u$ and $\hat R_d$ are labeled as `inert'. Similarly, 
the scalar component of $\hat N^c$ does not acquire any nonzero $vev$. The Dirac gaugino 
masses can be constructed with the introduction of chiral superfields, living in the adjoint 
representation of the SM gauge group. A singlet $\hat S$ is needed to form a Dirac mass for 
the $U(1)$ gaugino, a triplet $\hat T$ under $SU(2)_L$ is required to have a Dirac mass for 
the $SU(2)_L$ gauginos and similarly an octet $\hat O$ under $SU(3)_C$ must be there to 
generate the Dirac gluino mass. The $U(1)_R$ charges of the chiral superfields along with 
their SM gauge quantum numbers are shown in table \ref{R-charges}.
%%%%%%%%%%%%%%%%%%%%%%%%%%%%%%%%%%%%%%%%%%%%
\begin{table}[h!]
\begin{center}
 \begin{tabular}{| l | c | r | }
   \hline
   Superfields & $SU(3)_C$, $SU(2)_L$, $U(1)_Y$ & $U(1)_R$ \\ \hline
   ~~~~~$\hat Q$ & (3, 2, $\frac{1}{3}$) &  1 \\
   ~~~~~$\hat U_i^c$ & ($\bar 3$, 1, $-\frac{4}{3}$) & 1 \\
   ~~~~~$\hat D_i^c$ & ($\bar 3$, 1, $\frac{2}{3}$) & 1 \\ \hline
   ~~~~~$\hat L_i$ & ($1$, 2, $-1$) & 0 \\
   ~~~~~$\hat E_i^c$ & ($1$, 1, $2$) & 2 \\ \hline
   ~~~~~$\hat H_u$ & ($1$, 2, $1$) & 0 \\
   ~~~~~$\hat H_d$ & ($1$, 2, $-1$) & 0 \\ \hline
   ~~~~~$\hat R_u$ & ($1$, 2, $1$) & 2 \\
   ~~~~~$\hat R_d$ & ($1$, 2, $-1$) & 2 \\ \hline
   ~~~~~$\hat S$ & ($1$, 1, $0$) & 0 \\
   ~~~~~$\hat T$ & ($1$, 3, $0$) & 0 \\
   ~~~~~$\hat O$ & ($8$, 1, $0$) & 0 \\ \hline
   ~~~~~$\hat N^c$ & ($1$, 1, $0$) & 2 \\
   \hline
 \end{tabular}
\end{center}
\caption{Chiral superfields with 
the SM gauge quantum numbers and $U(1)_R$ charge assignments.}
\label{R-charges}
\end{table}
\vskip 0.2 cm
%%%%%%%%%%%%%%%%%%%%%%%%%%%%%%%%%%%%%%%%%%%%%%
Note that the scalar components transform in the same manner as their respective chiral 
superfields whereas the fermions have $R$-charge one less than that of the corresponding 
chiral superfields. Following reference \cite{Kumar-1}, we also identify the lepton numbers 
of the component fields to the negative of their $R$-charges. Such an identification leaves 
the lepton number assignments of the SM fermions unchanged from the usual ones while the 
superpartners acquire non-standard lepton numbers. As mentioned in the introduction, it is 
quite conspicuous that the left chiral sneutrino $vevs$ can be large since they do not 
become constrained by the lepton number violating Majorana neutrino masses~\cite{Kumar-1}. 
As a result, the sneutrino can play the role of a down type Higgs field. It is now possible 
to integrate out the superfields $\hat R_u$ and $\hat H_d$, which simplifies the superpotential 
and the scalar potential considerably. At this point all the three sneutrinos can acquire 
substantial $vev$s. However, without any loss of generality, one can always choose a basis 
in which only one of the sneutrinos get a non-zero $vev$, which we choose to be the electron 
type sneutrino\footnote{It should be noted at this point that the formulation, though 
independent of this kind of a choice of a particular basis, could have crucial impacts on 
the actual observables at the experiments. We justify our choice later in this paragraph.}, 
whereas, the $vev$s of the other two sneutrino fields are zero. Thus, the electron sneutrino 
($\widetilde \nu_a$, $a=1 (e)$) plays the role of a down type Higgs field. With this basis 
choice and the assumptions of~\cite{SC-1,SC-3}, the superpotential takes the following form;
%%%%%%%%%%%%%%%%%%%%%%%%%%%%%%%%%%%%%%%%%%%%%%%%%%%%%%%%%%%%
\begin{eqnarray}
W&=&y_{ij}^{u}\hat H_{u}\hat Q_{i}\hat U_{j}^{c}+\mu_{u}\hat
H_{u}\hat R_{d}+f\hat L_{a}\hat H_{u}\hat N^{c}+
\lambda_{S}\hat S\hat H_{u}\hat R_{d}
+2\lambda_{T}\hat H_{u}
\hat T\hat R_{d} \nonumber \\
&&-M_{R}\hat N^{c}\hat S + W^{\prime},
\label{final-superpotential}
\end{eqnarray}
%%%%%%%%%%%%%%%%%%%%%%%%%%%%%%%%%%%%%%%%%%%%%%%%%%%%%%%%%%%%
\begin{eqnarray}
&&W^{\prime}=\sum_{b=2,3} f^l_b {\hat L}_a {\hat L^\prime}_b
{\hat E^{\prime c}}_b + \sum_{k=1,2,3} f^d_k {\hat L}_a
{\hat Q^\prime}_k {\hat D^{\prime c}}_k 
+ \sum_{k=1,2,3} \dfrac{1}{2}
{\tilde \lambda}_{23k}{\hat L^\prime}_2 {\hat L^\prime}_3
{\hat E^{\prime c}}_k \nonumber \\
&&+ \sum_{j,k=1,2,3;b=2,3}{\tilde \lambda}^\prime_{bjk}
{\hat L^\prime}_b {\hat Q^\prime}_j {\hat D^{\prime c}}_k, 
\label{W-diag}
\end{eqnarray}
%%%%%%%%%%%%%%%%%%%%%%%%%%%%%%%%%%%%%%%%%%%%%%%%%%%%%%%%%%%%
where $y^u_{ij}$ is the up-type Yukawa coupling, $\mu_u$ is the Higgsino mass parameter 
consistent with the $R$-charge assignments. `$f$' represents the neutrino Yukawa coupling, 
$\lambda_S$ and $\lambda_T$ are the trilinear couplings and finally $M_R$ is the coefficient 
of the bilinear term $\hat N^c\hat S$. 

Note that, for simplicity in this work we have neglected the terms $\kappa\hat N^c\hat S\hat S$, 
$\eta\hat N^c$, $\xi\hat N^c Tr(\hat T\hat T)$ and $\zeta \hat N^c Tr(\hat O\hat O)$ from the 
superpotential. As long as $\eta\sim M^2_{\text{SUSY}}$, $\kappa$, $\xi$, $\zeta\sim 1$
and $vev$-s of the scalar components of $\hat S$ and $\hat T$ are very small 
($\sim\mathcal O(10^{-4})$ GeV, as we shall consider later), we do not expect any significant 
changes in the analysis and the results presented in this work. 

The prime in $W^{\prime}$ indicates the mass basis for the down type quarks and charged 
leptons. When the electron sneutrino gets a $vev$, the first two terms in $W^{\prime}$ give 
masses to the down type charged leptons and quarks. However, $SU(2)$ invariance ensures that 
the $LLE$ operator cannot generate a mass for the electron. The electron mass can be generated 
from higher dimensional operators suppressed by a mass scale as discussed in~\cite{Kumar-1}. 
Such operators would, in principle, contribute to the masses of other charged leptons as well. 
However, these contributions will be subdominant compared to the contribution to their masses 
from the $LLE$ operators. This makes it a natural choice for the electron sneutrino to 
acquire a non-zero $vev$. The other two terms in $W^{\prime}$ include all the trilinear 
$R$-parity violating but {\it lepton number preserving} terms in this model. 

In a realistic supersymmetric model, one needs to incorporate soft SUSY breaking terms such 
as the gaugino and the scalar mass terms. The Lagrangian consisting of the Dirac gaugino 
mass terms~\cite{Kumar-1} can be written as
%%%%%%%%%%%%%%%%%%%%%%%%%%%%%%%%%%%%%%%%%%%%%%%%%%%
\begin{eqnarray}
{\cal L}^{\rm Dirac}_{\rm gaugino} &=& \int d^2 \theta
\dfrac{W^\prime_\alpha}{\Lambda}[\sqrt{2} \kappa_1 ~W_{1 \alpha} {\hat S}
+ 2\sqrt{2} \kappa_2 ~{\rm tr}(W_{2\alpha} {\hat T})
+ 2\sqrt{2} \kappa_3 ~{\rm tr}(W_{3\alpha} {\hat O})] + h.c., \nonumber \\
\label{dirac-gaugino}
\end{eqnarray}
%%%%%%%%%%%%%%%%%%%%%%%%%%%%%%%%%%%%%%%%%%%%%%%%%%%
where $W^{\prime}_{\alpha}= \lambda_{\alpha} + \theta_{\alpha} D^{\prime}$ 
is a spurion superfield parametrising $D$-type SUSY breaking. $W_{i\alpha}$'s 
are the field strength tensors containing the gauginos of the MSSM vector 
superfields. The $D$-term $vev$ generates Dirac gaugino masses which can be 
schematically written as $M_i^D = \kappa_i \frac{<D^{\prime}>}{\Lambda}$, 
where $\kappa_i$'s are the order one coefficients and $\Lambda$ is the scale
of SUSY mediation.

Similarly, $R$-conserving but soft SUSY breaking terms in the scalar sector 
can be generated from a spurion superfield $\hat X$, where $\hat X = x + 
\theta^2 F_X$~\cite{Kumar-1}. The non-zero $vev$ of the $F$-term generates the 
scalar soft terms. In the rotated basis where only the electron type sneutrino 
acquires a $vev$, the soft SUSY breaking terms are given by
%%%%%%%%%%%%%%%%%%%%%%%%%%%%%%%%%%%%%%%%%%%%%%%%%%%%%%%%%%%%
\begin{eqnarray}
V_{soft}&=& m^{2}_{H_{u}}H_{u}^{\dagger}H_{u}+m^{2}_{R_{d}}
R_{d}^{\dagger}R_{d}+m^{2}_{\tilde L_{a}} \tilde L_{a}^{\dagger}
\tilde L_{a}
+\sum_{b=2,3} m^{2}_{\tilde L_{b}} \tilde L_{b}^{\dagger}
{\tilde L_{b}}+M_{N}^{2}{\tilde N}^{c\dagger} {\tilde N}^{c} \nonumber \\
&+&m^2_{{\tilde R}_i}{{\tilde l}^\dagger_{Ri} {\tilde l}_{Ri}} 
+ m^{2}_{{\tilde Q}_{i}}{\tilde Q}_{i}^{\dagger}{\tilde Q}_{i}
+ m^2_{{\tilde u}_i}{{\tilde u}^\dagger_{Ri} {\tilde u}_{Ri}}
+ m^2_{{\tilde d}_i}{{\tilde d}^\dagger_{Ri} {\tilde d}_{Ri}}
+m_{S}^{2} S^{\dagger}S \nonumber \\
&+&2m_{T}^{2} {\rm tr}(T^{\dagger}T) +2m_O^2 {\rm tr}(O^\dagger O)
- (B\mu_L H_u {\tilde L}_a + {\rm h.c.}) +(t_{S}S+{\rm h.c.}) \nonumber \\
&+&\frac{1}{2} b_{S}(S^{2}+{\rm h.c.}) +b_{T} ({\rm tr}(TT) + {\rm h.c.}) 
+B_O({\rm tr}(OO) + {\rm h.c.}). 
\label{final-softsusy-terms}
\end{eqnarray}
%%%%%%%%%%%%%%%%%%%%%%%%%%%%%%%%%%%%%%%%%%%%%%%%%%%%%%%%%%%%
It is important to note that the scalar singlet tadpole term ($t_S S$) is suppressed
~\cite{Goodsell:2012fm} in the scenarios with Dirac gaugino masses and that is what we will
consider in the present context. With this short description of the theoretical framework 
we now proceed to describe the scalar and the fermionic sectors of the model in appropriate 
details.
%%%%%%%%%%%%%%%%%%%%%%%%%%%%%%%%%%%%%%%%%%%%%%%%%%%%%%%%%%
\section{The neutral scalar sector and the Standard Model-like Higgs boson}
\label{scalar}
%%%%%%%%%%%%%%%%%%%%%%%%%%%%%%%%%%%%%%%%%%%%%%%%%%%%%%%%%%
In this section we discuss the CP-even scalar sector, followed
by a rather important discussion on the lightest CP-even mass eigenstate.
The scalar potential receives contributions from the $F$-term, the $D$-term,
the soft-SUSY breaking terms and the dominant quartic terms generated at one loop
and can be written down as
%%%%%%%%%%%%%%%%%%%
\begin{eqnarray}
V = V_{\rm F} + V_{\rm D} + V_{\rm soft} + V_{\rm one-loop}.
\end{eqnarray}
%%%%%%%%%%%%%%%%%%%
From the scalar potential and the subsequent minimization equations,
one can now write down the CP-even scalar mass matrix in the basis
$(h_R, \tilde \nu_R, S_R, T_R)$, where the subscript $R$ indicates the
real parts of the corresponding superfields. Both $R_d$ and $\widetilde N^c$ carry
$R$-charges of two units and hence gets decoupled from the CP-even
scalar mass matrix. In the 
$R$-symmetric scenario the elements of CP-even $4\times 4$ scalar mass 
matrix are given by~\cite{SC-1}
%%%%%%%%%%%%%%%%%%%
\begin{eqnarray}
\label{CP-even}
(M_S^2)_{11}&=&\frac{(g^{2}+g^{\prime 2})}{2}v^{2}\sin^{2}\beta+
(fM_{R}v_{S}-B\mu_{L}^{a})(\tan\beta)^{-1}
+2\delta\lambda_{u}v^{2}\sin^{2}\beta,\nonumber \\
(M_S^2)_{12}&=&f^{2}v^{2}\sin2\beta+B\mu_{L}^{a}-
\frac{(g^{2}+g^{\prime 2}-2\delta\lambda_{3})}{4}v^{2}
\sin2\beta
- fM_{R}v_{S},\nonumber \\
(M_S^2)_{13}&=& 2\lambda_{S}^{2}v_{S}v\sin\beta+2\mu_{u}
\lambda_{S}v\sin\beta+2\lambda_{S}\lambda_{T}v v_{T}\sin\beta
+\sqrt2 g^{\prime}M_{1}^{D}v\sin\beta-fM_{R}v\cos\beta,\nonumber \\
(M_S^2)_{14}&=& 2\lambda_{T}^{2}v_{T}v\sin\beta+2\mu_{u}\lambda_{T}
v\sin\beta+2\lambda_{S}\lambda_{T}v_{S}v\sin\beta
- \sqrt 2 gM_{2}^{D} v\sin\beta,\nonumber \\
(M_S^2)_{22}&=&\frac{(g^{2}+g^{\prime 2})}{2}v^{2}\cos^{2}\beta+
(fM_{R}v_{S}-B\mu_{L}^{a})\tan\beta
+ 2\delta\lambda_{\nu}v^{2}\cos^{2}\beta,\nonumber \\
(M_S^2)_{23}&=& -\sqrt 2 g^{\prime}M_{1}^{D}v\cos\beta
-fM_{R}v\sin\beta,\nonumber \\
(M_S^2)_{24}&=&\sqrt 2 g M_{2}^{D} v\cos\beta,\nonumber \\
(M_S^2)_{33}&=&-\mu_{u}\lambda_{S}\frac{v^{2}\sin^2\beta}{v_{S}}
-\frac{\lambda_{S}\lambda_{T}v_{T}v^{2}\sin^2\beta}{v_{S}}
-\frac{t_S}{v_{S}}+\frac{g^{\prime}M_{1}^{D}v^{2}\cos2\beta}{\sqrt 2 v_{S}}
+\frac{f M_{R}v^{2}\sin2\beta}{2v_{S}},\nonumber \\
(M_S^2)_{34}&=&\lambda_{S}\lambda_{T}v^{2}\sin^2\beta, \nonumber \\
(M_S^2)_{44}&=&-\mu_{u}\lambda_{T}\frac{v^{2}}{v_{T}}\sin^2\beta
-\lambda_{S}\lambda_{T}v_{S}\frac{v^{2}}{v_{T}}\sin^2\beta
-\frac{g M_{2}^{D}}{\sqrt 2} \frac{v^{2}}{v_{T}}\cos2\beta.
\end{eqnarray}
%%%%%%%%%%%%%%%%%%%%%%%%%%%%%%%%%%%%%%%%%%%%%%%%%%%%%%%%%%
The $\delta$'s appearing only in $(M_S^2)_{11}, (M_S^2)_{12}$
and $(M_S^2)_{22}$ quantify the dominant one-loop 
radiative corrections\footnote{See 
reference~\cite{Belanger,Kumar-1} for a detailed discussion.} to the 
quartic potential coming from the terms $\frac{1}{2}\delta\lambda_u 
(|H_u|^2)^2$, $\frac{1}{2}\delta\lambda_3|H_u^0|^2|\tilde\nu_a|^2$ and
$\frac{1}{2}\delta\lambda_{\nu}(|\tilde\nu_a|^2)^2$ where
%%%%%%%%%%%%%%%%%%%%%%%%%%%%%%%%%%%%%%%%%%%%%%%%%%%%%%%%%%
\begin{eqnarray}
\delta\lambda_u&=&\frac{3 y_t^4}{16\pi^2}\ln\Big(\frac{m_{\widetilde t_1}
m_{\widetilde t_2}}{m_t^2}\Big)+\frac{5\lambda_T^4}{16\pi^2}\ln\Big(
\frac{m_T^2}{v^2}\Big)+\frac{\lambda_S^4}{16\pi^2}\ln\Big(\frac{m_S^2}{v^2}\Big)
+... \nonumber \\
\delta\lambda_3&=&\frac{5\lambda_T^4}{32\pi^2}\ln\Big(\frac{m_T^2}{v^2}\Big)
+\frac{\lambda_S^2}{32\pi^2}\ln\Big(\frac{m_S^2}{v^2}\Big)+... \nonumber \\
\delta\lambda_{\nu}&=&\frac{3 y_b^4}{16\pi^2}\ln\Big(\frac{m_{\widetilde b_1}
m_{\widetilde b_2}}{m_t^2}\Big)+\frac{5\lambda_T^4}{16\pi^2}\ln\Big(
\frac{m_T^2}{v^2}\Big)+\frac{\lambda_S^4}{16\pi^2}\ln\Big(\frac{m_S^2}{v^2}\Big)
+... 
\label{quartic}
\end{eqnarray}
%%%%%%%%%%%%%%%%%%%%%%%%%%%%%%%%%%%%%%%%%%%%%%%%%%%%%%%%%
$m_S, m_T$ are the singlet and triplet soft masses while the singlet and the triplet $vev$s 
are denoted by $v_S$ and $v_T$, respectively \cite{Belanger}.
$g^{\prime}, g$ are the $U(1)_Y$ and $SU(2)_L$ gauge coupling constants, 
$M_1^D, M_2^D$ are the
Dirac bino and wino masses, respectively. $\tan\beta=v_u/v_a$, where
$v_u$ is the $vev$ of the up-type neutral Higgs field and $v_a$
represents the $vev$ of the electron type sneutrino.
The ellipsis at the end of each expression stands for missing subdominant terms. 
%%%%%%%%%%%%%%%%%%%%%%%%%%%%%%%%%%%%%%%%%%%%%%%%%%%%%%%%%%

In this work, we study a simplified scenario in which the singlet and 
the triplet $vev$s are very small. These effectively decouple the corresponding 
scalar fields. Thus, the CP-even scalar mass-squared matrix turns out to be a 
$2\times 2$ one and
can be written down in a compact form as
%%%%%%%%%%%%%%%%%%%%%%%%%%%%%%%
\begin{eqnarray}
M_{11}^{2}&=&M_{Z}^{2}\sin^2\beta+\xi \cot\beta,\nonumber \\
M_{12}^{2}&=&-\xi+\frac{1}{2}M_{Z}^{2}(\alpha-1)
\sin 2\beta = M_{21}^{2},\nonumber \\
M_{22}^{2}&=&\xi \tan\beta+M_{Z}^{2}\cos^2\beta,
\end{eqnarray}
%%%%%%%%%%%%%%%%%%%%%%%%%%%%%%
where $\alpha=\frac{2f^{2}v^{2}}{M_{Z}^{2}}$ and
$\xi=fM_{R}v_{S}-B\mu_{L}$. As long as $M_A^2 > M_Z^2$, where 
$M_A^2\equiv \frac{2(-B\mu_L + f M_R v_S)}{\sin 2\beta}$ is the CP-odd
Higgs mass, we
find that the tree level upper bound on the lightest CP-even Higgs 
boson mass is~\cite{SC-1} 
%%%%%%%%%%%%%%%%%%%%%%%%%%%%%%
\begin{eqnarray}
m_{h}^{2}\leqslant \left[M_{Z}^{2}\cos^2 2\beta+f^{2}
v^{2}\sin^2 2\beta\right].
\label{tree-higgs}
\end{eqnarray}
%%%%%%%%%%%%%%%%%%%%%%%%%%%%%
Clearly, this result is very interesting since for a large neutrino
Yukawa coupling, $f\sim\mathcal O(1)$, the Higgs boson mass receives a large
tree level enhancement. 
%%%%%%%%%%%%%%%%%%%%%%%%%%%%
This additional tree level contribution 
$(\Delta m_h^2)_{\rm Tree}=f^2 v^2 \sin^2 2\beta$ grows at low $\tan\beta$ 
and becomes significant for order one neutrino Yukawa coupling `$f$'~\cite{SC-1,SC-3}. 
The resulting enhancement could play a significant role in lifting the Higgs 
boson mass to 125 GeV. Furthermore, this additional contribution ameliorates 
the `naturalness' (pertaining to the mass of the Higgs boson)
issue in the MSSM. However, this tree level contribution 
gets diluted at large values of $\tan\beta$. There, the one loop quartic 
corrections~\cite{Belanger,Kumar-1} can come into play and can substantially enhance 
the Higgs boson mass in the presence of order one couplings, $\lambda_S$ and 
$\lambda_T$, as shown in equation \ref{quartic}. 
Thus, even for larger values of $\tan\beta$, one can easily
find a Higgs boson as heavy as observed at the LHC experiments, when the top
squarks are relatively light. 
%%%%%%%%%%%%%%%%%%%%%%%%%%%
\begin{figure}[ht]
\centering
\includegraphics[width=9cm]{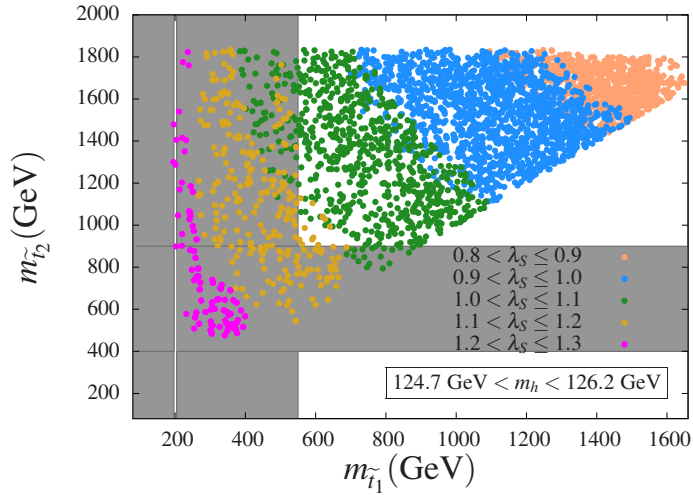}
\caption{Allowed region in $(m_{\widetilde t_1}$-$m_{\widetilde t_2})$ plane
compatible with $124.7~\text{GeV}<m_h<126.2~\text{GeV}$ Higgs 
mass after taking into consideration the full one
loop corrections. The grey bands indicate the values of $m_{\widetilde t_1}$ and
$m_{\widetilde t_2}$ ruled out by the LHC (see text for more details). The
narrow vertical strip over 
197 GeV $\lesssim m_{\widetilde t_1} \lesssim 205$ GeV refers to the
`stealth' top squark regime.
}
\label{fig:stop}
\end{figure}
%%%%%%%%%%%%%%%%%%%%%%%%%%%
Figure~\ref{fig:stop} illustrates the region in the plane of $m_{\widetilde t_1}$
and $m_{\widetilde t_2}$ compatible with 124.7 GeV $< m_h <$ 126.2 GeV
and various slices of $\lambda_S$ over the ranges shown. 
This takes into account the one-loop corrections computed in the effective potential 
approach to the neutral scalar potential as implemented 
in {\tt SARAH (v4.4.1)}~\cite{Staub,Staub1,Staub2}.
For this scattered plot we use $\tan\beta$ 
= 23, $M^D_1=M^D_2$ = 1.2 TeV, $M^D_3$ = 1.5 TeV, $\mu_u$ = 200 GeV, $f$ = 1, 
$v_S=v_T = 10^{-4}$ GeV, $B\mu_L = -(200~\text{GeV})^2$, $t_S = (174~\text{GeV})^3$ 
and vary $\lambda_S$ in the range $0.8<\lambda_S\leq 1.3$. We also vary 
the soft scalar masses $(m^2_Q)_{33}, (m^2_u)_{33}$ in the range $-8\times 10^{6}$ 
(GeV)$^2$ to $8\times 10^{6}$ (GeV)$^2$ to vary the top squark mass. We observe 
that higher values of the superpotential coupling $\lambda_S$ (hence larger 
$\lambda_T$, as $\lambda_T=\lambda_S\tan\theta_W$ ($\theta_W$ being the usual
weak mixing angle), chosen to fit the neutrino mass 
at the tree level) provide larger corrections to the Higgs boson mass at the loop 
level and hence the requirement of having multi-TeV top squarks can be avoided. 
The largest values considered for $m_{\widetilde t_1}$ and $m_{\widetilde t_2}$
in this plot are close to 1660 GeV and 1830 GeV, respectively. 

As can be seen from this figure, relatively light top squarks with 
$m_{\widetilde t_1}\simeq m_{\widetilde t_2}$ may be generic to our scenario 
(though such a situation could attract more aggressive constraints from the LHC). 
Furthermore, both of them can have sub-TeV masses simultaneously and can still be
consistent with the observed value of the SM-like Higgs mass if $\lambda_S > 1$.
In the absence of any appreciable chiral mixing (left-right ($L$-$R$) mixing),
top squarks as heavy as values considered here, could only 
raise the Higgs boson mass up to around 114 GeV in the MSSM. Hence the model under 
consideration has a very interesting and a distinct feature where light to 
moderately heavy ($\sim 1$ TeV) top squarks with negligible $L$-$R$ mixing can be 
compatible with observed mass of the Higgs boson. Similar possibilities are discussed 
earlier in generic setups~\cite{Curtin} and more recently, in a specific 
SUSY scenario like the Next-to-Minimal SUSY extension of the SM (NMSSM)~\cite{Beuria}. 

Taking advantage of this situation, we explore relatively light top squarks
satisfying the relevant direct search constraints. We further note that 
a top squark within a narrow mass-window between  197 GeV and 205 
GeV~\cite{Aad:2015pfx}, which is still allowed by 
data from the 8 TeV run of the LHC, can be obtained in our scenario 
for $\lambda_S\geq 1.2$. Such a mass-window with $m_{\tilde{t}} \gtrsim m_t$ is 
known in the literature as the `stealth' window~\cite{stealth-stop,Fan,Fan1,Csaki1,Weiler} 
for the top squark when the carriers of MET 
are extremely light (as with the neutrinos in the cascade of the SM top quark). 
This renders the signal characteristics of the top squark rather similar to that 
of the top quark thus making the presence of the former in the data difficult to
recognise. Given that the model we consider indeed bears super-light carriers of MET
(as we will discuss in section \ref{neutralinos}), such a `stealth' scenario can 
be easily accommodated in our scenarios. 
Precise measurements of the SM top quark properties 
such as its pair-production cross-section could offer probe to the `stealth' top squark 
scenario. Equivalently, presence of a light top squark may also have some
impact on the measurements of the top quark mass \cite{Till}. 
We briefly touch upon the situation with such `stealth' top
squarks in section \ref{stealth}.

%%%%%%%%%%%%%%%%%%%%%%%%%%%%%%%%%%%%%%%%%%%%%%%%%%%%%%%%%%%%%%%%%%%%%%%%%%%%%%%%%%%%%%%%%%%%
\section{The electroweak gauginos}
\label{gauginos} 
%%%%%%%%%%%%%%%%%%%%%%%%%%%%%%%%%%%%%%%%%%%%%%%%%%%%%%%%%%%%%%%%%%%%%%%%%%%%%%%%%%%%%%%%%%%%
In this section we discuss issues pertaining to neutrino masses and the 
fermionic sector of the scenario. These will help us understand the effect of large 
neutrino Yukawa coupling (`$f$') in the present context.
The electroweak gauginos are comprised of the neutralinos and charginos. 
However, contrary to the MSSM scenario, the neutralinos are Dirac fermions in the 
$R$-preserving case. In the $R$-breaking case they take a pseudo Dirac form,
in general.
%%%%%%%%%%%%%%%%%%%%%%%%%%%%%%%%%%%%%%%%%%%%%%%%%%%%%%%%%%
\subsection{The neutralino sector}
\label{neutralinos} 
%%%%%%%%%%%%%%%%%%%%%%%%%%%%%%%%%%%%%%%%%%%%%%%%%%%%%%%%%%
The decay branching fractions in various available final states of the 
top squarks depend crucially on the neutralino
sector. The neutralino sector in this model differs from that of the MSSM 
due to the presence of additional fermionic fields such as $\widetilde S,
\widetilde T^0, \widetilde R_d^0$. In addition, due to the non-vanishing
sneutrino $vev$, the active neutrino mixes with the neutralinos. Let us
first discuss the neutralino mass matrix in the $R$-preserving scenario,
although, ultimately we carry out our analysis by considering a 
mild $R$-symmetry violation. This opens up several new and interesting
phenomenological issues, which we will mention in due course.
%%%%%%%%%%%%%%%%%%%%%%%%%%%%%%%%%%%%%%%%%%%%%%%%%%%%%%%%%%
\subsubsection{The $R$-conserving case}
\label{R-conserving}
%%%%%%%%%%%%%%%%%%%%%%%%%%%%%%%%%%%%%%%%%%%%%%%%%%%%%%%%%%
The part of the Lagrangian that corresponds 
to the neutral fermion mass matrix is given by $\mathcal{L}=(\psi^{0+})^{T}
M_{\chi}^D(\psi^{0-})$ where $\psi^{0+}=(\tilde b^{0},\tilde w^{0},
\tilde R_{d}^{0},N^{c})$ and $\psi^{0-}=(\tilde S,\tilde T^{0},
\tilde H_{u}^{0},\nu_{e})$. The superscript ($\pm$) indicates the respective
$R$-charges which are $+1$ and $-1$. 
The neutral Dirac fermion mass matrix $M_{\chi}^D$ is given by~\cite{SC-1}
%%%%%%%%%%%%%%%%%%%%%%%%%%%
\begin{eqnarray}
M_{\chi}^D=\left(
\begin{array}{cccc}
M_{1}^{D} & 0 & \frac{g^{\prime}v_{u}}{\sqrt 2}& -\frac{g^{\prime}v_{a}}{\sqrt 2}\\
0 & M_{2}^{D} & -\frac{gv_{u}}{\sqrt 2}& \frac{gv_{a}}{\sqrt 2}\\
\lambda_{S}v_{u} & \lambda_{T}v_{u} & \mu_{u}+\lambda_{S}v_{S}
+\lambda_{T}v_{T} & 0\\
M_{R} & 0 & -fv_{a} & -fv_{u}
\end{array} \right).
\label{neutrino-neutralino-matrix}
\end{eqnarray}
%%%%%%%%%%%%%%%%%%%%%%%%%%
The Dirac neutralino mass matrix can be diagonalised by a bi-unitary transformation
involving two unitary matrices $V^N$ and $U^N$. The resulting four Dirac mass
eigenstates are ${\widetilde \chi}^{0+}_i \equiv \left( \begin{array}{c} 
{\widetilde \psi}^{0+}_i \\ \overline {{\widetilde \psi}^{0-}_i}\end{array} \right)$,
with $i=1, 2, 3, 4$ and ${\widetilde \psi}^{0+}_i = V^N_{ij} \psi^{0+}_j$, 
${\widetilde \psi}^{0-}_i = U^N_{ij} \psi^{0-}_j$.
With certain simplifying assumptions \cite{SC-1} and with
\begin{eqnarray}
\lambda_T &=& \lambda_S \tan\theta_W, \nonumber \\
M_R &=& \frac{\sqrt 2 f M_1^D \tan\beta}{g\tan\theta_W},
\label{rightmass} 
\end{eqnarray}
the expression for the mass of the lightest neutralino
state $(\widetilde\chi^0_1)$, i.e., the Dirac neutrino reduces to~\cite{SC-1} 
%%%%%%%%%%%%%%%%%%%%%%%%%
\begin{eqnarray}
m^D_{\nu_e}&=& \frac{v^3 f g \sin\beta}{\sqrt 2 (\mu_u + \lambda_S v_S 
+ \lambda_T v_T)}\lambda_T \frac{(M_2^D-M_1^D)}{M_1^D M_2^D}.
\label{nu-dirac-mass}
\end{eqnarray}
%%%%%%%%%%%%%%%%%%%%%%%%%
Note that only $\nu_e$ acquires a mass at the tree level since only the
electron type sneutrino gets a non-zero $vev$. By choosing nearly degenerate Dirac 
masses for the electroweak gauginos, i.e., ($M_2^D-M_1^D\simeq0.1$ GeV) one can 
find from equation~(\ref{nu-dirac-mass}) that the Dirac neutrino mass can be in the 
right ballpark of 0.1 eV even when $f\sim\mathcal O(1)$ and assuming $M_1^D, 
M_2^D$ and $\mu_u$ to be close to a few hundred GeV. The requirement of such a 
degeneracy between the Dirac masses of the electroweak gauginos, however, could be 
relaxed if one can consider an appropriately small  $\lambda_T\sim 10^{-6}$~\cite{SC-1}.  
However, a small $\lambda_T$ is not so interesting for our purpose since this results 
in a diminished contribution to $m_h$ and thus, brings back the scenario with 
multi-TeV top squarks to have the Higgs boson mass at the right ballpark.
At colliders, a direct attempt to probe this connection would inevitably
involve the heavy right handed neutrino ($N_R$) coupling to an active neutrino
($\nu_e$) and the SM-like Higgs boson. However, while a larger value of `$f$'
enhances this coupling, this also pushes up the mass of the right handed
neutrino, as conspicuous from eq.~(\ref{rightmass}). We find that it is rather 
difficult to obtain $m_{N_R}\leq 1.5$ TeV in a consistent manner while keeping
$f\sim\mathcal O(1)$. An immediate probe to this at the LHC can be the electroweak
process $pp\rightarrow N_R e^+ h$ followed by $N_R$ decaying to $h\nu_e$
thus giving rise to a final state $\nu_e e^+ h h$. Our preliminary study
reveals that the corresponding rate could barely reach an attobarn level
at the 13 TeV run of the LHC and thus one needs to wait for a very high integrated
luminosity. A detailed analysis in this context is beyond the scope of this paper
and we postpone it for a future work~\cite{Future}.
%%%%%%%%%%%%%%%%%%%%%%%%%%%%%%%%%%%%%%%%%%%%%%%%%%%%%%%%%%
\subsubsection{The $R$-breaking case}
%%%%%%%%%%%%%%%%%%%%%%%%%%%%%%%%%%%%%%%%%%%%%%%%%%%%%%%%%%
In the context of a supergravity theory broken spontaneously in the hidden sector, 
we consider $R$-symmetry to be broken by a non-zero gravitino mass. The 
$R$-breaking information has to be communicated from the hidden sector to the 
visible sector. We choose anomaly mediated supersymmetry breaking (AMSB) to play 
the role of a messenger~\cite{Bertuzzo,SC-1}. The $R$-breaking Lagrangian contains 
the following terms
%%%%%%%%%%%%%%%%%%%%%%%%%%%%%%%%%%%%%%%
\begin{eqnarray}
\mathcal{L}_{\cancel{R}}&=& M_{1}\widetilde b^0 \widetilde b^0 +M_{2}\widetilde
w^0 \widetilde w^0 +M_{3}\widetilde g\widetilde g +
\sum_{b=2,3} A^l_b {\tilde L}_a {\tilde L}_b {\tilde E^{c}}_b
+ \sum_{k=1,2,3} A^d_k {\tilde L}_a
{\tilde Q}_k {\tilde D^{c}}_k \nonumber \\
&+& \sum_{k=1,2,3} \dfrac{1}{2}
A^{\lambda}_{23k}{\tilde L}_2 {\tilde L}_3
{\tilde E^{c}}_k 
+ \sum_{j,k=1,2,3;b=2,3} A^{{\lambda^\prime}}_{bjk}{\tilde L}
_b {\tilde Q}_j {\tilde D^{c}}_k
+A^{\nu}H_{u}\tilde L_{a}\tilde N^{c}
+H_{u} \tilde Q A^u \tilde U^{c},\nonumber \\
\end{eqnarray}
%%%%%%%%%%%%%%%%%%%%%%%%%%%%%%%%%%%%%%%
where, $M_1$, $M_2$ and $M_3$ are the Majorana masses corresponding to $U(1)$,
$SU(2)$ and $SU(3)$ gauginos, respectively. The `$A$' terms are the standard 
trilinear scalar couplings. For the rest of this work, we will consider the 
$R$-breaking effects to be small, parametrised in terms of the gravitino mass, 
considered to be roughly around $\sim 20$ GeV. 

The Majorana neutralino mass matrix containing $R$-breaking effects can be written
in the basis $\psi^0 = (\tilde b^0, \tilde S, \tilde w^0, \tilde T,
\tilde R_d, \tilde H_u^0, N^c, \nu_e)^T$ as 
\begin{eqnarray}
{\cal L}^{\rm mass}_{{\tilde \chi}^0} = \dfrac{1}{2} (\psi^0)^T M_{\chi}^{M}
\psi^0 + h.c.
\end{eqnarray}
%%%%%%%%%%%%%%%%%%%%%%%%%%%%%%%%%%%%%%%
The symmetric $8\times 8$ neutralino mass matrix $M_{\chi}^M$ is given by
%%%%%%%%%%%%%%%%%%%%%%%%%%%%%%%%%%%%%%%
\begin{eqnarray}
M_{\chi}^{M}=\left(
\begin{array}{cccccccc}
M_{1} & M_{1}^{D} & 0 & 0 & 0 & \frac{g^{\prime}v_{u}}{\sqrt 2} & 0
& -\frac{g^{\prime}v_{a}}{\sqrt 2}\\
M_{1}^{D} & 0 & 0 & 0 & \lambda_{S}v_{u} & 0 & M_{R} & 0\\
0 & 0 & M_{2} & M_{2}^{D} & 0 & -\frac{g v_{u}}{\sqrt 2} & 0 &
\frac{g v_{a}}{\sqrt 2}\\
0 & 0 & M_{2}^{D} & 0 & \lambda_{T}v_{u} & 0 & 0 & 0 \\
0 & \lambda_{S}v_{u} & 0 & \lambda_{T}v_{u} & 0 & \mu_{u}+
\lambda_{S}v_{S}+\lambda_{T}v_{T} & 0 & 0\\
\frac{g^{\prime}v_{u}}{\sqrt 2} & 0 & -\frac{gv_{u}}{\sqrt 2}& 0 &
\mu_{u}+\lambda_{S}v_{S}+\lambda_{T}v_{T} & 0 & -fv_{a} & 0\\
0 & M_{R} & 0 & 0 & 0 & -fv_{a} & 0 & -fv_{u} \\
-\frac{g^{\prime}v_{a}}{\sqrt 2}& 0 & \frac{g v_{a}}{\sqrt 2} & 0
& 0 & 0 & -fv_{u} & 0
\end{array} \right).\nonumber \\
\label{majorana-neutralino}
\end{eqnarray}
%%%%%%%%%%%%%%%%%%%%%%%%%%%%%%%%%%%%%%%
This can be diagonalised by a unitary transformation 
\begin{eqnarray}
N^\star M_{\chi}^{M} N^\dagger = (M_{\chi})_{\rm diag}.
\end{eqnarray}
We define the two-component mass eigenstates as,
\begin{eqnarray}
\chi^0_i = N_{ij} \psi^0_j, ~~~~~~i,j = 1,...,8.
\end{eqnarray}
Finally the four-component Majorana spinors in terms of the two-component states 
are defined as
\begin{eqnarray}
{\tilde \chi}^0_i = \left(
\begin{array}{c}
\chi^0_i \\
{\bar \chi}^0_i
\end{array} \right), ~~~~~~i = 1,...8.
\end{eqnarray}
The lightest state ($\widetilde \chi_1^0$), which is the neutrino, 
becomes a Majorana particle. Similar to the Dirac case, the lightest 
eigenvalue of this Majorana neutralino mass matrix resembles  
the mass of the active neutrino. Other two active neutrinos remain 
massless at this stage.  
Using the relationships between $\lambda_S$ and $\lambda_T$
as well as $M_R$ and $f$ as shown in section \ref{R-conserving},
the mass of the active neutrino can be expressed as~\cite{SC-1} 
%%%%%%%%%%%%%%%%%%%%%%%%%%%%%%%%%%%%%%
\begin{eqnarray}
(m_\nu)_{\rm Tree} &=&-v^{2}\frac{\left[g \lambda_{T} v^{2}
(M_{2}^{D}-M_{1}^{D})\sin\beta\right]^{2}}{\left[M_{1}\alpha^{2}
+M_{2}\delta^{2}\right]},
\label{neutrino_majorana}
\end{eqnarray}
where
\begin{eqnarray}
\alpha&=&\frac{2 M_{1}^{D} M_{2}^{D}(\mu_u+\lambda_S v_S+\lambda_T v_T)
\tan\beta}{g\tan\theta_{w}}
+\sqrt 2  v^{2}\lambda_{S}\tan\beta(M_{1}^{D}\sin^{2}\beta+
M_{2}^{D}\cos^{2}\beta),\nonumber \\
\delta&=&\sqrt2  M_{1}^{D}v^{2}\lambda_{T}\tan\beta.
\end{eqnarray}
%%%%%%%%%%%%%%%%%%%%%%%%%%%%%%%%%%%%%%
It is noteworthy that the parameter $\lambda_T$ not only gives a tree level
mass to the neutrinos but also help lift the Higgs mass through quartic terms 
generated at one loop level. For example, we have observed that the tree level
Majorana mass of the active neutrino varies from 0.09 eV to 0.23 eV when $\lambda_T$
varies in the range 0.8-1.3, used in Fig.~\ref{fig:stop}, which gives correct SM-like
Higgs mass ($\sim$ 125 GeV) when top squarks are also not too heavy. Also note that the neutrino
Majorana mass given in eq.~(\ref{neutrino_majorana}) does not depend on the neutrino Yukawa
coupling `$f$'. This is because the expression has the functional form $M_R/f$. Since $M_R\sim f$
therefore, the neutrino Majorana mass becomes devoid of `$f$'. Similar to the case of 
Dirac neutrino mass, an appropriately small Majorana mass of the active neutrino requires 
highly degenerate Dirac gaugino masses. Note that (equation~(\ref{neutrino_majorana})) 
the neutrino Majorana mass is independent of the neutrino Yukawa coupling `$f$'. This is 
in clear contrast to the Dirac case. However, by choosing appropriate values of other 
parameters one can obtain a light active neutrino ($\widetilde\chi_1^0$) with mass 0.1 eV. 
Nevertheless, some interesting observations can be made for various sizes of `$f$'. For 
example, $f\sim\mathcal O(10^{-4})$ gives a sterile neutrino with mass around a few keV 
\cite{SC-2} which can be accommodated as a dark matter candidate. On the other hand, 
for $f\sim \mathcal O(1)$, where we obtain a large tree level correction to the Higgs boson 
mass for low values of $\tan\beta$, a light bino-like neutralino ($\widetilde\chi_2^0$) with 
mass around a few hundred MeV (mass of this neutralino is mostly controlled by the $R$-breaking 
parameter $M_1$). This MeV neutralino LSP could decay to SM fermions via $R$-parity violating 
modes. The probable decay modes could be $\widetilde\chi_2^0\rightarrow q \bar{q} \nu, 
e^+e^-\nu, \nu\nu\nu, q \bar{q}^{\prime} e^-$, where $q, q^{\prime}$ are the SM light quark 
states from the first two generations. However, as these involve very small couplings thus 
resulting in small total decay widths, the decay lengths happen to be much larger than the 
collider dimension \cite{Romao}. As a result, the LSP neutralino contributes to MET signals.
Furthermore, the gravitino NLSP ($\sim 20$ GeV) would decay to a photon and the bino-like 
LSP neutralino. This affects the light element abundances which is strongly constrained 
observationally and results in an upper bound on the reheating temperature of the universe
$T_R \leq 10^8$ GeV~\cite{SSHD}. In our scenario, one of the active neutrinos acquire 
a mass at the tree level. In addition, there are one-loop contributions to the neutrino 
Majorana mass matrix. We observe, that in this model and for our benchmark points, the 
one-loop contributions involving the $b-\widetilde b$, $\tau-\widetilde\tau$ loop interfere 
destructively with the Higgs-neutralino loop resulting in a somewhat relaxed bound 
($\sim$ 20 GeV) on the gravitino mass. This cancellation occurs for $(m_{\nu})_{11}$ where 
the Higgs-neutralino loop is present. For other elements in the neutrino mass matrix we 
obtain a bound on the relevant RPV operators~\cite{SC-1}.
This cancellation was not considered previously ~\cite{Gregoire, Bertuzzo} and as a 
result a stronger bound on gravitino mass was obtained. We have also implemented the model 
in {\tt SARAH-4.4.1}, which performs a full one-loop correction to the neutralino mass matrix, 
and we have cross-checked with the spectrum file that gravitino mass in the ballpark of 
$\mathcal O(10)$ GeV is consistent with light neutrino masses and mixing. 
%It is important to note that ref.~\cite{Gregoire,
%Bertuzzo} for the first time gave an order of magnitude estimatation of the
%neutrino-mass generated at one loop because of anomaly mediated $R$-breaking.
%The Majorana neutrino mass is an artifact of $R$-breaking and the gravitino mass
%happens to be an order parameter, whose mass is constrained from neutrino masses
%and are roughly of the order of $\sim 100$ MeV. In our scenario, one of the
%active neutrino acquires a mass at the tree level. We also showed in ref~\cite{SC-1},
%that for our case the one loop contributions to the Majorana neutrino mass originating
%from the top-stop loop and neutralino-Higgs loops cancels destructively. This
%further relaxes the bound on the gravitino mass. In this work also, we have
%implemented the model in {\tt SARAH-4.4.1}, which performs a one loop correction
%to the whole neutralino mass matrix. In this scenario, we also observe that appropriate
%active neutrino mass can be obtained when gravitino mass is in the ballpark of $\sim 10$
%GeV or so.
%%%%%%%%%%%%%%%%%%%%%%%%%%%%%%%%%%%%%%%%%%%%%%%%%%%%%%%%%%%%%%%%%%%%%%%%%%%%%%%%%%%%
\subsection{The chargino sector}
\label{subsec:charginos}
%%%%%%%%%%%%%%%%%%%%%%%%%%%%%%%%%%%%%%%%%%%%%%%%%%%%%%%%%%%%%%%%%%%%%%%%%%%%%%%%%%%%
Just like the neutrino and the neutralinos would mix, the charged lepton 
mixes\footnote{The mixing between the charged leptons and the charginos gives rise to 
deviation of the $Z$ to charged lepton couplings. Such a deviation is very much constrained 
from the electroweak precision measurements leading to a lower bound on $\tan\beta \geq 
2.7$~\cite{Gregoire,Kumar-1}. It is also pertinent to mention that an upper bound on $\tan\beta$ 
comes from $\tau$ Yukawa coupling contributing to the ratio $R_{\tau}\equiv \Gamma 
(\tau\rightarrow e \bar\nu_e \nu_{\tau})/\Gamma (\tau\rightarrow \mu\bar\nu_{\mu}
\nu_{\tau})$~\cite{Gregoire,Kumar-1}. Choosing $m_{\widetilde\tau_R}$ to be around 1 TeV 
corresponds to $\tan\beta \leq 70$.} with the charginos. This results in an extended chargino 
mass matrix~\cite{SC-3} compared to the MSSM. The relevant Lagrangian after $R$-breaking 
consists of the following terms
\begin{eqnarray}
\mathcal L_{ch} &=& M_2 \widetilde w^+ \widetilde w^- + \left(M_2^D-g 
v_T\right)~\widetilde T_d^- \widetilde w^+ +\sqrt 2 \lambda_T v_u 
\widetilde T_u^+ \widetilde R_d^- +g v_u \widetilde H_u^+\widetilde 
w^- -\mu_u \widetilde H_u^+ \widetilde R_d^- \nonumber \\
&+&\lambda_T v_T \widetilde H_u^+ \widetilde R_d^- 
- \lambda_S v_S \widetilde H_u^+ \widetilde R_d^- +g v_a \widetilde w^+ e_L^-
+ \left(M_2^D+g v_T\right)~\widetilde T_u^+ \widetilde w^- + m_e e_R^c e_L^- 
+ h.c. \nonumber \\
\end{eqnarray}
%%%%%%%%%%%%%%%%%%%%%%%%%%%%%%%%%%%%%%%%%%%%%%%%%%%%%%%%%%%%%%%%%%%%%%%%%%%%%%%%%%%%
The $4 \times 4$ chargino mass matrix written in the basis $\psi_i^{+}=(\widetilde w^+, 
\widetilde T_u^+, \widetilde H_u^+, e^c_R)^T$ and $\psi_i^{-}=(\widetilde w^-, \widetilde T_d^-, 
\widetilde R_d^-, e_L^-)^T$ is given by
%%%%%%%%%%%%%%%%%%%%%%%%%%%%%%%%%%%%%%%%%%%%%%%%%%%%%%%%%%%%%%%%%%%%%%%%%%%%%%%%%%%%
\begin{eqnarray}
M_{c}=\left(
\begin{array}{cccc}
M_{2} & M_{2}^{D}-g v_T & 0 & gv_a \\
M_{2}^{D}+g v_T & 0 & \sqrt 2 v_u \lambda_T & 0 \\
g v_u & 0 & -\mu_u-\lambda_S v_S+\lambda_T v_T & 0\\
0 & 0 & 0 & m_e \\
\end{array} \right).
\label{chargino}
\end{eqnarray}
%%%%%%%%%%%%%%%%%%%%%%%%%%%%%%%%%%%%%%%%%%%%%%%%%%%%%%%%%%%%%%%%%%%%%%%%%%%%%%%%%%%%
Again, this can be diagonalised by a bi-unitary transformation $U M_c V^T = M_D^{\pm}$.
The chargino mass eigenstates (two-component) are written in terms of the gauge eigenstates 
in a compact form as
%%%%%%%%%%%%%%%%%%%%%%%%%%%%%%%
\begin{eqnarray}
\chi_i^- &=& U_{ij}\psi_j^-, \nonumber \\
\chi_i^+ &=& V_{ij}\psi_j^+.
\end{eqnarray}
%%%%%%%%%%%%%%%%%%%%%%%%%%%%%%%
The four-component Dirac spinors written in terms of the two-component
spinors take the form
%%%%%%%%%%%%%%%%%%%%%%%%%%%%%%%
\begin{eqnarray}
{\widetilde \chi}_i^+ = \left(
\begin{array}{c}
\chi_i^+ \\
\overline \chi^-_i 
\end{array}\right), ~~~~~~~~~~~~~~~~~~(i = 1,...,4).
\end{eqnarray}
%%%%%%%%%%%%%%%%%%%%%%%%%%%%%%%
The lightest chargino $(\widetilde \chi_1^-)$ corresponds to the electron,
$\widetilde\chi_2^-$ is the lightest chargino state reminiscent of the lighter chargino 
in the MSSM with mass of ${\cal O}$(100) GeV. It is also pertinent to mention that 
$\psi_i^-$ and $\psi_i^+$ would also include $\mu_L^-$, $\tau_L^-$ and $\mu_R^c$, $\tau_R^c$, 
respectively. However, as discussed in section~\ref{model} only the electron type sneutrino 
acquires a $vev$, and the $vev$ of the other two sneutrinos can be rotated away without any 
loss of generality. Therefore, $\mu$ and $\tau$ do not mix with the chargino states.
%%%%%%%%%%%%%%%%%%%%%%%%%%%%%%%%%%%%%%%%%%%%%%%%%%%%%%%%%%%%%%%%%%%%%%%%%%%%%%%%%%%%
\section{The top squark sector}
\label{top}
%%%%%%%%%%%%%%%%%%%%%%%%%%%%%%%%%%%%%%%%%%%%%%%%%%%%%%%%%%%%%%%%%%%%%%%%%%%%%%%%%%%%
In this work we concentrate on the third generation squarks, mainly the top squarks,
which play important roles in lifting the Higgs boson mass. Scenarios with light top squarks 
draw their motivations from the `naturalness' argument. They also provide rich and
interesting collider signatures. As discussed in section~\ref{scalar}, the model which 
we consider here gives us the opportunity to study such light top squarks. Furthermore, 
$R$-symmetry prohibits any trilinear scalar couplings (the `$A$' terms) and 
Higgsino mass parameter (the $\mu$ term). Therefore, we investigate a situation where
both the top squarks are light ($\sim$500 GeV) and have negligible chiral mixing,
which originates from small $R$-breaking. 

The relevant terms in the top squark mass matrix are generated from the $F$-term,
the $D$-term and the soft terms. The $SU(2)_L$ and $U(1)_Y$ contributions to the $D$-fields
are given by
%%%%%%%%%%%%%%%%%%%%%%%%%%%%%%%%%%%%%%%%%
\begin{eqnarray} 
D^a &=& g\Big[H_u^{\dagger}\tau^a H_u+\widetilde L_i^{\dagger}\tau^a \widetilde L_i
+\widetilde q_{iL}^{\dagger}\tau^a\widetilde q_{iL}+T^{\dagger}\lambda^a T\Big]
+\sqrt 2 \Big[M_2^D T^a + M_2^D T^{a\dagger}\Big]\nonumber \\
D^Y &=& -\frac{1}{2}g^{\prime}\Big[H_u^{\dagger}H_u-\widetilde L_i^{\dagger}
\widetilde L_i+2\widetilde e_{iR}^{*}e_{iR}+\frac{1}{3}\widetilde q_{iL}^{\dagger}
\widetilde q_{iL}-\frac{4}{3}\widetilde u_{iR}^{\dagger}u_{iR}+\frac{2}{3}
\widetilde d_{iR}^{\dagger}\widetilde d_{iR}\Big] \nonumber \\
&-&\sqrt 2 M_1^D\Big[S+S^{\dagger}\Big].
\label{D-term}
\end{eqnarray}
%%%%%%%%%%%%%%%%%%%%%%%%%%%%%%%%%%%%%%%%%
The $\tau$ and $\lambda$ matrices are the generators of $SU(2)_L$ group in the
fundamental and adjoint representation, respectively. From equation~(\ref{D-term}) it 
is straightforward to calculate the elements of the mass-squared matrix in the top 
squark sector, which in the basis $(\widetilde t_{L}, \widetilde t_{R})$ turn out to be
%%%%%%%%%%%%%%%%%%%%%%%%%%%%%%%%%%%%%%%%
\begin{eqnarray}
(M_{\tilde t}^2)_{11}&=& m^2_{\tilde Q_3} + m_t^2 + m_Z^2 \cos 2\beta 
~\bigg(\frac{1}{2}-\frac{2}{3}\sin^2\theta_W\bigg)+\frac{\sqrt 2}{3}v_S M_1^D+
\sqrt 2 g M_2^D v_T, \nonumber \\
(M_{\tilde t}^2)_{12}&=& (M_{\tilde t}^2)_{21} = 0, \nonumber \\
(M_{\tilde t}^2)_{22}&=& m^2_{\tilde u_3} + m_t^2 +\frac{2}{3} m_Z^2
\sin^2 \theta_W \cos 2\beta - \frac{4\sqrt 2}{3}g^{\prime} M_1^D v_S.
\end{eqnarray}
%%%%%%%%%%%%%%%%%%%%%%%%%%%%%%%%%%%%%%%%
Note that in absence of `$A$' terms and the $\mu$ term, the off-diagonal entries
vanish and hence the top squark sector is devoid of any chiral mixing.
Thus, the left- and right-chiral states are equivalent to the
mass eigenstates. Such a `zero' mixing situation can be contrasted with the
MSSM, in which, a substantial mixing is generally required to obtain  the
observed value of the Higgs mass.
Motivated by the recent collider bounds on the masses of the top
squarks (depending on its various decay modes), to be justified
in some detail in section \ref{bounds}, we
choose $\widetilde t_1 \approx \widetilde t_R$ and $\widetilde t_2\approx
\widetilde t_L$.
%%%%%%%%%%%%%%%%%%%%%%%%%%%%%%%%%%%%%%%%

At the LHC, top squarks are being searched in their direct production,
$p p \rightarrow \tilde t\tilde t^*$, followed by their subsequent decays 
in various possible modes. Out of these, the decays that are relevant 
to our scenario~\cite{LHC-stop,LHC-s1,LHC-s2,LHC-s3} 
are the $R$-parity conserving ones
%%%%%%%%%%%%%%%%%%%%%%%%%%%%%%%%%%%%%%%%%%%%%%%%%%%%%
\begin{eqnarray}
\widetilde t\rightarrow b\widetilde\chi^+ \qquad {\mathrm {and}} \qquad
\widetilde t\rightarrow t\widetilde\chi^0,
\end{eqnarray}
and the modes that violate $R$-parity when a top squark could decay to a bottom 
quark and a charged lepton
\cite{ATLAS-RPV,CMS-RPV}
\begin{eqnarray}
\widetilde t\rightarrow b\ell^+.
\end{eqnarray}
%%%%%%%%%%%%%%%%%%%%%%%%%%%%%%%%%%%%%%%%%%%%%%%%%%%%%
These channels are of major relevance in the context of our model. The reasons 
are twofold: first, in the large `$f$' scenario, we obtain a light bino-like 
neutralino with mass around a few hundred MeV, in addition to an active neutrino. 
Therefore, the channels with top squark decaying to a top quark and a bino-like 
neutralino and/or an active neutrino open up. Secondly, top squark decaying to a bottom quark and 
a chargino is also important. Additionally, 
top squark decaying to a bottom quark and an electron becomes an interesting 
channel to look for. This decay mode is predominantly controlled by the $R$-parity violating 
operator $\lambda^{\prime}_{133}$. In the framework of the MSSM with $R$-parity 
violation, strong limit on this particular coupling exists from the neutrino Majorana
mass, $|\lambda^{\prime}_{133}|\sim 3.4\times 10^{-3}\sqrt{\frac{m_b}
{m_{\widetilde b}}}$~\cite{Barbier}. Hence, in such a scenario, the resulting decay 
rate becomes highly suppressed. However, in the present context, 
$\lambda^{\prime}_{133}$ is identified with the bottom Yukawa coupling $y_b$. 
The smallness of the neutrino mass is then explained through 
small $R$-breaking effect, parametrised in terms of a small gravitino mass~\cite{SC-1}.
Thus, a large decay rate for $\widetilde t\rightarrow b e^+$ becomes a 
generic feature in our model. In principle, $\widetilde t$ could also
decay to $b\mu^{+}$ and $b\tau^{+}$ via RPV couplings $\lambda^{\prime}_{233}$
and $\lambda^{\prime}_{333}$. However, these decays are subdominant compared
to $\widetilde t\rightarrow b e^+$ because of the stringent constraints
on the relevant couplings as discussed later. We note in passing that 
$\widetilde t\rightarrow t\widetilde G$ is also a possibility but highly 
suppressed~\cite{Ambrosanio} for a gravitino of mass $\sim 20$ GeV in the present context.

The relevant Lagrangians are worked out in the four component notation following 
\cite{Gunion,Low} and are given by
%%%%%%%%%%%%%%%%%%%%%%%%%%%%%%%%%%%%%%%%%%%%%%%%%%%%%
\begin{eqnarray}
\mathcal L_{\widetilde t t\widetilde \chi_0} &=& -\bar t\Bigg[y_t P_L N_{i6}
+\frac{1}{\sqrt 2}\Bigg\{g P_R N_{i3}
+\frac{g^{\prime}}{3}P_R N_{i1}\Bigg\}\Bigg]\widetilde t_L\widetilde\chi_i^0
+\bar t\Bigg[\frac{4g^{\prime}}{3\sqrt 2}P_L N_{i1}-y_t P_R N_{i6}\Bigg]
\widetilde t_R\widetilde\chi_i^0 \nonumber \\
&+& h.c.,
\label{stop-top-neutralino}
\end{eqnarray}
and
\begin{eqnarray}
\mathcal L_{\widetilde t b \widetilde \chi^+} &=& \bar b\Bigg[-g P_L U_{i1}
\Bigg]\widetilde t_L\widetilde \chi_i^c + \bar b\Bigg[
y_t P_R V_{i3}\Bigg]\widetilde t_R \widetilde \chi_i^c +\lambda^{\prime}_{133} \widetilde t_L
P_L U_{i4} \widetilde\chi_i^c \bar b + h.c.,
\end{eqnarray}
%%%%%%%%%%%%%%%%%%%%%%%%%%%%%%%%%%%%%%%%%%%%%%%%%%%%%
where, $\lambda^{\prime}_{133}=y_b=\frac{m_b}
{v\cos\beta}$, the bottom Yukawa coupling and $y_t=\frac{m_t}{v\sin\beta}$ is the
top Yukawa coupling and $m_t$ and $m_b$ are the top and the bottom quark masses,
respectively. The neutralino and the chargino mixing matrices $N_{ij}$, $U_{ij}$
and $V_{ij}$ are as defined earlier. Note that, for $i=1$ (corresponding to 
$\widetilde\chi_i^c\equiv e^-$) the mixing matrix elements $U_{11}$ and $V_{13}$
are suppressed.
In the following subsections we briefly
discuss the salient decay modes of the lighter ($\widetilde t_1\approx \widetilde t_R$)
and the heavier ($\widetilde t_2\approx\widetilde t_L$) top squarks.
%%%%%%%%%%%%%%%%%%%%%%%%%%%%%%%%%%%%%%%%%%%%%%%%%%%%%
\subsection{Decay rates of $\widetilde t_1~(\approx \widetilde t_R)$}
%%%%%%%%%%%%%%%%%%%%%%%%%%%%%%%%%%%%%%%%%%%%%%%%%%%%%
The partial decay widths of $\widetilde t_1$ in the $t\widetilde\chi_i^0$
and $b\widetilde\chi_i^+$ modes are given by
\begin{eqnarray}
\Gamma (\widetilde t_1\rightarrow t\widetilde\chi_i^0) &=& \frac{1}{16\pi 
m_{\widetilde t_1}^3}\Bigg[\bigg(\eta^{2}_{Ri}+\zeta^2_{Ri}\bigg)\bigg(m^2_{\widetilde t_1}-
m_t^2-m^2_{\widetilde \chi_i^0}\bigg)-4 \eta_{Ri}\zeta_{Ri}
m_t m_{\widetilde\chi_i^0}\Bigg]\nonumber \\
&&\times \bigg[m^4_{\widetilde t_1}+m^4_{\widetilde\chi_i^0}+m^4_t-2 m_{\widetilde t_1}^2
m_t^2 - 2 m_{\widetilde\chi_i^0}^2 m_t^2 - 2 m_{\widetilde \chi_i^0}^2
m_{\widetilde t_1}^2\bigg]^{\frac{1}{2}},
\end{eqnarray}
and 
\begin{eqnarray}
\Gamma (\widetilde t_1\rightarrow b\widetilde\chi_i^+) &=& \frac{1}{16\pi 
m_{\widetilde t_1}^3}\Bigg[\bigg(\alpha^{2}_{Ri}+\beta^2_{Ri}\bigg)\bigg(m^2_{\widetilde t_1}-
m_b^2-m^2_{\widetilde \chi_i^+}\bigg)-4 \alpha_{Ri}\beta_{Ri}
m_b m_{\widetilde\chi_i^+}\Bigg]\nonumber \\
&&\times \bigg[m^4_{\widetilde t_1}+m^4_{\widetilde\chi_i^+}+m^4_b-2 m_{\widetilde t_1}^2
m_b^2 - 2 m_{\widetilde\chi_i^+}^2 m_b^2 - 2 m_{\widetilde \chi_i^+}^2
m_{\widetilde t_1}^2\bigg]^{\frac{1}{2}},
\end{eqnarray}
%%%%%%%%%%%%%%%%%%%%%%%%%%%%%%%%%%%%%%%%%%%%%%%%%%%%%
where, 
\begin{eqnarray}
\eta_{Ri} &=& \frac{4 g^{\prime}}{3\sqrt 2} N_{i1}, \nonumber \\
\zeta_{Ri} &=& y_t N_{i6}, \nonumber \\
\alpha_{Ri} &=& 0, \nonumber \\
\beta_{Ri} &=& y_t V_{i3}.
\label{couplings}
\end{eqnarray}
%%%%%%%%%%%%%%%%%%%%%%%%%%%%%%%%%%%%%%%%%%%%%%%%%%%%%
We note down a few important observations below.
\vskip -0.15cm
%%%%%%%%%%%%%%%%%%%%%%%%%%%%%%%%%%%%%%%%%%%%%%%%%%%%%
\begin{itemize}
\item In the large `$f$' case we obtain a light ($\sim$ few hundred MeV, governed by 
the $R$-breaking Majorana mass $M_1$) bino-like neutralino ($\widetilde\chi_2^0$).
This is because of the presence of $M_R N^c \widetilde S$ term in the Lagrangian, where 
the coefficient $M_R$ becomes very large ($\sim 10^5$ GeV) for an order one `$f$'. This
results in forming a heavy pseudo-Dirac pair with mass $\sim M_R$ and makes the
lightest eigenvalue very small and predominantly bino-like. 
\item The Dirac wino mass $M_2^D$ 
is considered to be heavy to evade bounds from $Z$ boson coupling to electrons
~\cite{Kumar-1} (see also section \ref{subsec:charginos}). 
The $\mu_u$ parameter, which controls the mass of the Higgsino 
(both neutral and charged) can vary between the electroweak scale ($\sim 200$ GeV) 
and a much larger value, i.e., a few TeV.
\item Based on the above discussion and with the help of 
equations~(\ref{stop-top-neutralino})-(\ref{couplings}), we find that $\widetilde t_1$ 
would decay into $t\widetilde\chi_{2,3,4}^0$ and $b\widetilde\chi_2^+$. The neutralino 
can be both bino- or Higgsino-like whereas the chargino would only be Higgsino-like, 
assuming the Higgsino mass parameter 
$\mu_u (< m_{\widetilde t_1}) <<M_2^D$.  
\item We expect the dominant decay modes of $\widetilde t_1$ to have 
the Higgsino-like neutralinos or chargino as the decay products
rather than the bino-like neutralino. 
This is because of the enlarged couplings for the former which are proportional
to the top Yukawa coupling, $y_t$.
\item Finally, in the limit when $\mu_u > m_{\widetilde t_1}$, the top squark
cannot decay to an on-shell top quark and a Higgsino-like neutralino or a bottom
quark and a Higgsino-like chargino due to phase space constraints. Therefore, 
the dominant channel would only be $\widetilde t_1\rightarrow t\widetilde 
\chi_2^0$, where $\widetilde\chi_2^0$ is the bino-like MeV neutralino.
Moreover, $\widetilde t_1\rightarrow
t\nu_e$ would also contribute to MET, although the branching
is suppressed due to the small neutralino-neutrino mixing.
\end{itemize}
%%%%%%%%%%%%%%%%%%%%%%%%%%%%%%%%%%%%%%%%%%%%%%%%%%%%%%%%%%%%%%%%%%%%%%%%%%%%%%%%%%%%
\subsection{Decay rates of $\widetilde t_2~(\approx\widetilde t_L)$}
\label{subsec:stop2-decay}
%%%%%%%%%%%%%%%%%%%%%%%%%%%%%%%%%%%%%%%%%%%%%%%%%%%%%%%%%%%%%%%%%%%%%%%%%%%%%%%%%%%%
The partial decay widths of $\widetilde t_2$ are given by
%%%%%%%%%%%%%%%%%%%%%%%%%%%%%%%%%%%%%%%%%%%%%%%%%
\begin{eqnarray}
\Gamma (\widetilde t_2\rightarrow t\widetilde\chi_i^0) &=& \frac{1}{16\pi 
m_{\widetilde t_2}^3}\Bigg[\bigg(\eta^{2}_{Li}+\zeta^2_{Li}\bigg)\bigg(m^2_{\widetilde t_2}-
m_t^2-m^2_{\widetilde \chi_i^0}\bigg)-4 \eta_{Li}\zeta_{Li}
m_t m_{\widetilde\chi_i^0}\Bigg]\nonumber \\
&&\times \bigg[m^4_{\widetilde t_2}+m^4_{\widetilde\chi_i^0}+m^4_t-2 m_{\widetilde t_2}^2
m_t^2 - 2 m_{\widetilde\chi_i^0}^2 m_t^2 - 2 m_{\widetilde \chi_i^0}^2
m_{\widetilde t_2}^2\bigg]^{\frac{1}{2}},
\end{eqnarray} 
and
\begin{eqnarray}
\Gamma (\widetilde t_2\rightarrow b\widetilde\chi_i^+) &=& \frac{1}{16\pi 
m_{\widetilde t_2}^3}\Bigg[\bigg(\alpha^{2}_{Li}+\beta^2_{Li}\bigg)\bigg(m^2_{\widetilde t_2}-
m_b^2-m^2_{\widetilde \chi_i^+}\bigg)-4 \alpha_{Li}\beta_{Li}
m_b m_{\widetilde\chi_i^+}\Bigg]\nonumber \\
&&\times \bigg[m^4_{\widetilde t_2}+m^4_{\widetilde\chi_i^+}+m^4_b-2 m_{\widetilde t_2}^2
m_b^2 - 2 m_{\widetilde\chi_i^+}^2 m_b^2 - 2 m_{\widetilde \chi_i^+}^2
m_{\widetilde t_2}^2\bigg]^{\frac{1}{2}},
\end{eqnarray} 

%%%%%%%%%%%%%%%%%%%%%%%%%%%%%%%%%%%%%%%%%%%%%%%%%%%%%
where, 
\begin{eqnarray}
\eta_{Li} &=& y_t N_{i6}, \nonumber \\
\zeta_{Li} &=& \frac{1}{\sqrt 2} \Bigg(g N_{i3} + \frac{g^{\prime}}{3} N_{i1}\Bigg), \nonumber \\
\alpha_{Li} &=& - g U_{i1}, \nonumber \\
\beta_{Li} &=& 0.
\label{stopL}
\end{eqnarray}
In addition, for $\widetilde t_2$ we also have the interesting possibility of 
$\widetilde t_2\rightarrow b e^+$. The corresponding partial decay width is given by
\begin{eqnarray}
\Gamma (\widetilde t_2 \rightarrow b e^+) = \frac{y_b^2 |U_{i4}|^2}{16\pi} m_{\widetilde t_2}.
\end{eqnarray}
%%%%%%%%%%%%%%%%%%%%%%%%%%%%%%%%%%%%%%%%%%%%%%%%%%%%%
Some features of $\widetilde t_2$ decays are as follows:
%%%%%%%%%%%%%%%%%%%%%%%%%%%%%%%%%%%%%%%%%%%%%%%%%%%%%
\begin{itemize}
\item The decay $\widetilde t_2 \to b e^+$ is an interesting possibility. 
This faces no suppression from the phase space and the decay rate is proportional 
to the bottom Yukawa coupling, $y_b$ which grows with $\tan\beta$. Hence a 
substantial branching fraction in this mode is expected at large $\tan\beta$ 
and for a fixed top squark mass.  
\item When $\mu_u < m_{\widetilde t_2}$, 
$\widetilde t_2$ would decay to Higgsino-like chargino and neutralinos. 
Also, decay to a bino-like neutralino is a possibility.
However, a quick look at the couplings in equation~(\ref{stopL}) would
suggest that the decay to Higgsino-like neutralinos 
($\widetilde\chi_3^0,\widetilde\chi_4^0$) is $\eta_{Li}\sim y_t$-enhanced 
and hence, is more probable than a decay to a bino-like neutralino (suppressed by
$g^{\prime}/3$ in the coupling) or to a Higgsino-like chargino 
(suppressed by $g$ times the wino component of the lighter chargino, $U_{i1}$).
\item Again, for $\mu_u >  m_{\widetilde t_2}$, decays of 
$\widetilde t_2$ to Higgsino-like neutralinos and charginos 
are kinematically barred.
Under such a circumstance, $\widetilde t_2$ mainly 
decays to a bottom quark and an electron (positron). 
The decay mode $\widetilde t_2 (\widetilde t_L)
\rightarrow t \widetilde\chi_2^0$ is again suppressed because of a (comparatively)
small involved coupling. Finally, $\alpha_{Ri}=\beta_{Li}=0$ reflects the absence 
of $\hat H_d$ in the Lagrangian, which has been integrated out from the theory.
\end{itemize}

%%%%%%%%%%%%%%%%%%%%%%%%%%%%%%%%%%%%%%%%%%%%%%%%%%%%%%%%%%%%%%%%%%%%%%%%%%%%%%%%%%%%
\subsection{Bounds on top squarks}
\label{bounds}
%%%%%%%%%%%%%%%%%%%%%%%%%%%%%%%%%%%%%%%%%%%%%%%%%%%%%%%%%%%%%%%%%%%%%%%%%%%%%%%%%%%%
\begin{itemize}
\item Recently ATLAS measured the spin correlation in the top-antitop quark events
and searched for top squark pair production~\cite{ATLAS-191} in the pp 
collisions at $\sqrt s = 8$ TeV centre of mass energy and integrated luminosity
($\cal L$) 
of 20.3 fb$^{-1}$. This particular search has ruled out top squarks with masses 
between the top quark mass and 191 GeV with 95$\%$ confidence level. 
A very recent study \cite{Aad:2015pfx} reveals that the window of
$197~\text{GeV} \lesssim m_{\widetilde t_1} \lesssim 205~\text{GeV}$, in the so-called 
`stealth' regime (i.e., with vanishing LSP mass), cannot yet be ruled out.
\item Dedicated searches for pair-produced top squarks decaying 100\% 
of the time to bottom quarks and lighter charginos have been performed 
\cite{ATLAS-470,CMS-stop} within the framework of the MSSM.
For a chargino with mass close to 200 GeV,
the top squark below 470 GeV has been ruled out at the $\sqrt s=8$ TeV 
run of the LHC. In our scenario, $\widetilde t_R$
decays to this particular channel if $\mu_u < m_{\widetilde t_R}$. Although
in our model the corresponding branching fraction is less than 100\%,
we take a conservative approach and respect this bound. In addition, this search 
gives the most relaxed bound on the mass
of the top squark which is relevant to our analysis. Hence in the present study
we choose $\widetilde t_R$ to be the lighter top squark, i.e., 
$\widetilde{t_1} \approx \widetilde{t_R}$.

\item Another decay mode of the top squark relevant to our scenario is
$\widetilde t\rightarrow t\widetilde\chi_1^0$, where $\widetilde\chi_1^0$
implies the lightest supersymmetric particle in the MSSM 
(in our work, however, $\widetilde\chi_1^0$
is identified with the active neutrino and $\widetilde\chi_2^0$ represents the 
lightest bino-like neutralino). At the 8 TeV run of the LHC, top squark with mass below
550 GeV is ruled out at 95$\%$ confidence level~\cite{ATLAS-470,CMS-stop} with
the assumption $\text{BR}~(\widetilde t_1\rightarrow t\widetilde\chi_1^0)=100\%$. 
This bound applies for a massless neutralino ($m_{\widetilde\chi_1^0}=0$). For heavier
neutralinos in the final states, this lower bound on the top squark
mass can be relaxed further. Note that in the large `$f$' scenario we 
find a light super-light bino-like neutralino with mass around a few hundred MeV
which thus attracts this bound on the mass of the lighter top squark. 
\item A top squark decaying via $R$-parity violating mode has also been
probed by the LHC experiments. If a top squark undergoes an $R$-parity violating
decay only to a bottom quark and an electron, a stringent lower bound 
\cite{ATLAS-RPV,CMS-RPV} exists
on the top squark mass with $m_{\widetilde t} > 900$ 
GeV.\footnote{Note however, that if $\widetilde t\rightarrow b\tau^+$ opens up, 
the corresponding
lower bound on the top squark mass can be relaxed.
The decay $\widetilde t\rightarrow b\tau^+$ is mostly controlled by the
$R$-parity violating coupling $\lambda^{\prime}_{333}$. 
The existing bound on this particular coupling is much relaxed: 
$\lambda^{\prime}_{333}<1.4\cos\beta$~\cite{Kumar-1} and can 
be saturated for small values of $\tan\beta \, (\leq 5)$.  
However, in the present scenario we confine ourselves in the limit where
$\tan\beta\gsim 20$, which renders the decay $\widetilde t\rightarrow b\tau^+$ 
insignificant. On the other hand, the decay $\widetilde t\rightarrow b\mu^+$
is also negligible because of the strong constraint $|\lambda^{\prime}_{233}|<6.8
\times 10^{-3}\cos\beta$~\cite{DF}.} 
Accommodating an even lighter top squark, which is
central to our present work, thus requires a situation where such a bound is 
preferentially applicable to the heavier top squark state ($\widetilde{t_2}$) 
of the scenario.
As described in section \ref{subsec:stop2-decay}, only $\widetilde t_L$ could decay 
to a bottom quark and an electron (positron).
Hence we choose the heavier top squark $\widetilde t_2$ to be the
$\widetilde t_L$, i.e., $\widetilde{t_2} \approx \widetilde{t_L}$.
Note again that our consideration is pretty conservative and, as we would find
in section \ref{benchmarks}, for generic scenarios where such a decay can have a branching 
fraction below 50\%, the bounds can get considerably weaker thus allowing for an
even lighter $\widetilde{t_2}$.
Phenomenological discussions on top squarks undergoing such an 
$R$-parity violating decay can be found in references 
\cite{Chun, Marshall, Sujoy,Datta, Datta:2006ak,Datta:2009dc,Bose:2014vea}.
\end{itemize}

The squarks from the third generation have understandably attracted a lot of attention
in the recent times. The flavor changing decay of the top squark, $\widetilde t_1\rightarrow
c\widetilde\chi_1^0$ has been analysed in great detail in
~\cite{Bedo,Grober,Beuria,Ferretti,Belangert}. Recent searches performed by both 
ATLAS and CMS collaborations
have looked into this channel extensively and ruled out top squark masses below
300 GeV~\cite{ATLAS-FV,CMS-FV,Aad:2015pfx}. Top squarks decaying to a top quark
along with a neutralino (LSP or NLSP) has also been probed in various SUSY models. 
A lower limit close to 1 TeV for the top squark mass can be obtained at $\sqrt{s}=$14~TeV 
and with the high luminosity option~\cite{Amit1,Amit2,Eckel}.
In addition, thorough phenomenological studies have also been performed in the decay
of top squark into a bottom quark and a chargino~\cite{Abe,Kim}.
%%%%%%%%%%%%%%%%%%%%%%%%%%%%%%%%%%%%%%%%%%%%%%%%%%%%%%%%%%%%%%%%%%%%%%%%%%%%%%%%%%%%

We note in passing that interesting final state signatures can be obtained
for the decays of bottom squarks as well~\cite{Han:2015tua}. For example 
$\widetilde b_L$ decays to a bottom quark and the bino-like neutralino with a 
branching ratio close to 71\% and
$\widetilde b_R$ decays to a top quark and an electron with a branching
close to 47\% in the BP-I scenario. Such a branching would imply
2~$b$-jets+$\cancel{E_T}$ or 2~$b$-jets+4~leptons+$\cancel{E}_T$ in the final states, 
respectively. The most stringent limit on the mass of the bottom squark comes from the 
search where it decays to a bottom quark and the lightest neutralino (LSP) 
with BR ($\widetilde b\rightarrow b\widetilde\chi_1^0$)=100\%.
Bottom squark mass up to 700 GeV has been excluded at 95\% confidence level
for neutralino mass less than 50 GeV~\cite{CMS-PAS}. We note that all the benchmark points
are chosen in a way such that they satisfy existing bounds on top squark mass under
various circumstances pertaining to its decay. 

%%%%%%%%%%%%%%%%%%%%%%%%%%%%%%%%%%%%%%%%%%%%%%%%%%%%%%%%%%%%%%%%%%%%%%%%%%%%%%%%%%%%
%%%%%%%%%%%%%%%%%%%%%%%%%%%%%%%%%%%%%%%%%%%%%%%%%%%%%%%%%%%%%%%%%%%%%%%%%%%%%%%%%%%%
\section{The benchmarks and the final states}
\label{benchmarks}
%%%%%%%%%%%%%%%%%%%%%%%%%%%%%%%%%%%%%%%%%%%%%%%%%%%%%%%%%%%%%%%%%%%%%%%%%%%%%%%%%%%%
In this section we discuss a few benchmark scenarios that would be broadly representative
of the phenomenology that is expected of the framework under consideration.
As mentioned earlier in section~\ref{scalar}, we embed the model in 
{\tt SARAH (v4.4.1)}~\cite{Staub,Staub1,Staub2}. We use the low energy output of 
{\tt SARAH (v4.4.1)} 
and generate the SUSY spectrum using {\tt SPheno (v3.3.3)}~\cite{Porod,Porod1}.  
{\tt FlavorKit}~\cite{Vicente} is used to ensure benchmark 
points are consistent with all relevant flavor violating constraints. Higgs boson
cross-sections and signal strengths are computed using 
{\tt HiggsBounds}~\cite{Higgsbounds,Higgsbounds1,Higgsbounds2,Higgsbounds3} and {\tt HiggsSignals}
~\cite{HiggsSignals1,HiggsSignals2}. 
As discussed earlier, we will mainly consider two regimes, viz., $\mu_u > m_{\widetilde t_{1,2}}$ 
and $\mu_u < m_{\widetilde t_{1,2}}$. For each case, we point out the dominant decay
modes of both $\widetilde t_1$ and $\widetilde t_2$. 
These dictate the types of interesting signatures at the LHC for each
of these cases.
%%%%%%%%%%%%%%%%%%%%%%%%%%%%%%%%%%%%%%%%%%%%%%%%%%%%%%%%%%%%%%%%%%%%%%%%%%%%%%%%%%%%
\subsection{\textbf{\textit{Case 1:}} $\mu_u < m_{\widetilde t_{1,2}}$}
\label{case1}
\begin{figure}[t!]
    \centering
    \begin{subfigure}[t]{0.5\textwidth}
        \centering
        \includegraphics[height=2.0in]{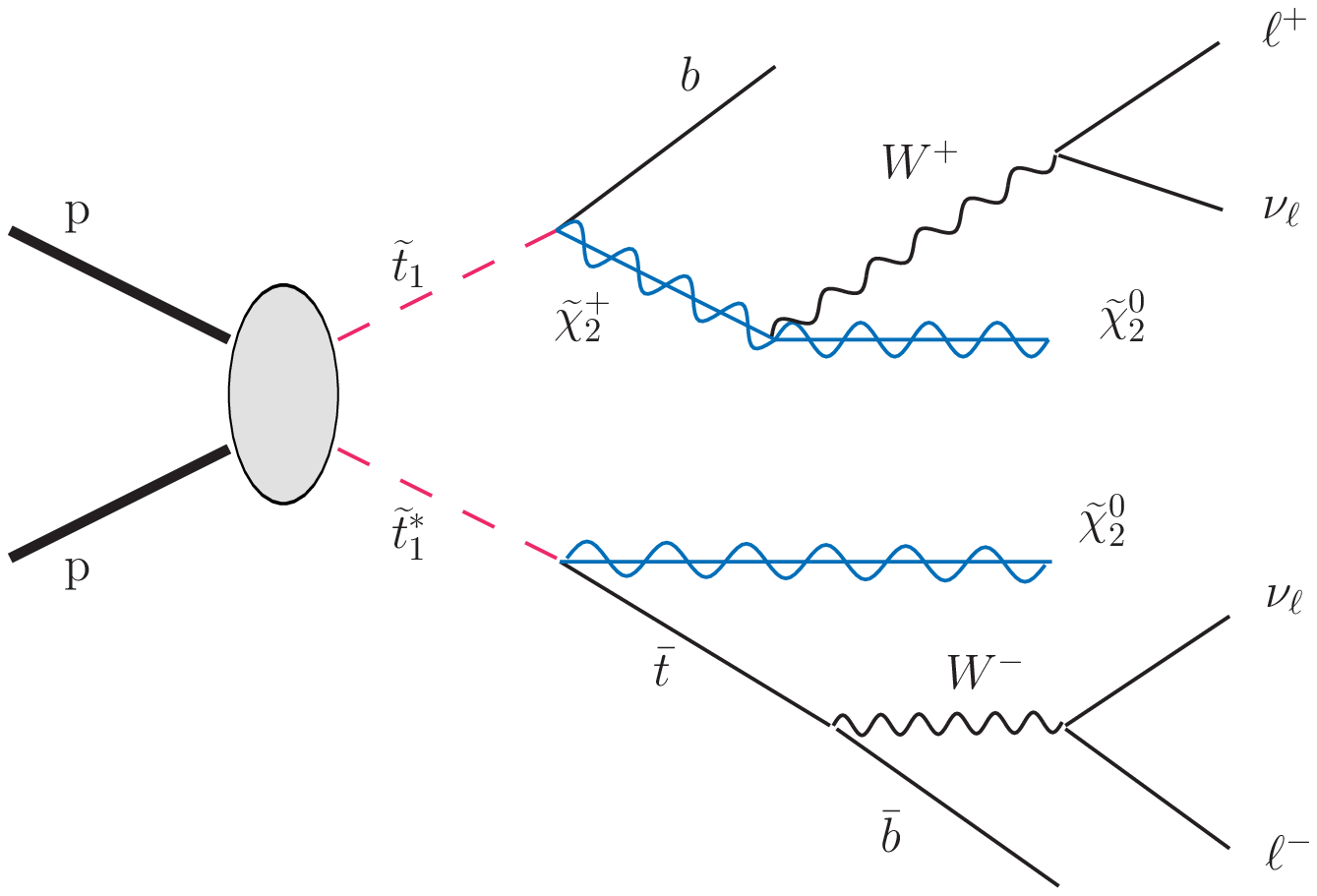}
        \caption{Asymmetric decays of the top squarks with one decaying to 
        $b\widetilde\chi_2^+$ while the other decaying to $t\widetilde\chi_2^0$,
        thus leading to $2b+2\ell+\cancel{E}_T$ final state.
         }
        \label{subfig:1}
    \end{subfigure}%
    ~~ 
    \begin{subfigure}[t]{0.5\textwidth}
        \centering
        \includegraphics[height=2.0in]{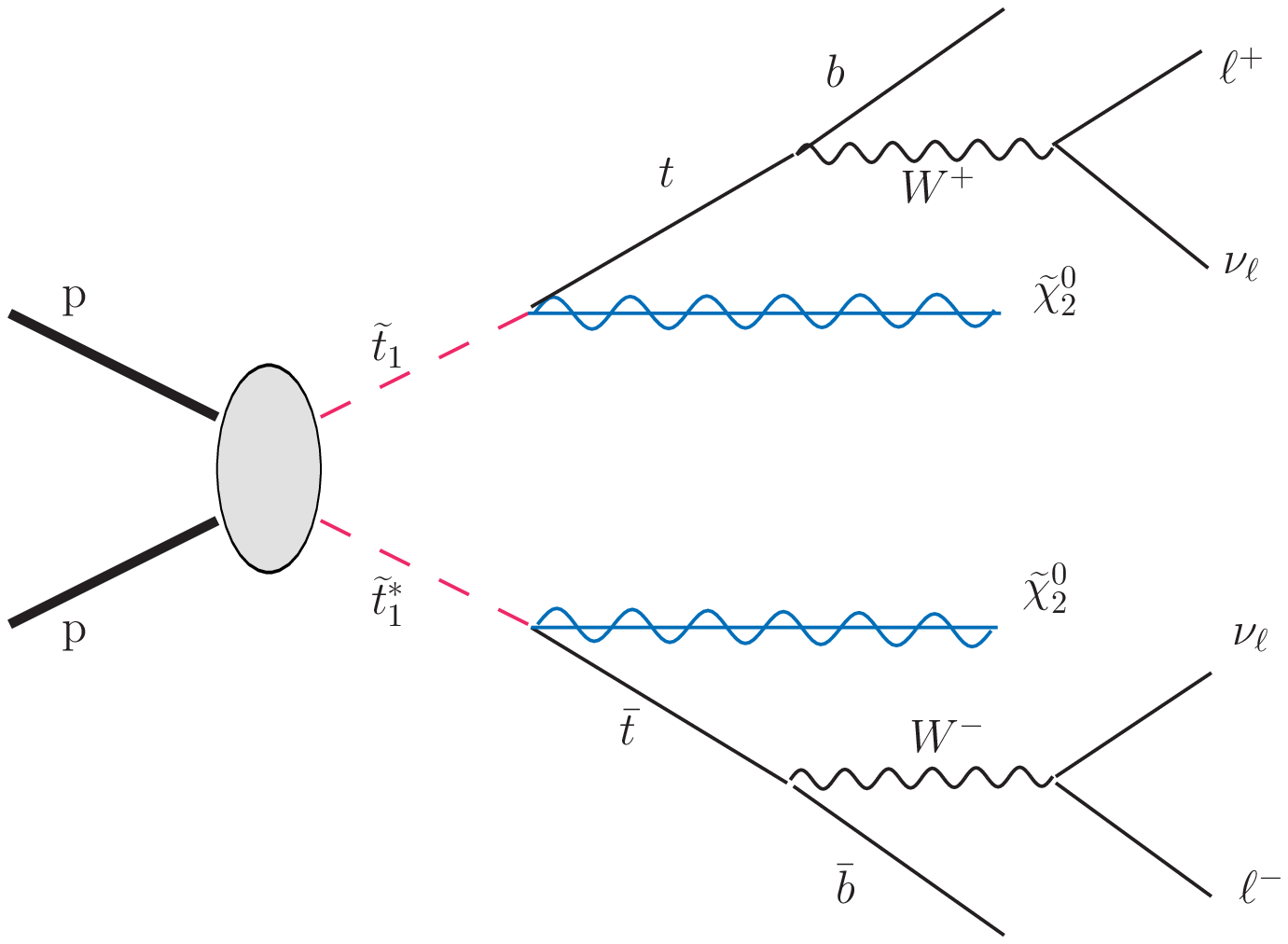}
        \caption{Symmetric decays of top squarks with both decaying to  $t\widetilde\chi_2^0$,
        thus leading to $2b+2\ell+\cancel{E}_T$ final state.}
        \label{subfig:2}
    \end{subfigure}\\
  \vspace{1.5cm}
    \begin{subfigure}[t]{0.5\textwidth}
        \centering
        \includegraphics[height=2.0in]{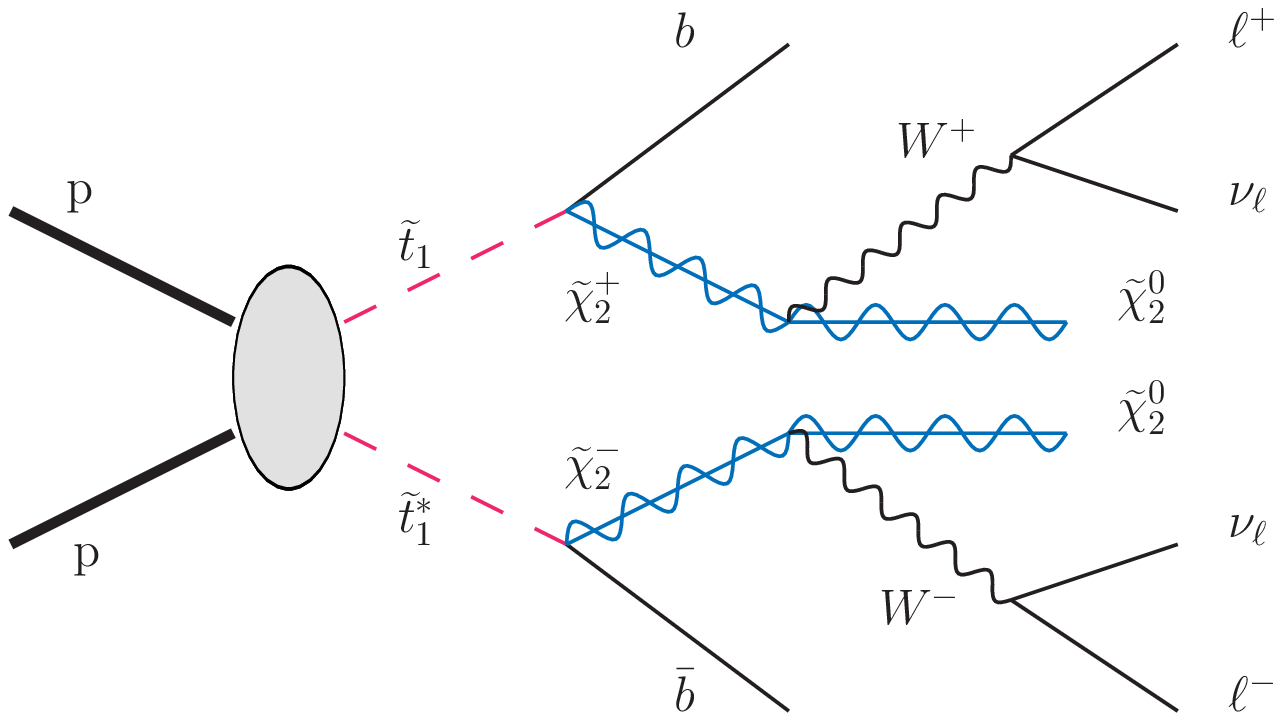}
        \caption{Symmetric decays of top squarks with both decaying to $b\widetilde\chi^{\pm}$,
        thus leading to $2b+2\ell+\cancel{E}_T$ final state.}
        \label{subfig:3}
    \end{subfigure}
    \caption{Final state topologies yielding $2b+2\ell+\cancel{E}_T$ in the 
       }
        \label{fig:bjet-lep-1}
\end{figure}

Two benchmark points for this case are shown in table~\ref{Tab:Spec1}. 
A relatively low value of $\mu_u (=200$ GeV) is chosen for both the
benchmark points. The masses of the Higgsino-like chargino and the neutralinos
are mainly controlled by $\mu_u$.
We assume the singlet and the triplet $vevs$ to
be small; roughly to be around $10^{-4}$ GeV. The Dirac gluino mass 
($M_3^D$) is considered
to be 1.5 TeV. Since we are considering a small amount of $R$-breaking,
the Majorana gaugino masses are roughly around a few hundred MeV. Fixing
the order parameter of $R$-breaking, i.e., the gravitino mass $\cal{O}$(10 GeV), 
fixes these
soft SUSY breaking parameters. Both $\lambda_S$ and $\lambda_T$ are considered
to be large, which for large $\tan\beta$ ($> 20$), provide significant radiative 
corrections to the Higgs
boson mass through one loop quartic terms. Such a choice allows us to
have a situation where both the top squarks are moderately light. 
Note that the chosen values of $m_{\widetilde t_1} (\sim 470\text{GeV})$
for BP-1 and BP-2 are expected to be consistent with the latest LHC bounds
discussed in section~\ref{bounds}. The bound assuming BR$(\widetilde t_1\rightarrow
b\widetilde\chi_1^+)=100\%$ is evidently satisfied while the one ($m_{\widetilde t_1}
> 550$ GeV) that assumes BR$(\widetilde t_1\rightarrow t\widetilde\chi_1^0)=100\%$
is not applicable here (see table  \ref{Br:Tabl1}). 
Furthermore, the neutrino Yukawa coupling `$f$'
is chosen to be 1. Hence, to have the active neutrino mass in the right ballpark, 
we need to consider the Dirac bino and wino masses to be almost degenerate.
As conspicuous from equation~(\ref{neutrino_majorana}), this degeneracy
provides a suitable suppression to the neutrino Majorana mass when 
the Dirac gaugino masses themselves are roughly around a TeV or so. Some of the 
low energy flavor violating branching ratios (which satisfy the respective experimental
constraints) are also shown in table~\ref{Tab:Spec1}.
%%%%%%%%%%%%%%%%%%%%%%%%%%%%%%%%%%%%%%%%%%%%%%%%%%%%%%%%%%%%%%%%%%%%%%%%%%%%%%%%%%%%
%\newpage
\begin{table}[ht]
\centering
\begin{tabular}{|c|c|c|c|}
\hline
Parameters              & BP-1                      & BP-2 \\ [0.6ex]
\hline
$M_1^D$                 & 1200 GeV                  & 800 GeV   \\
$M_2^D$                 & 1200.1 GeV                & 800.1 GeV \\ 
$M_3^D$                 & 1500 GeV                  & 1500 GeV  \\
$\mu_u$                 & 200 GeV                   & 200 GeV   \\
$m_{3/2}$               & 20 GeV                    & 20 GeV    \\
$\tan\beta$             & 23                        & 35        \\
$(m^2_{u})_{33}$        & 2.3$\times 10^5$ GeV$^2$         & $2.5\times 10^5$ GeV$^2$\\
$(m^2_{Q})_{33}$        & 5.5$\times 10^5$ GeV$^2$         & $6.1\times 10^5$ GeV$^2$\\
$f$                     & 1                         & 1\\
$v_S$                   & $2\times 10^{-4}$ GeV     & $1.5\times 10^{-4}$ GeV\\
$v_T$                   & $10^{-4}$ GeV                & $10^{-4}$ GeV\\
$\lambda_S$             & 1.130                     & 1.116\\
$B\mu_{L}$              & $-(200~\text{GeV})^2$      & $-(200~\text{GeV})^2$\\
$t_S$                   & $(174~\text{GeV})^3$         & $(174~\text{GeV})^3$\\
\hline
Observables              & BP-1                     & BP-2       \\ [0.6ex]
\hline
$m_h$                   & 124.9 GeV                & 125.7 GeV\\
$m_{\tilde t_1}$        & 566.2 GeV                & 580.5 GeV\\
$m_{\tilde t_2}$        & 918.5 GeV                & 904.8 GeV\\
$m_{\widetilde\chi_1^0}\equiv m_{\nu_e}$& 0.01 eV                  & 0.04 eV\\
$m_{\widetilde\chi_2^0} (\text{bino-like})$& 167.9 MeV                & 168.3 MeV\\
$m_{\widetilde\chi_3^0}$& 211.5 GeV                & 213.8 GeV\\
$m_{\widetilde\chi_4^0}$& 211.5 GeV                & 213.8 GeV\\
$m_{\widetilde\chi_1^+}\equiv m_e$& 0.51 MeV                  & 0.51 MeV\\
$m_{\widetilde\chi_2^+}$& 243.8 GeV                & 247.1 GeV\\
\hline
Flavor Observables       & BP-1                              & BP-2\\[0.6ex]
\hline
BR($B\rightarrow X_S \gamma$)    & $3.4\times 10^{-4}$     & $3.3\times 10^{-4}$\\
BR($B_S^0\rightarrow \mu\mu$)    & $2.4\times 10^{-9}$     & $2.5\times 10^{-9}$\\
BR($\mu\rightarrow e\gamma$)     & $3.8\times 10^{-24}$    & $4.9\times 10^{-24}$\\
BR($\mu\rightarrow 3e$)          & $3.0\times 10^{-26}$    & $4.0\times 10^{-26}$\\
\hline
$\mu_{\gamma\gamma}$    & 1.05   & 1.06     \\
\hline
\end{tabular}
\caption{\label{Tab:Spec1} Benchmark sets of input parameters in the
large neutrino Yukawa coupling (`$f$') scenario and the
resulting mass-values for some relevant excitations for $\mu_u < m_{\widetilde t_{1,2}}$
(case 1). $M_3^D$ denotes the Dirac gluino mass. Also indicated are
some of the relevant flavor observables and their values, all of which are currently
allowed by experiments. The corresponding values of $\mu_{\gamma\gamma}$ 
(the estimated Higgs di-photon rate compared to its SM expectation) are also mentioned.
}
\end{table}
%%%%%%%%%%%%%%%%%%%%%%%%%%%%%%%%%%%%%%%%%%%%%%%%%%%%%%%%%%%%%%%%%%%%%%%%%%%%%%%%%%%%
\begin{itemize}
\item \underline{Decay branching fractions of $\widetilde t_1$}
\end{itemize}
%%%%%%%%%%%%%%%%%%%%%%%%%%%%%%%%%%%%%%%%%%%%%%%%%%%%%%%%%%%%%%%%%%%%%%%%%%%%%%%%%%%%
The dominant decay branching fractions of $\widetilde t_1$ for BP-1 and BP-2 are indicated
in table~\ref{Br:Tabl1}.
\begin{table}[ht]
\centering
\begin{tabular}{|c|c|c|}
\hline
Decay modes & BR for BP-1 & BR for BP-2 \\
\hline
$\widetilde t_1 \rightarrow b~\widetilde\chi_2^{+}$ &  $52.5\%$ & $51.7\%$ \\
$\widetilde t_1 \rightarrow t~\widetilde\chi_3^{0}$ & $20.0\%$   & $20.1\%$ \\
$\widetilde t_1 \rightarrow t~\widetilde\chi_4^{0}$ & $20.0\%$   & $20.1\%$ \\
$\widetilde t_1 \rightarrow t~\widetilde\chi_2^{0}$ & $6.0\%$   & $6.6\%$ \\
\hline
\end{tabular}
\caption{Decay branching fractions of $\widetilde t_1$ in BP-1 and BP-2 for 
$\mu_u < m_{\widetilde t_1}$.}
\label{Br:Tabl1}
\end{table}
%%%%%%%%%%%%%%%%%%%%%%%%%%%%%%%%%%%%%%%%%%%%%%%%%%%%%%%%%%%%%%%%%%%%%%%%%%%%%%%%%%%%
$\widetilde \chi_2^+$ is the Higgsino-like chargino and $\widetilde\chi_{3,4}^0$
are the Higgsino-like neutralinos. $\widetilde\chi_2^0$ is the bino-like neutralino
with mass in the ballpark of a few hundred MeV. As can be seen from table~\ref{Br:Tabl1},
the top squarks, once produced in pairs, can undergo both symmetric as well as asymmetric decays.
Table~\ref{fin:Tabl1} lists all possible final state topologies. However, in the
context of this work, we will mainly consider the dilepton final states accompanied by $b$-jets
and MET. Such a final state might arise when the top squarks, on being pair-produced, 
undergo the decay $\widetilde t_1\rightarrow b\widetilde\chi_2^+$.
$\widetilde\chi_2^+$ in turn, decays to a $W^+$ and a $\widetilde\chi_2^0~(\widetilde\chi_1^0)$
with a branching ratio close to 90\% (10\%) followed by $W$-s decaying leptonically.
Although a semileptonic ($\ell\nu jj$) final state from $W$-decays is a good compromise
between the rate and the cleanliness of the signal,
we go for a cleaner channel where both the $W$ bosons decay leptonically. 
As shown in figure~\ref{subfig:3}, such a topology leads to a final state 
2 $b$-$\textrm{jets}+2$ $\textrm{leptons}+\cancel{E_T}$.
A similar final state could also arise from other decays of the lighter top squark, such as
those involving $\widetilde t_1\rightarrow t\widetilde\chi_2^0$ as shown in figures~\ref{subfig:1}
and~\ref{subfig:2}. However, the effective branching ratio is rather suppressed.

We note in passing that the various decay combinations shown in table~\ref{Br:Tabl1}
could also provide exotic multilepton and multijet final states
depending on the leptonic or hadronic decays of both $W^{\pm}$ or $Z$ boson. For 
example, decays such as $\widetilde t_1\rightarrow t\widetilde\chi_{3/4}^0$
could give rise to a 2 $b$-$\textrm{jets}+6$ $\textrm{leptons}+\cancel{E_T}$ final state.
Some relevant final states arising from the decays of $\widetilde t_1$ are tabulated
in table~\ref{fin:Tabl1}. The branching fractions in BP-1 and BP-2 are rather similar
since we are dealing with similar top squark masses. Also, top Yukawa coupling
is practically insensitive to larger values of $\tan\beta$, as considered in our study. 
This results in similar branching fractions in the $b\widetilde\chi^+$
mode in BP-1 and BP-2. The dynamics of other decays are essentially controlled by the 
gauge couplings and therefore, they remain similar.
%%%%%%%%%%%%%%%%%%%%%%%%%%%%%%%%%%%%%%%%%%%%%%%%%%%%%%%%%%%%%%%%%%%%%%%%%%%%%%%%%%%%
\begin{table}[h!]
\centering
 \begin{tabular}{|c|} 
 \hline
 $\mu_u < m_{\widetilde t_1}$: Decays of ${\widetilde t}_1$ \\ [0.5ex] 
 \hline\hline
 $pp\rightarrow \widetilde t_1 \widetilde{t_1^{*}} \rightarrow \widehat{b\widetilde\chi_2^+}
 ~~\widehat{\bar b \widetilde\chi_2^-} \rightarrow \widehat{b W^+ \widetilde\chi_2^0}  
 ~~\widehat{\bar{b} W^- \widetilde\chi_2^0}\rightarrow 2b+2W+\cancel{E_T}$\\
 $pp\rightarrow \widetilde t_1 \widetilde{t_1^{*}} \rightarrow \widehat{t\widetilde\chi_2^0}
 ~~\widehat{\bar t \widetilde\chi_2^0}\rightarrow \widehat{b W^+ \widetilde\chi_2^0}
 ~~\widehat{b W^- \widetilde\chi_2^0}\rightarrow 2b+2W+\cancel{E_T}$\\
% \vskip 0.5cm
  $pp\rightarrow \widetilde t_1 \widetilde{t_1^{*}} \rightarrow \widehat{b\widetilde\chi_2^+}
 ~~\widehat{\bar{t}\widetilde\chi_2^0}+\textrm{h.c.}\rightarrow \widehat{b W^+ \widetilde\chi_2^0}
 ~~\widehat{\bar{b} W^- \widetilde\chi_2^0}+\textrm{h.c.}\rightarrow 2b+2W+\cancel{E_T}$\\
 \hline
 $pp\rightarrow \widetilde t_1 \widetilde{t_1^{*}} \rightarrow \widehat{t\widetilde\chi_{3/4}^0}
 ~~\widehat{\bar{t}\widetilde\chi_2^0}+\textrm{h.c.}\rightarrow \widehat{b W^+ Z\widetilde\chi_2^0}
 ~~\widehat{\bar{b} W^- \widetilde\chi_2^0}+\textrm{h.c.}\rightarrow 2b+2W+Z+\cancel{E_T}$\\
 $pp\rightarrow \widetilde t_1 \widetilde{t_1^{*}} \rightarrow \widehat{b\widetilde\chi_2^+}
 ~~\widehat{\bar{t}\widetilde \chi_{3/4}^0}+\textrm{h.c.}\rightarrow \widehat{b W^+ \widetilde\chi_2^0}
 ~~\widehat{\bar{b} W^- Z\widetilde \chi_2^0}+\textrm{h.c.}\rightarrow 2b+2W+Z+\cancel{E_T}$\\
 \hline
 $pp\rightarrow \widetilde t_1 \widetilde{t_1^{*}} \rightarrow \widehat{t\widetilde\chi_{3/4}^0}
 ~~\widehat{\bar{t}\widetilde\chi_{3/4}^0}\rightarrow \widehat{b W^+ Z \widetilde\chi_2^0}
 ~~\widehat{\bar{b} W^- Z \widetilde\chi_2^0}\rightarrow 2b+2W+2Z+\cancel{E_T}$\\
 \hline
\end{tabular}
\caption{Possible final states arising out of various decay modes of $\widetilde t_1$ 
when $\mu_u < m_{\widetilde t_1}$.}
\label{fin:Tabl1}
\end{table}
%%%%%%%%%%%%%%%%%%%%%%%%%%%%%%%%%%%%%%%%%%%%%%%%%%%%%%%%%%%%%%%%%%%%%%%%%%%%%%%%%%%%
%%%%%%%%%%%%%%%%%%%%%%%%%%%%%%%%%%%%%%%%%%%%%%%%%%%%%%%%%%%%%%%%%%%%%%%%%%%%%%%%%%%%
\begin{itemize}
\item \underline{Decay branching fractions of $\widetilde t_2$}
\end{itemize}
%%%%%%%%%%%%%%%%%%%%%%%%%%%%%%%%%%%%%%%%%%%%%%%%%%%%%%%%%%%%%%%%%%%%%%%%%%%%%%%%%%%%
The dominant decay branching fractions of $\widetilde t_2$ for BP-1 and BP-2 are 
shown in table~\ref{Br:Tabl2}. The pattern can be justified following the 
discussion in section \ref{subsec:stop2-decay}. 
\begin{table}[ht]
\centering
\begin{tabular}{|c|c|c|}
\hline
Decay modes & BR for BP-1 & BR for BP-2 \\
\hline
$\widetilde t_2 \rightarrow b~e^{+}$ &  $27.8\%$ & $47.2\%$ \\
$\widetilde t_2 \rightarrow t~\widetilde\chi_3^{0}$ & $35.7\%$   & $26.0\%$ \\
$\widetilde t_2 \rightarrow t~\widetilde\chi_4^{0}$ & $35.7\%$   & $26.0\%$ \\
\hline
\end{tabular}
\caption{Decay branching fractions of $\widetilde t_2$ in BP-1 and BP-2 for which 
$\mu_u < m_{\widetilde t_2}.$}
\label{Br:Tabl2}
\end{table}
%%%%%%%%%%%%%%%%%%%%%%%%%%%%%%%%%%%%%%%%%%%%%%%%%%%%%%%%%%%%%%%%%%%%%%%%%%%%%%%%%%%%
\begin{figure}[ht]
\centering
\includegraphics[width=6cm]{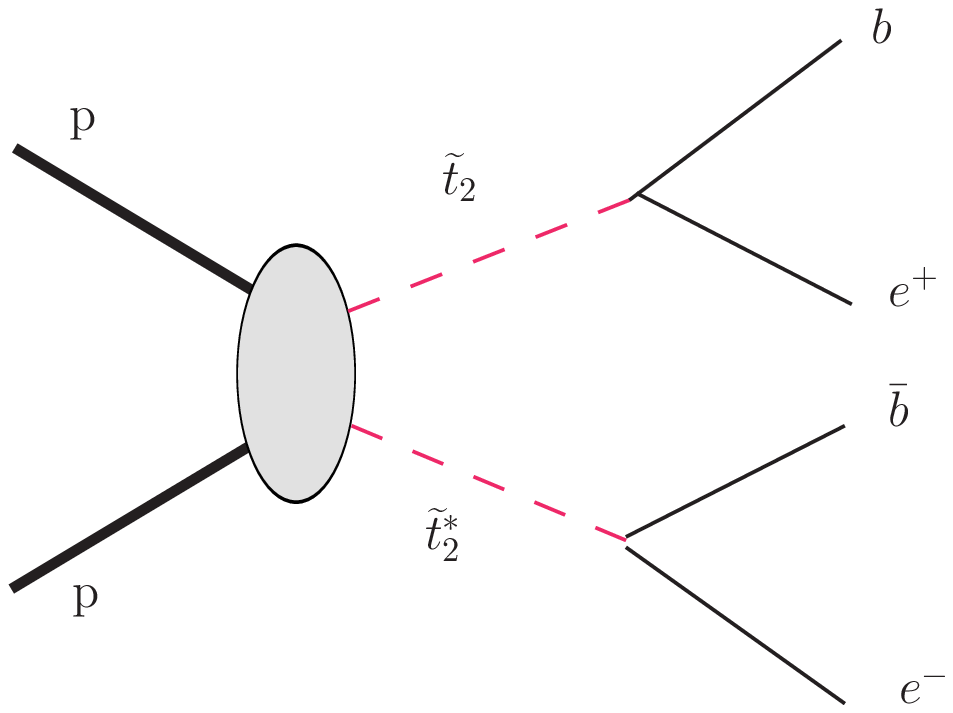}
\caption{2$b+$$e^+e^-$ final state arising from both $\widetilde t_2$-s decaying directly to
        a bottom (anti-bottom) quark and a positron (electron).}
\label{fig:bjet2}
\end{figure}
%%%%%%%%%%%%%%%%%%%%%%%%%%%%%%%%%%%%%%%%%%%%%%%%%%%%%%%%%%%%%%%%%%%%%%%%%%%%%%%%%%%%
The branching fractions of 
$\widetilde{t_2}$ to the three modes indicated are comparable. These lead to distinct 
final state signatures with appreciable strength. 
Possible final states arising from the decays of $\widetilde t_2$ are listed 
in table~\ref{fin:Tabl2}. The decay channels $\widetilde t_2 \rightarrow \widetilde t_1 Z(h)$
are absent due to negligibly small mixing between the left and the right chiral states
of the top squark.
%%%%%%%%%%%%%%%%%%%%%%%%%%%%%%%%%%%%%%%%%%%%%%%%%%%%%%%%%%%%%%%%%%%%%%%%%%%%%%%%%%%%
\begin{table}[h!]
\centering
 \begin{tabular}{|c|} 
 \hline
 $\mu_u < m_{\widetilde t_2}$: Decays of $\widetilde t_2$ \\ [0.5ex] 
 \hline\hline
 $pp\rightarrow \widetilde t_2 \widetilde{t_2^{*}} \rightarrow \widehat{b e^+}
 ~~\widehat{\bar{b} e^-} \rightarrow 2b+e^+ e^-$\\  
 \hline
  $pp\rightarrow \widetilde t_2 \widetilde{t_2^{*}} \rightarrow \widehat{b e^+}
 ~~\widehat{\bar{t}\widetilde \chi_{3/4}^0}+\textrm{h.c.}\rightarrow \widehat{b e^+}
 ~~\widehat{\bar{b} W^- Z\widetilde \chi_2^0}+\textrm{h.c.}\rightarrow 2b+W+Z+e^+ +\cancel{E_T}$\\
 \hline
  $pp\rightarrow \widetilde t_2 \widetilde{t_2^{*}} \rightarrow \widehat{t\widetilde\chi_{3/4}^0}
 ~~\widehat{\bar{t}\widetilde\chi_{3/4}^0}\rightarrow \widehat{b W^+ Z\widetilde\chi_2^0}
 ~~\widehat{\bar{b} W^- Z\widetilde\chi_2^0}\rightarrow 2b+2W+2Z +\cancel{E_T}$\\
 \hline
\end{tabular}
\caption{Possible final states arising out of various decay modes of
$\widetilde t_2$ when $\mu_u < \widetilde m_{t_2}$.}
\label{fin:Tabl2}
\end{table}
%%%%%%%%%%%%%%%%%%%%%%%%%%%%%%%%%%%%%%%%%%%%%%%%%%%%%%%%%%%%%%%%%%%%%%%%%%%%%%%%%%%%
A remarkable point to note here is the significant decay branching fraction of
$\widetilde t_2$ to a bottom quark and an electron. 
To reiterate, this decay rate is proportional to $\lambda^{\prime}_{133}$,
 which is identified with $y_b$. Therefore, the corresponding decay rate is large.
Also, because of the large difference between the mass of the decaying particle ($\widetilde t_2$)
and the total mass of the particles in the final state ($m_b+m_e$), the final state
electron is expected to be hard. 
The schematic diagram for such a process is presented in figure~\ref{fig:bjet2}.
Other decay modes, presented in table~\ref{fin:Tabl2}, are similar to the previous 
case where the decay products of $\widetilde t_2$ are a top quark and a Higgsino-like neutralino. These
would further decay to give a final state comprising of 6 leptons+2 $b$-jets
+$\cancel E_T$. The variations in the branching fractions as we go from BP-1 to BP-2,
as can be seen in table~\ref{Br:Tabl2}, are due to changing bottom Yukawa coupling
as $\tan\beta$ changes.
%%%%%%%%%%%%%%%%%%%%%%%%%%%%%%%%%%%%%%%%%%%%%%%%%%%%%%%%%%%%%%%%%%%%%%%%%%%%%%%%%%%%
\subsection{\textbf{\textit{Case 2:}} $\mu_u > m_{\widetilde t_{1,2}}$}
\label{case2}
\begin{figure}[ht]
        \centering
        \includegraphics[height=2.2in]{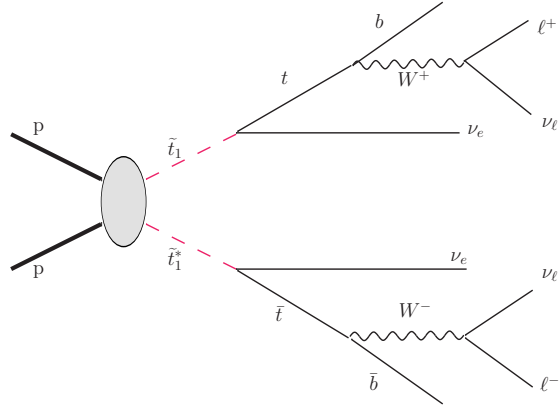}
    \caption{Symmetric decays of $\widetilde t_1$ and $\widetilde t_1^*$ 
             (via $t\nu_e$) mode leading
             to 2$b$+2$\ell+\cancel{E}_T$.}
    \label{fig:p5}
\end{figure}

%%%%%%%%%%%%%%%%%%%%%%%%%%%%%%%%%%%%%%%%%%%%%%%%%%%%%%%%%%%%%%%%%%%%%%%%%%%%%%%%%%%%
As opposed to the previous case, we consider the situation where 
$\mu_u > m_{\widetilde t_{1,2}}$. 
%Table~\ref{Tab:Spec2}. However,
To have the Higgs boson mass in the right range we tweak $\lambda_S$. The 
soft masses $(m_u)^2_{33}$ and $(m_Q)^2_{33}$ are modified to get
different top squark masses satisfying relevant LHC constraints. $v_T$ just takes a
different in sign when compared to BP-1 and BP-2 only
to exclude tachyonic states. All the other parameters are kept fixed to their values
in table~\ref{Tab:Spec1}.
Due to such a choice of $\mu_u$, a top squark cannot decay to 
Higgsino-like chargino and neutralinos.
Note that, in the present case as well, 
the masses of the top squarks are chosen in such a manner that they satisfy
the present experimental bounds. Values of flavor observables are checked to satisfy
experimental constraints. However, those are not shown explicitly this time.
%%%%%%%%%%%%%%%%%%%%%%%%%%%%%%%%%%%%%%%%%%%%%%%%%%%%%%%%%%%%%%%%%%%%%%%%%%%%%%%%%%%%
\begin{table}[ht]
\centering
\begin{tabular}{|c|c|c|c|}
\hline
Parameters              & BP-3                      & BP-4 \\ [0.6ex]
\hline
$\mu_u$                 & 1500 GeV                  & 1100 GeV   \\
$v_T$                   & $-10^{-4}$ GeV               & $-10^{-4}$ GeV\\
$(m^2_u)_{33}$          & 4$\times 10^{5}$~(GeV)$^2$        & 5$\times 10^{5}$~(GeV)$^2$\\
$(m^2_Q)_{33}$          & 5.2$\times 10^{5}$~(GeV)$^2$         & 6$\times 10^{5}$~(GeV)$^2$\\
$\lambda_S$             & 1.09                     & 1.06\\
\hline
Observables              & BP-3                     & BP-4       \\ [0.6ex]
\hline
$m_h$                   & 126.6 GeV                & 126.1 GeV\\
$m_{\tilde t_1}$        & 728.7 GeV                & 802.8 GeV\\
$m_{\tilde t_2}$        & 909.5 GeV                & 908.8 GeV\\
$m_{\widetilde\chi_1^0}\equiv m_{\nu_e}$& 0.03 eV                  & 0.12 eV\\
$m_{\widetilde\chi_2^0}~(\text{bino-like})$& 175.8 MeV                & 175.6 MeV\\
$m_{\widetilde\chi_3^0}$& 1202.1 GeV                & 804.3 GeV\\
$m_{\widetilde\chi_4^0}$& 1202.2 GeV                & 804.3 GeV\\
$m_{\widetilde\chi_1^+}\equiv m_e$& 0.51 MeV                  & 0.51 MeV\\
$m_{\widetilde\chi_2^+}$& 1304.2 GeV                & 877.4 GeV\\
\hline
\end{tabular}
\caption{\label{Tab:Spec2} Same as in \ref{Tab:Spec1} but for an extra sign
on $v_T$ and for BP-3 and BP-4 for
both of which
$\mu_u > m_{\widetilde t_{1,2}}$ (case 2). Values of flavor observables,
not shown here explicitly, satisfy all the experimental constraints. 
}
\end{table}
%%%%%%%%%%%%%%%%%%%%%%%%%%%%%%%%%%%%%%%%%%%%%%%%%%%%%%%%%%%%%%%%%%%%%%%%%%%%%%%%%%%%
\begin{itemize}
\item \underline{Decay branching fractions of $\widetilde t_1$}
\end{itemize}
The decay branching fractions for $\widetilde t_1$ when $\mu_u > m_{\widetilde t_{1,2}}$
are shown in table~\ref{Tabl3} for the benchmark points BP-3 and BP-4. 
An interesting point to note here is that the top squark decays only to a top quark
accompanied either by a $\widetilde\chi_2^0$ (figure \ref{subfig:2}) or a $\nu_e$ 
(figure \ref{fig:p5}) both of which are carriers of MET while the former being the 
dominant one.
\begin{table}[ht]
\centering
\begin{tabular}{|c|c|c|}
\hline
Decay modes & BR for BP-3 & BR for BP-4 \\
\hline
$\widetilde t_1\rightarrow t\widetilde\chi_2^0$ &  $87.8\%$ & $94.6\%$ \\
$\widetilde t_1\rightarrow t \nu_e$ & $12.2\%$   & $5.3\%$ \\
\hline
\end{tabular}
\caption{Decay branching fractions of $\widetilde t_1$ in BP-3 and BP-4
for which $\mu_u > m_{\widetilde t_1}$.}
\label{Tabl3}
\end{table}
%
%%%%%%%%%%%%%%%%%%%%%%%%%%%%%%%%%%%%%%%%%%%%%%%%%%%%%%%%%%%%%%%%%%%%%%%%%%%%%%%%%%%%
Again, both symmetric
and asymmetric decays of the pair produced $\widetilde t_1$-s are possible. These would
lead to 2 $b$-jets+2$\ell+\cancel{E}_T$ final states. Note that 
more exotic final states with a
larger lepton multiplicity would be absent as heavier Higgsino-like neutralino(s) will now
be missing in the cascades of $\widetilde t_1$. This is in sharp contrast with what is 
expected for $\mu_u < m_{\widetilde t_1}$ (case 1) as discussed in section~\ref{case1}
and thus, may be exploited to distinguish between these two broad scenarios. 
%%%%%%%%%%%%%%%%%%%%%%%%%%%%%%%%%%%%%%%%%%%%%%%%%%%%%%%%%%%%%%%%%%%%%%%%%%%%%%%%%%%%
\begin{itemize}
\item \underline{Decay branching fractions of $\widetilde t_2$}
\end{itemize}

Similarly, the absence of a light Higgsino-like chargino
and neutralinos implies $\widetilde t_2$ would dominantly decay to a bottom quark 
and an electron (positron).
The branching fractions of $\widetilde t_2$ under such a circumstance are presented
in table~\ref{Tabl4}.
%%%%%%%%%%%%%%%%%%%%%%%%%%%%%%%%%%%%%%%%%%%%%%%%%%%%%%%%%%%%%%%%%%%%%%%%%%%%%%%%%%%%
\begin{table}[ht]
\centering
\begin{tabular}{|c|c|c|}
\hline
Decay modes & BR for BP-3 & BR for BP-4 \\
\hline
$\widetilde t_2 \rightarrow b~e^{+}$ &  $97.8\%$ & $98.7\%$ \\
$\widetilde t_2 \rightarrow t~\widetilde\chi_2^{0}$ & $1.8\%$   & $0.9\%$ \\
\hline
\end{tabular}
\caption{Decay branching fractions of $\widetilde t_{2}$ for BP-3 and BP-4 
for which $\mu_u > m_{\widetilde t_2}$.}
\label{Tabl4}
\end{table}
%
%%%%%%%%%%%%%%%%%%%%%%%%%%%%%%%%%%%%%%%%%%%%%%%%%%%%%%%%%%%%%%%%%%%%%%%%%%%%%%%%%%%%
\section{Collider (LHC) analysis}
\label{collider}
%%%%%%%%%%%%%%%%%%%%%%%%%%%%%%%%%%%%%%%%%%%%%%%%%%%%%%%%%%%%%%%%%%%%%%%%%%%%%%%%%%%%
In this section we present the setup and the results of the simulation we 
carry out at the 13 TeV LHC for pair-produced top squarks that eventually cascade
to the final states discussed in section~\ref{benchmarks}.
%%%%%%%%%%%%%%%%%%%%%%%%%%%%%%%%%%%%%%%%%%%%%%%%%%%%%%%%%%%%%%%%%%%%%%%%%%%%%%%%%%%%
\subsection{The simulation setup and reconstructing the physics objects}
%%%%%%%%%%%%%%%%%%%%%%%%%%%%%%%%%%%%%%%%%%%%%%%%%%%%%%%%%%%%%%%%%%%%%%%%%%%%%%%%%%%%
We have implemented the model in {\tt MadGraph5\_aMC@NLO}~\cite{Madgraph}. 
Events for both signals and backgrounds are generated using the same. 
We use the parton distribution function {\tt CTEQ6L1}~\cite{Pumplin:2002vw}
and a factorisation/renormalisation scale of $\sqrt{m_{\widetilde t_1}
m_{\widetilde t_2}}$ for generating 
events at the lowest order. The inclusive rates are then normalised to their
respective values obtained after higher order corrections as given by 
{\tt MadGraph5\_aMC@NLO} in the cases for the SM background and {\tt Prospino2 (v2.1)}
\cite{P1,Prospino} for the case of the SUSY productions. Appropriate branching 
fractions are obtained from
the spectrum generator {\tt SPheno} \cite{Porod} which, in the first place,
is generated by {\tt SARAH} \cite{Staub,Staub1,Staub2}. 
We note in passing that the production cross-section
for the top squarks (at the tree level) in this model is same as in the MSSM, 
considering only the dominant strong interaction.

Events in the LHE format are fed into 
{\tt Pythia-6.4.28}~\cite{Pythia} for showering, hadronisation and jet formation. 
Clustering of jets is performed with 
the built-in {\tt Pythia} module {\tt PYCELL} which employs a cone
algorithm and incorporates appropriate smearing of the momenta. 
In {\tt PYCELL} we allowed for an angular coverage of $|\eta|< 5$ for the 
hadron calorimeter with a cell-segmentation of 
$\Delta \eta \times \Delta \phi = 0.1 \times 0.1$
which resembles a generic LHC detector. A cell is required to have a minimum value of deposited 
$E_T=1$ GeV for it to be considered. A jet-cone radius of $\Delta R(j,j)=0.4$ is employed for 
finding jets. A minimum
summed $E_T$ of 20 GeV is required within such a geometry for the configuration to be considered 
as a jet. Ultimately, formed jets within $|\eta| < 2.5$ are considered in our analysis.
Care has been taken to isolate final state leptons by imposing the 
following cuts and isolation criteria:
\begin{itemize}
\item To select leptonic events we have used $p_T^{\ell} > 10$ GeV and $|\eta_{\ell}|<2.4$.
\item Lepton-lepton separation has been done by choosing $\Delta 
R(\ell^{\prime}, \ell)>0.2$, where $\Delta R=\sqrt{(\Delta \eta)^2+(\Delta\phi)^2}$.
\item Subsequently, to separate leptons from jets we have used $\Delta 
R(j, \ell)>0.5$.
\item Finally, the sum of the energy deposits of the hadrons which fall
within a cone of $\Delta R\leq 0.2$ around a lepton, must be less
than 10 GeV. 
\end{itemize}
In this work, by leptons we mean only electrons and muons for which the
detection efficiencies are generally very high, unlike the $\tau$ lepton.
We have used a minimum $p_T$ cut of 10 GeV and 17 GeV to isolate muons and electrons,
respectively. To estimate the number of $b$-jets in the final state, a flat 
(but somewhat conservative) $b$-tagging
efficiency of $60\%$ has been used.  

%%%%%%%%%%%%%%%%%%%%%%%%%%%%%%%%%%%%%%%%%%%%%%%%%%%%%%%%%%%%%%%%%%%%%%%%%%%%%%%%%%%%
\subsection{Top squark pair-production cross section}
\label{subsec:pairprod}

\begin{figure}[ht]
\centering
\includegraphics[width=10cm]{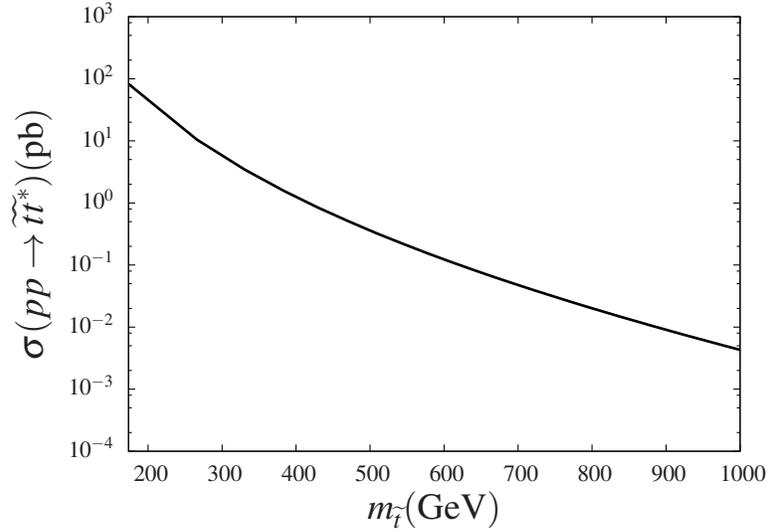}
\caption{Variation of production cross section for a pair of top squarks
at the 13 TeV LHC. Other parameters
are kept fixed at values given for BP-$4$ in table~\ref{Tab:Spec2}
(see text). Rates include appropriate $K$-factors obtained
from the package {\tt Prospino2}. 
}
\label{fig:top_stop}
\end{figure}

The phenomenology we discuss in this work crucially depends on the rate of
pair-production of the top squarks. It is to be noted that at the lowest
order these rates are the same as in the MSSM. Considering the dominant
strong contributions in the rates, the variation of the same is only
dependent on the mass of the top squark, irrespective of its chiral content.
As a quick reckoner for this basic rate, we present the same as a function
of $m_{\tilde t}$ in figure \ref{fig:top_stop} for the 13 TeV run of the LHC.
Appropriate $K$-factors as obtained from the package {\tt Prospino2} 
\cite{P1,Prospino} are already folded in.
To this end, we fix all the parameters as given in BP-4 except for 
the right handed soft squark mass, which we
varied from $-4\times 10^4~\text{GeV}^2<(m^2_u)_{33}<8\times 10^5~\text{GeV}^2$. 
Such a choice would surely move the Higgs mass away from the allowed range. However,
we are here merely concentrating on study of the production cross-section 
for $\widetilde t\widetilde t^*$. The parameters such as $\lambda_S, \lambda_T$
can be adjusted to fit the Higgs mass, which is unlikely to affect the production
rate any significantly (via unknown higher order effects).

%%%%%%%%%%%%%%%%%%%%%%%%%%%%%%%%%%%%%%%%%%%%%%%%%%%%%%%%%%%%%%%%%%%%%%%%%%%%%%%%%%%%
%%%%%%%%%%%%%%%%%%%%%%%%%%%%%%%%%%%%%%%%%%%%%%%%%%%%%%%%%%%%%%%%%%%%%%%%%%%%%%%%%%%%
%%%%%%%%%%%%%%%%%%%%%%%%%%%%%%%%%%%%%%%%%%%%%%%%%%%%%%%%%%%%%%%%%%%%%%%%%%%%%%%%%%%%
\subsection{The Standard Model backgrounds}
\label{subsec:smbg}
%%%%%%%%%%%%%%%%%%%%%%%%%%%%%%%%%%%%%%%%%%%%%%%%%%%%%%%%%%%%%%%%%%%%%%%%%%%%%%%%%%%%
As mentioned earlier, we would mostly concentrate on the final states with 
2 $b$-jets+2 leptons+$\cancel{E}_T$ and 2 $b$-jets+$e^+e^-$. In the first case, 
we consider only the most dominant background coming from $t\bar t$ production when 
both the top quarks decay leptonically. In order to have a realistic normalisation 
of this background, $t\bar t$ events generated at the lowest order (LO) using 
{\tt MadGraph5\_aMC@NLO} and the size of the event sample is scaled appropriately 
to correspond to its next-to-leading order (NLO) + next-to-next-to-leading log (NNLL) 
cross-section ($\approx$ 816 pb)~\cite{ttsm}.
The heavier top squark undergoes a significant decay to a bottom quark and an electron.
The dominant background comes from the direct production of a pair of bottom quarks 
with one of them radiating a $Z$ or $\gamma^{*}$ which subsequently produces a 
pair of $e^+ e^-$. This background can be largely suppressed by putting an 
on-shell $Z$ veto for the $e^+ e^-$ pair. 
To be conservative and for the robustness of the estimate, the NLO computation
~\cite{Campbell,FebresCordero:2008ci,Dawson:2003zu,Frederix}
is done with two additional jets (10 GeV $< p_T^{\text{jet}}<60$ GeV) in the
final state. The SM cross-section for $pp\rightarrow b\bar{b}e^{+}e^{-}+$jets 
we used is 9.43 pb.
%%%%%%%%%%%%%%%%%%%%%%%%%%%%%%%%%%%%%%%%%%%%%%%%%%%%%%%%%%%%%%%%%%%%%%%%%%%%%%%%%%%%
\subsection{Event selection}
\label{subsec:selection}
%%%%%%%%%%%%%%%%%%%%%%%%%%%%%%%%%%%%%%%%%%%%%%%%%%%%%%%%%%%%%%%%%%%%%%%%%%%%%%%%%%%%
To optimise the signal to background ratios, we now have to adopt a set of event
selection criteria. Towards this, various appropriate kinematic distributions
are studied for both signals and the backgrounds. We present our study for two broad
scenarios discussed in section~\ref{benchmarks}, i.e., for $\mu_u < 
m_{\widetilde t_{1,2}}$ and $\mu_u > m_{\widetilde t_{1,2}}$. For each of these
cases, two different final states are considered, viz., 2 $b$-jets+2 leptons
+$\cancel{E}_T$ and $b\bar b e^+ e^-$, arising from $\widetilde t_1$ and 
$\widetilde t_2$ decays, respectively.
%%%%%%%%%%%%%%%%%%%%%%%%%%%%%%%%%%%%%%%%%%%%%%%%%%%%%%%%%%%%%%%%%%%%%%%%%%%%%%%%%%%%
\subsubsection{\textbf{\textit{Case 1:}} $\mu_u < m_{\widetilde t_{1,2}}$}
%%%%%%%%%%%%%%%%%%%%%%%%%%%%%%%%%%%%%%%%%%%%%%%%%%%%%%%%%%%%%%%%%%%%%%%%%%%%%%%%%%%%
Here we discuss the decays of both the top squarks pertaining to the case
where $\mu_u < m_{\widetilde t_{1,2}}$.
%%%%%%%%%%%%%%%%%%%%%%%%%%%%%%%%%%%%%%%%%%%%%%%%%%%%%%%%%%%%%%%%%%%%%%%%%%%%%%%%%%%%
\begin{itemize}
\item \underline{$p p\rightarrow \widetilde t_1\widetilde t_1^*\rightarrow b\widetilde\chi_2^+/
t\widetilde\chi_2^0\rightarrow 2$ $b$-$\text{jets}+2$ $\text{leptons}+ \cancel{E}_T$
(figure~\ref{fig:bjet-lep-1})}
%%%%%%%%%%%%%%%%%%%%%%%%%%%%%%%%%%%%%%%%%%%%%%%%%%%%%%%%%%%%%%%%%%%%%%%%%%%%%%%%%%%%

Such a final state could arise from top squarks decaying
to $b\widetilde\chi_2^{\pm}$ and/or $t\widetilde\chi_2^0$. 
The final state leptons arise from the decays of $W$ bosons. In 
addition, $\widetilde\chi_2^{\pm}$ is somewhat heavier than the top quark
for both BP-1 and BP-2. Hence, on an average, one would expect the leptons
to be a little harder compared to the background leptons originating in the cascades of the
top quarks. This can be seen from the left panel of figure~\ref{fig:tot1},
where the $p_T$ distributions of the harder lepton in the signal in both the benchmarks
have extended tails compared to a similar lepton originating from the SM background. 
%%%%%%%%%%%%%%%%%%%%%%%%%%%%%%%%%%%%%%%%%%%%%%%%%%%%%%%%%%%%%%%%%%%%%%%%%%%%%%%%%%%%
\begin{figure}[t!]
\centering
\begin{subfigure}[t]{0.5\textwidth}
\centering
\includegraphics[height=2.1in]{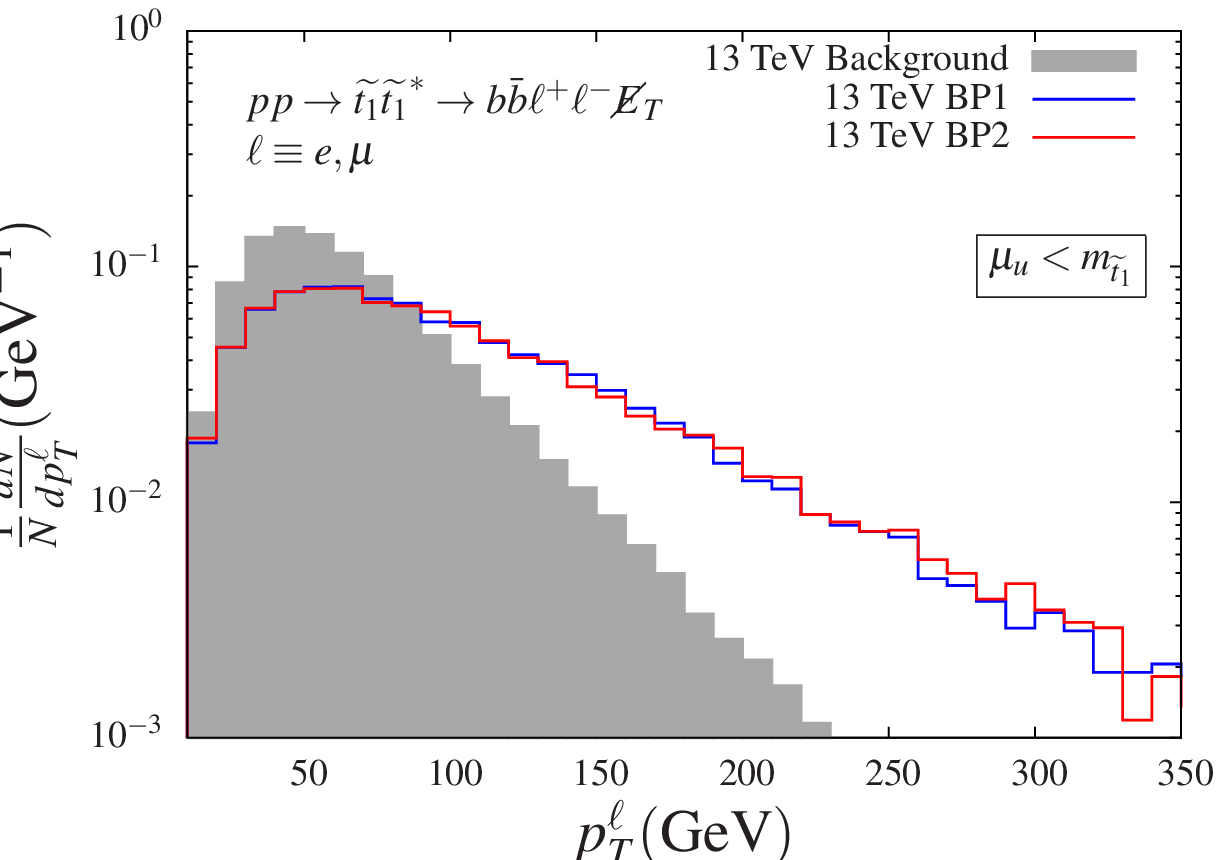}
\end{subfigure}%
~~
\begin{subfigure}[t]{0.5\textwidth}
\centering
\includegraphics[height=2.1in]{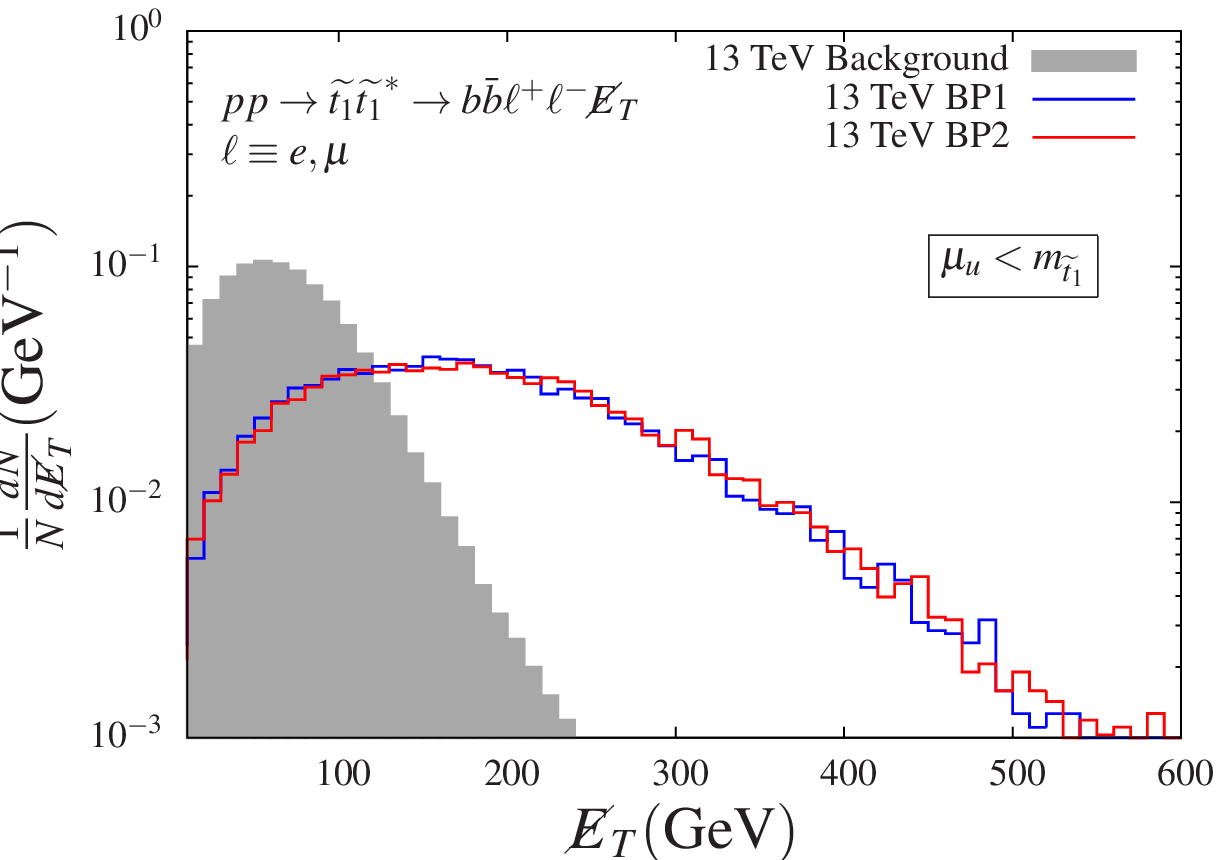}
\end{subfigure}
~~
\begin{subfigure}[t]{0.5\textwidth}
\centering
\includegraphics[height=2.1in]{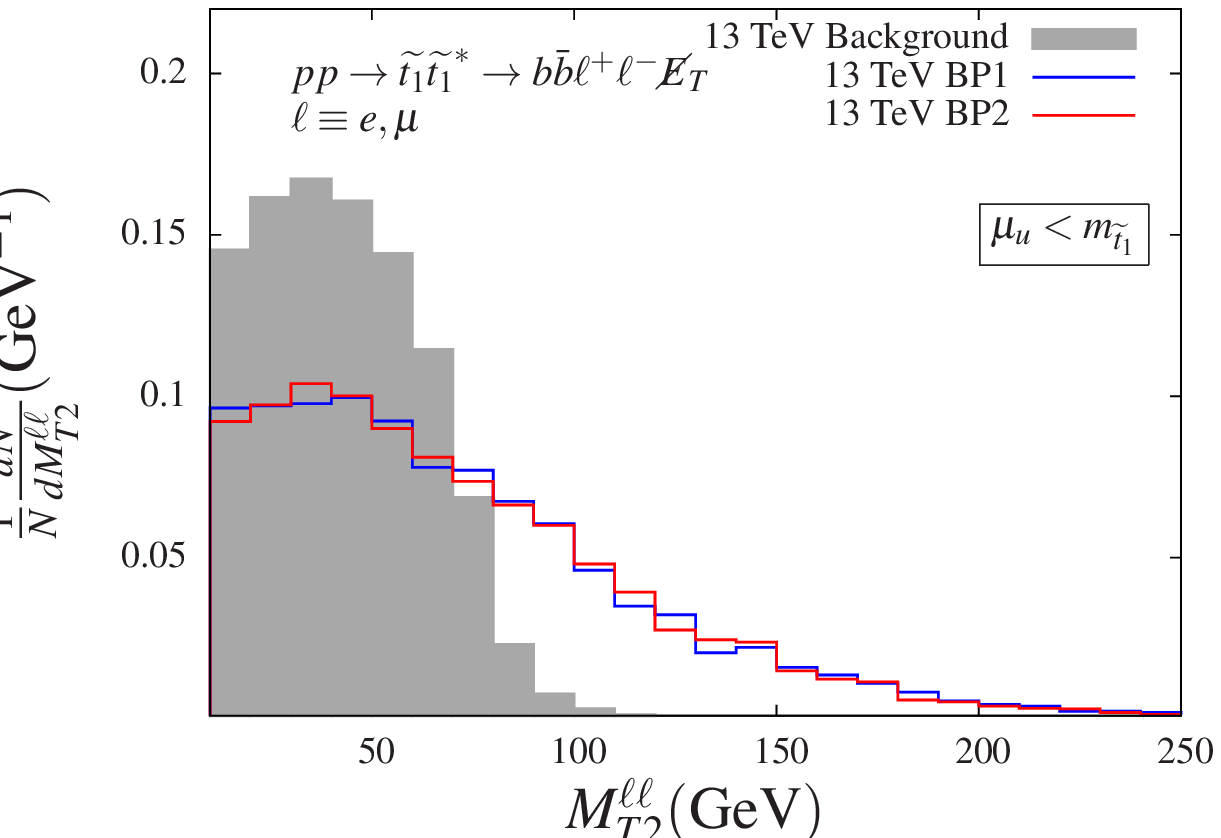}
\end{subfigure}\\
\caption{$p_T$ distributions of the harder lepton (left) and the
         $\cancel{E}_T$ distributions (right) for the background and the
         signals and dileptonic stransverse mass (in benchmark scenarios 
         BP-1 and BP-2) in the 2 $b$-jet+2 lepton+$\cancel{E}_T$ final state 
         arising from decays of $\widetilde t_1$ for the case $\mu_u < m_{\widetilde t_1}$.}
\label{fig:tot1}
\end{figure}
The signal distributions for BP-1 and BP-2 are similar because of similar values 
of top squark masses in the two benchmarks. The signal $\cancel{E}_T$ distributions 
are different from the corresponding distribution for the SM background. This 
may be attributed to the much larger mass of the top squark (compared to $m_t$) and 
the presence of extra carriers of $\cancel{E}_T$, i.e., the bino-like MeV neutralino 
($\widetilde\chi_2^0$) and the active neutrino ($\nu_e$) emerging from top squark 
decays. It is evident from the right panel of figure~\ref{fig:tot1} that the SM 
background can be effectively suppressed by applying a hard enough $\cancel{E}_T$ 
cut, viz., $\cancel{E}_T > 200$ GeV. We have also constructed the dileptonic
stransverse mass variable to see if the SM background can be suppressed further.
The stransverse mass is a kinematic variable which is used to measure the masses
of the pair produced semi-invisibly decaying heavy particles. The dileptonic stransverse
mass is defined as \cite{Ox}
%%%%%%%%%%%%%%%%%%%%%%%%%%%%%%%%5
\begin{eqnarray}
M_{T2}^{\ell\ell}(p_T^{\ell_1}, p_T^{\ell_2}, \cancel{p}_T)&=& 
\textrm{min}_{\cancel{p}_T=\cancel{p}_T^1+
\cancel{p}_T^2}\Bigg[\textrm{max}\Big\{M_T (p_T^{\ell_1}, \cancel{p}_T^1),
M_T (p_T^{\ell_2}, \cancel{p}_T^2)\Big\}\Bigg],
\end{eqnarray} 
%%%%%%%%%%%%%%%%%%%%%%%%%%%%%%%%5
where $\ell_1$ and $\ell_2$ are two isolated leptons and $\cancel{p}_T$ is the
total missing transverse momentum in the event and the transverse mass of the
system $M_T$ has its usual definition. Although, the stransverse mass is a standard
variable used in the recent top squark search, however, the yield with $\cancel{E}_T$
cut is more effective compared to the dileptonic $M_{T2}^{\ell \ell}$ cut for BP1 and BP2.
%%%%%%%%%%%%%%%%%%%%%%%%%%%%%%%%%%%%%%%%%%%%%%%%%%%%%%%%%%%%%%%%%%%%%%%%%%%%%%%%%%%%
\item \underline{$p p\rightarrow \widetilde t_2\widetilde t_2^{*} \rightarrow
b\bar b e^+ e^-$ (figure~\ref{fig:bjet2})}
%%%%%%%%%%%%%%%%%%%%%%%%%%%%%%%%%%%%%%%%%%%%%%%%%%%%%%%%%%%%%%%%%%%%%%%%%%%%%%%%%%%%

As has been pointed out earlier, $\widetilde t_2$ could have a significant
decay branching fraction to a bottom quark and an electron, which is a characteristic
of such a scenario. Along with the enlarged phase space available to this decay mode, 
a moderately large coupling ($\sim\lambda_{133}^{\prime}\equiv y_b$) does boost the decay rate.  
Naturally, we expect electrons (positrons) 
with high $p_T$. In the absence of a genuine carrier
of $\cancel{E}_T$ in such a final state, low or at most a 
moderate $\cancel{E}_T$ is expected from mis-measured momenta of the involved 
physics objects. The leptons are also
expected to have uncorrelated momenta. 
Such events are rare in the 
SM.
%%%%%%%%%%%%%%%%%%%%%%%%%%%%%%%%%%%%%%%%%%%%%%%%%%%%%%%%%%%%%%%%%%%%%%%%%%%%%%%%%%%%
\begin{figure}[t!]
\centering
\begin{subfigure}[t]{0.5\textwidth}
\centering
\includegraphics[height=2.1in]{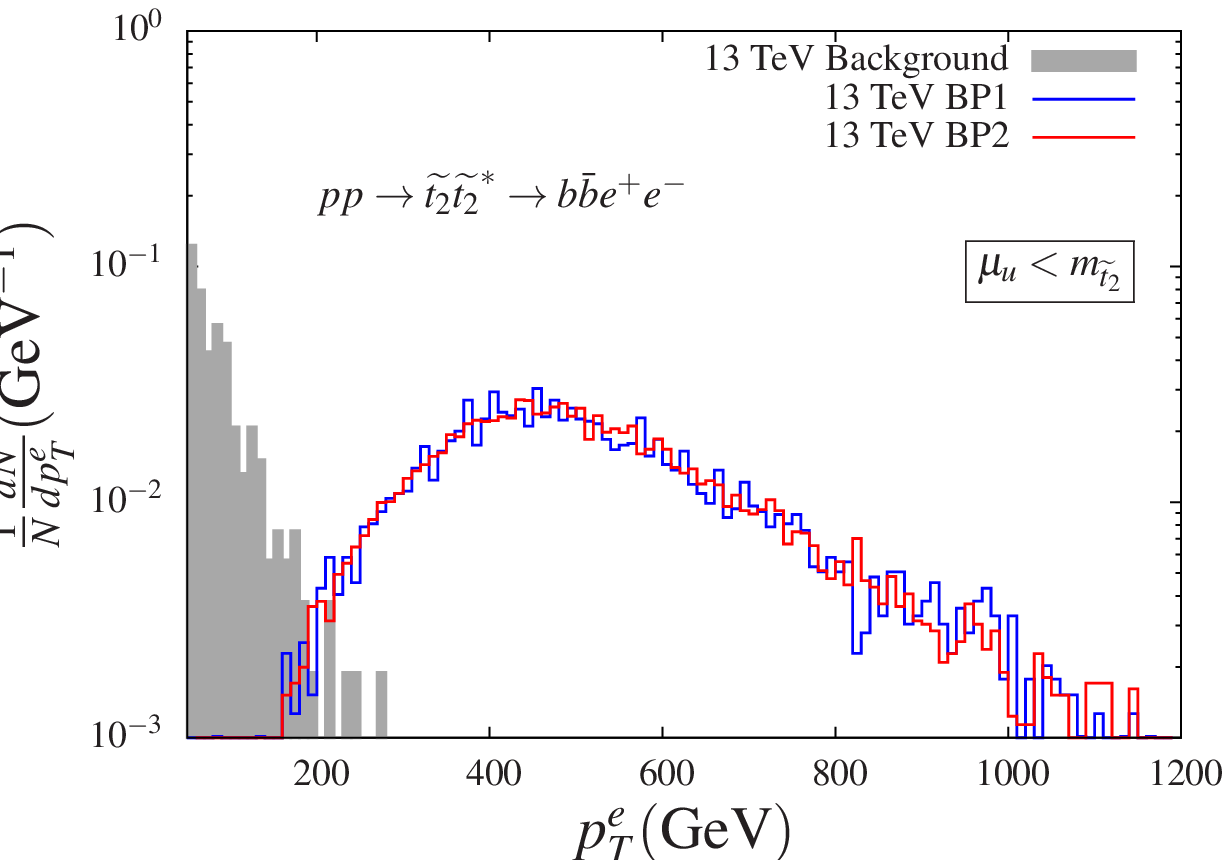}
\end{subfigure}%
~~
\begin{subfigure}[t]{0.5\textwidth}
\centering
\includegraphics[height=2.1in]{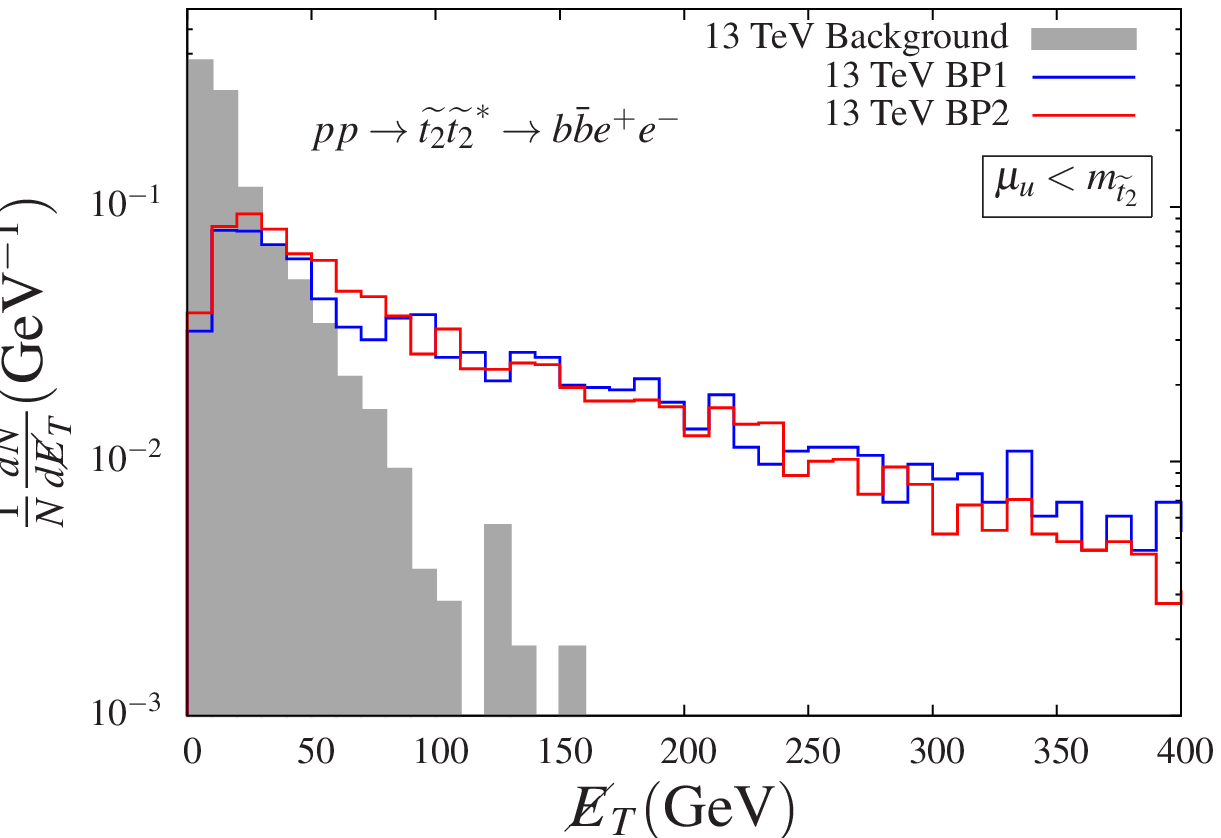}
\end{subfigure}\\
\caption{$p_T$ distributions of the harder electron (positron) from 
         both background and signal (left) and 
         $\cancel{E}_T$ ~distributions (right) 
         (in benchmark scenarios BP-1 and BP-2) 
         in the $b\bar b e^+ e^-$ final state arising from decays of a pair of
         $\widetilde t_2$-s for the case $\mu_u < m_{\widetilde t_2}$. 
}
\label{fig:tot2}
\end{figure}
The left of panel figure~\ref{fig:tot2} illustrates the hardest electron (positron) 
$p_T$ distribution in the scenario where $\widetilde t_2$ decays to a bottom 
quark and an electron (positron). We impose a minimum $p_T$ cut of 200 GeV to reduce the 
SM background substantially. Since $m_{\widetilde t_2}$ is very similar for 
BP-1 and BP-2 and so is its kinematics for these two benchmark points, the distributions 
look very similar. In the right panel of 
figure~\ref{fig:tot2}, we present the MET distribution which arises in this
case from mis-measurements of momenta of visible entities in the final state.
As expected, the MET distributions peak at small MET ($\approx 25$ GeV).

Note that eventually, one should be able to reconstruct $\widetilde{t_2}$-s
in the invariant mass spectra of appropriately chosen $b$-jet-electron (positron) 
systems which would show peaks at $m_{\widetilde t_2}$.
Clearly, the efficiency of reconstructing $\widetilde t_2$ would be limited by 
various detector effects and a close study of the kinematic distributions discussed
above would surely be of crucial help. Nonetheless, it appears that the peaks cannot be missed
and a reasonable estimation of $m_{\widetilde t_2}$ would thus be possible.
\end{itemize}
%%%%%%%%%%%%%%%%%%%%%%%%%%%%%%%%%%%%%%%%%%%%%%%%%%%%%%%%%%%%%%%%%%%%%%%%%%%%%%%%%%%%
\subsubsection{\textbf{\textit{Case 2:}} $\mu_u > m_{\widetilde t_{1,2}}$}
\begin{figure}[t!]
\centering
\begin{subfigure}[t]{0.5\textwidth}
\centering
\includegraphics[height=2.1in]{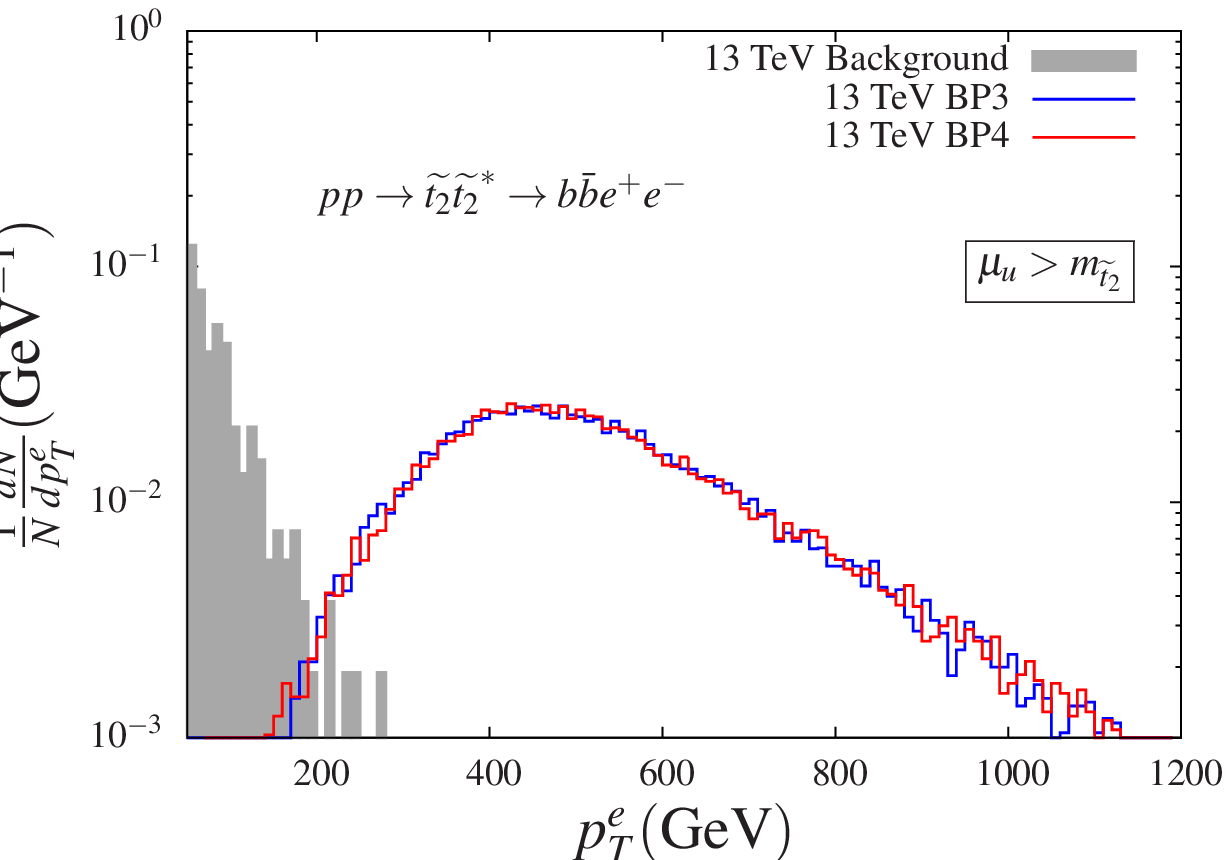}
\end{subfigure}%
~~
\begin{subfigure}[t]{0.5\textwidth}
\centering
\includegraphics[height=2.1in]{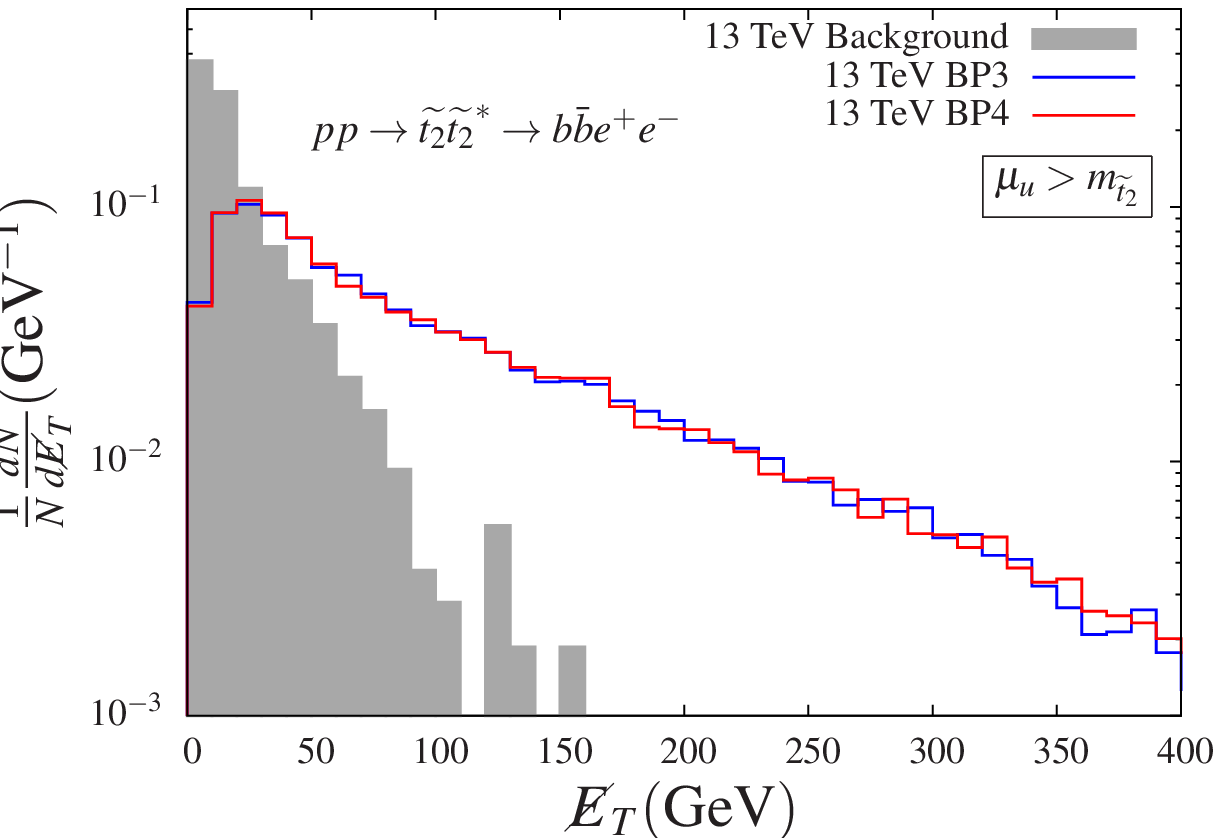}
\end{subfigure}\\
\caption{Same as in figure \ref{fig:tot2} but for benchmark scenarios BP-3 and BP-4
         and for the case $\mu_u > m_{\widetilde t_2}$. 
         } 
\label{fig:tot4}
\end{figure}

%%%%%%%%%%%%%%%%%%%%%%%%%%%%%%%%%%%%%%%%%%%%%%%%%%%%%%%%%%%%%%%%%%%%%%%%%%%%%%%%%%%%
For $\mu_u > m_{\widetilde t_{1,2}}$, the top squarks
decay mostly in a symmetric manner with $\widetilde t_1\rightarrow t\widetilde\chi_2^0$
and $\widetilde t_2\rightarrow b e^+$ as can be seen from table~\ref{Tabl4}.  
%%%%%%%%%%%%%%%%%%%%%%%%%%%%%%%%%%%%%%%%%%%%%%%%%%%%%%%%%%%%%%%%%%%%%%%%%%%%%%%%%%%%
\begin{itemize}
\item\underline{$p p\rightarrow \widetilde t_1\widetilde t_1^{*}\rightarrow
t\widetilde \chi^0_2 \; \bar{t} \widetilde \chi^0_2 \rightarrow 2$ $b$-$\text{jets}+2$ 
$\text{leptons}+\cancel{E}_T$
(figure~\ref{subfig:2})}
%%%%%%%%%%%%%%%%%%%%%%%%%%%%%%%%%%%%%%%%%%%%%%%%%%%%%%%%%%%%%%%%%%%%%%%%%%%%%%%%%%%%

For $\mu_u > m_{\widetilde t_1}$, $\widetilde t_1$ decays mostly to a top quark and
a bino-like neutralino (see table~\ref{Tabl3}). The top quark would subsequently 
decay to a $W$ boson and a $b$-jet via cascades. A pair of $W$'s can then decay 
leptonically, semi-leptonically or hadronically. We confine ourselves to leptonic 
decays of $W$-bosons for cleaner signals. The final state would then be comprised 
of 2 $b$-jets+2 leptons+$\cancel{E}_T$. The $p_T$ distributions for the harder of 
the final state leptons are shown in the left panel of figure~\ref{fig:tot3}. The 
presence of an additional source of $\cancel{E}_T$ and the heavier mass of 
$\widetilde t_1$ in the signal are behind harder $\cancel{E}_T$ distributions
(see right panel of figure (\ref{fig:tot3})) when compared to 
the SM background. To optimise the signal significance, we have observed
the dileptonic $M_{T2}^{\ell \ell}$ cut of 150 GeV works better compared to the $\cancel{E}_T$
cut of 200 GeV as used in Case 1.
%
%%%%%%%%%%%%%%%%%%%%%%%%%%%%%%%%%%%%%%%%%%%%%%%%%%%%%%%%%%%%%%%%%%%%%%%%%%%%%%%%%%%%
\begin{figure}[t!]
\centering
\begin{subfigure}[t]{0.5\textwidth}
\centering
\includegraphics[height=2.1in]{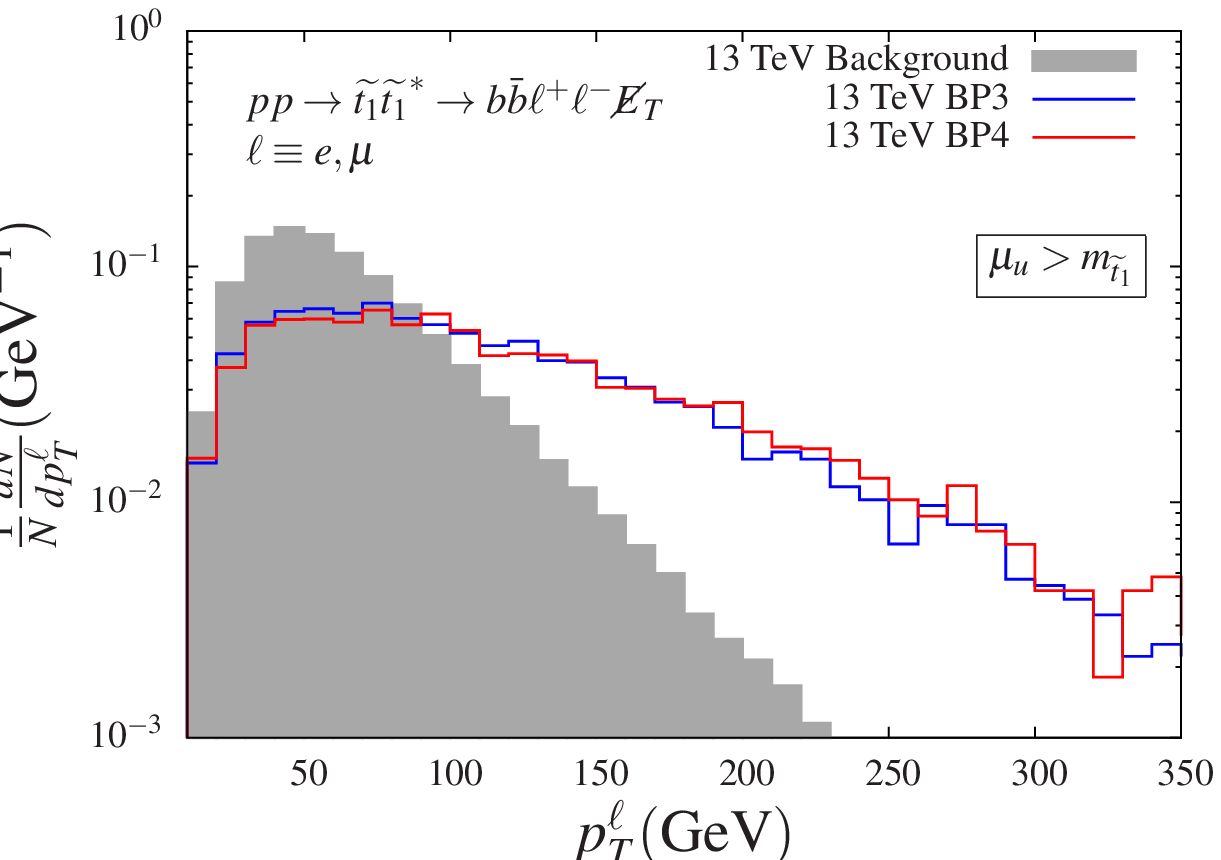}
\end{subfigure}%
~~
\begin{subfigure}[t]{0.5\textwidth}
\centering
\includegraphics[height=2.1in]{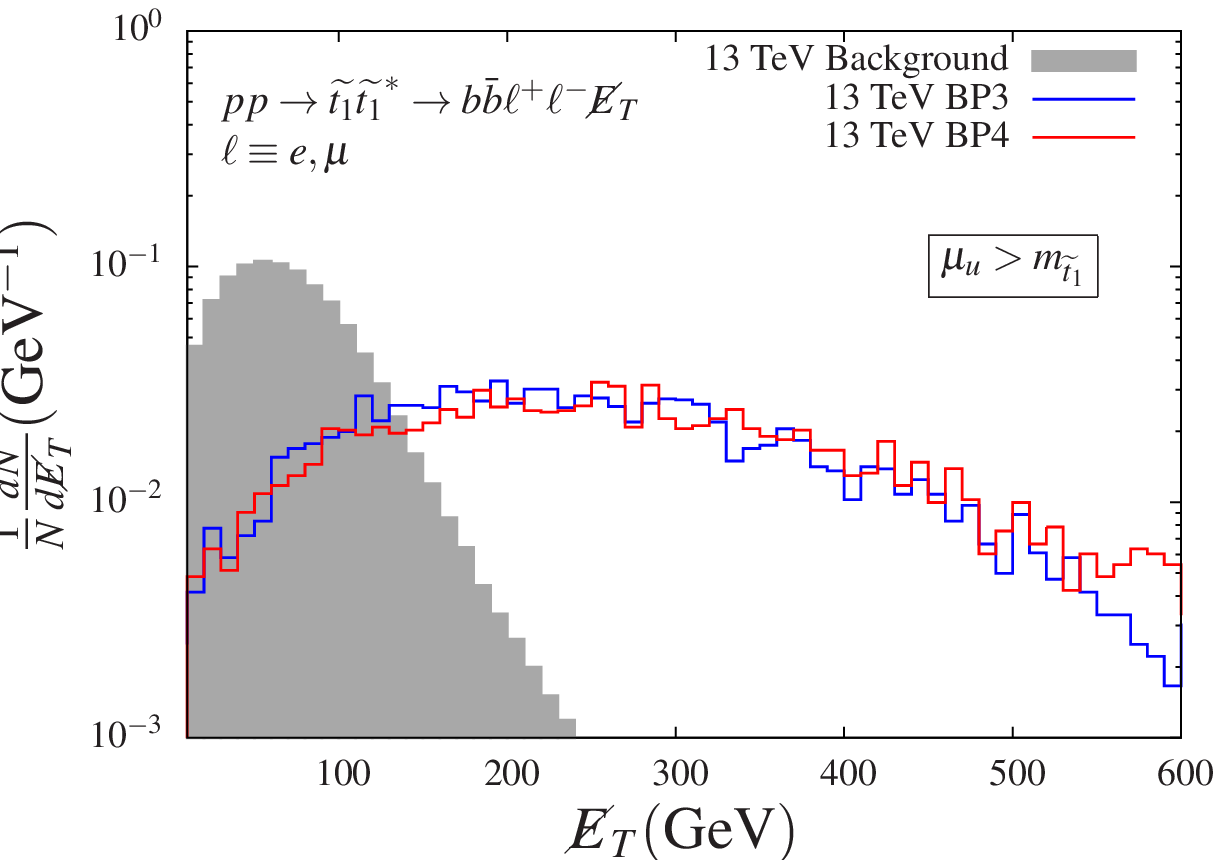}
\end{subfigure}
~~
\begin{subfigure}[t]{0.5\textwidth}
\centering
\includegraphics[height=2.1in]{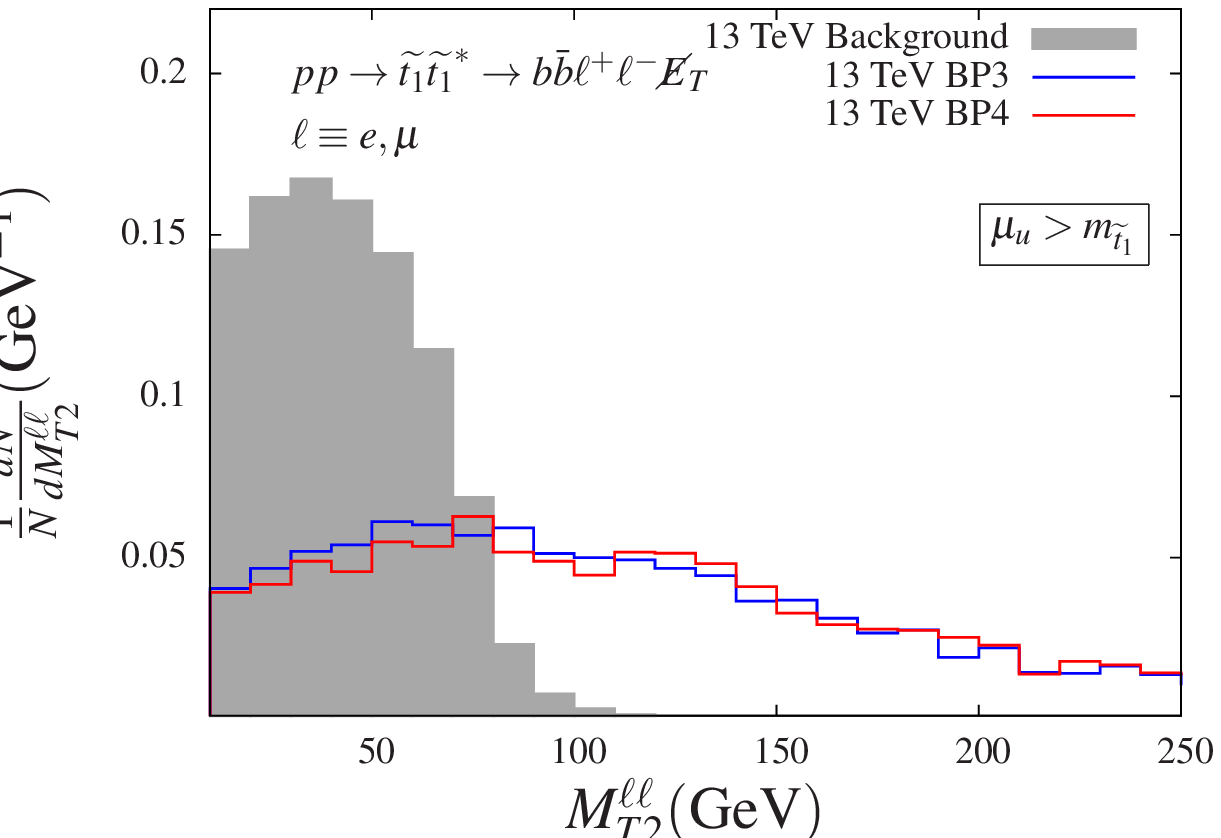}
\end{subfigure}\\
\caption{$p_T$ distributions of the harder lepton (left) and the
         $\cancel{E}_T$ distributions (right) and dileptonic stransverse mass
         for the background and the signals (in benchmark scenarios BP-3 and BP-4) 
         in the 2 $b$-jet +2 lepton+$\cancel{E}_T$ final state arising from decays of
         $\widetilde t_1$ for the case $\mu_u > m_{\widetilde t_1}$.
}
\label{fig:tot3}
\end{figure}
%%%%%%%%%%%%%%%%%%%%%%%%%%%%%%%%%%%%%%%%%%%%%%%%%%%%%%%%%%%%%%%%%%%%%%%%%%%%%%%%%%%%
\item\underline{$p p\rightarrow \widetilde t_2\widetilde t_2^{*}\rightarrow
b \bar b e^+e^-$ (figure~\ref{fig:bjet2})}
%%%%%%%%%%%%%%%%%%%%%%%%%%%%%%%%%%%%%%%%%%%%%%%%%%%%%%%%%%%%%%%%%%%%%%%%%%%%%%%%%%%%

In this case the overwhelmingly dominant decay mode is $\widetilde t_2\rightarrow
b e^{+}$. As mentioned earlier, the emitted electron (positron) could have a very 
high $p_T$ as is evident from the left panel of figure~\ref{fig:tot4}. A strong 
$p_T$ cut ($>$200 GeV) on the electron can thus be easily afforded to suppress the 
SM background effectively. 
Similar to the case of figure \ref{fig:tot2}, the distributions of MET (of
spurious origin) for the 
present case are presented in the right panel of figure \ref{fig:tot4}. Again, the 
MET distributions peak at small values ($\approx 25$ GeV), as expected and
explained earlier. Again, possible reconstructions of $\widetilde{t_2}$-s in the 
invariant mass distributions of suitable pairs of $b$-jet-electron (positron) 
systems are  on the cards. We would touch briefly on this issue later in this section.

It is also important to note that ATLAS has performed a search for RPV stops in 
this channel  \cite{ATLAS:2015jla}. Our analysis strategy is somewhat different 
from what they have chosen. ATLAS uses a cut on hadronic transverse momentum 
$H_{T}>1.1$~TeV and requires the invariant masses of the  $b$-lepton pairs to be 
within $20\%$ of each other. We find that the simple cut on electron momentum 
suppresses the background equally well and should be robust even at high pileup 
conditions.
\end{itemize}
%%%%%%%%%%%%%%%%%%%%%%%%%%%%%%%%%%%%%%%%%%%%%%%%%%%%%%%%%%%%%%%%%%%%%%%%%%%%%%%%%%%%

Before going into the assessment of the signal significance, we 
mention below some  
issues of interest/importance pertaining to possible final states 
in these two cases.
%%%%%%%%%%%%%%%%%%%%%%%%%%%%%%%%%%%%%%%%%%%%%%%%%%%%%%%%%%%%%%%%%%%%%%%%%%%%%%%%%%%%
\begin{itemize}
\item For both $\mu_u < m_{\widetilde t_1}$ (section~\ref{case1}) and 
$\mu_u > m_{\widetilde t_1}$ (section~\ref{case2}), we have only looked into the 2 
$b$-jet+2 lepton+$\cancel{E}_T$ final state arising from 
$\widetilde{t_1}$ pair production. However, the first scenario is 
phenomenologically richer 
as it can yield multi-lepton signals with 4-6 leptons in the final 
states when $\widetilde{t_1}$-s and $\widetilde{t_2}$-s decay via cascades 
involving the heavier neutralinos and charginos that in turn decay to SM $Z$ bosons
(see tables \ref{Br:Tabl1} and \ref{Br:Tabl2}).
Some corroborative analyses 
can take advantage of such inclusive final states comprising of 4 to 6 leptons 
along with $b$-jets and MET.
\item Furthermore, such a possibility could help differentiate 
$\widetilde t_2$ from the two distinct scenarios considered in this work. 
For $\mu_u < m_{\widetilde t_2}$, in addition 
to the 2 $b$-jets + $e^+ e^-$ final state out of which a pair of $\widetilde{t_2}$ could
be reconstructed, there would also be multi-lepton final states where 
leptons other than $e^+ (e^-)$ would appear. This is sharp contrast to the 
regime with $\mu_u > m_{\widetilde t_2}$.
\item A final state like 2 $b$-jets + $e^+ e^-$ arising from the decays of
$\widetilde{t_2}$-s would be ideally free from any MET. However, as pointed
out earlier, in reality, mis-measurements of various momenta may give rise to 
low to moderate amount of MET thus rendering the final state arising from a pair of
$\widetilde{t_2}$-s to be similar to that is obtained from $\widetilde{t_1}$ 
pair-production in a part of the phase space.
This gives rise to some legitimate concern as to how efficiently the signature
of $\widetilde{t_2}$-s could be deciphered, given the rates for such a final state 
originating in $\widetilde{t_1}$ pair production would be, in general, large 
thanks to smaller mass of $\widetilde{t_1}$. 

Such contaminations, however, can be avoided to a 
reasonable extent by imposing hard cuts on the minimum $p_T$ of the leading 
electrons as guided by the lepton $p_T$
distributions in the left panels of figures \ref{fig:tot1} and \ref{fig:tot2} 
(for $\mu_u < m_{\widetilde t_1}$) and figures \ref{fig:tot3} and \ref{fig:tot4} 
(for $\mu_u > m_{\widetilde t_1}$). In addition, imposition of a cut on the
maximum allowed MET 
could effectively restrict the contamination thus allowing for a more efficient
reconstruction of $\widetilde{t_2}$. 
By studying the MET distributions presented in the right panels of figures 
\ref{fig:tot2} and \ref{fig:tot4}, we find an optimal value of this cut to
be $\cancel{E}_T<$ 50 GeV that helps retain a healthy number of `signal' events with 
low $\cancel{E}_T$, a characteristic of such a final state originating in the decays
of $\widetilde{t_2}$.

%(* START HERE *)
%%%%%%%%%%%%%%%%%%%%%%%%%%%%%%%%%%%%%%%%%%%%%%%%%%%%%%%%%%%%%%%%%%%%%%%%%%%%%%%%%%%%
\begin{figure}[t!]
\centering
\begin{subfigure}[t]{0.5\textwidth}
\centering
\includegraphics[height=2.1in]{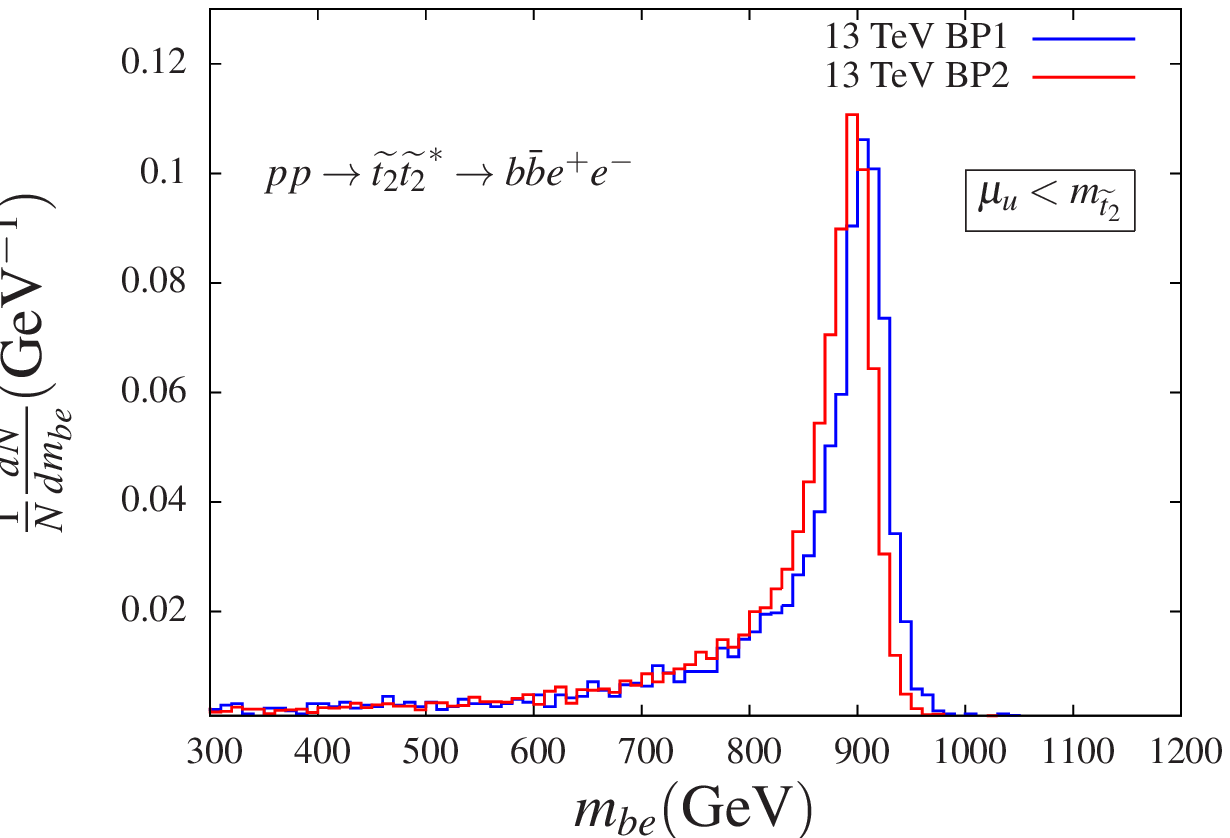}
\end{subfigure}%
~~
\begin{subfigure}[t]{0.5\textwidth}
\centering
\includegraphics[height=2.1in]{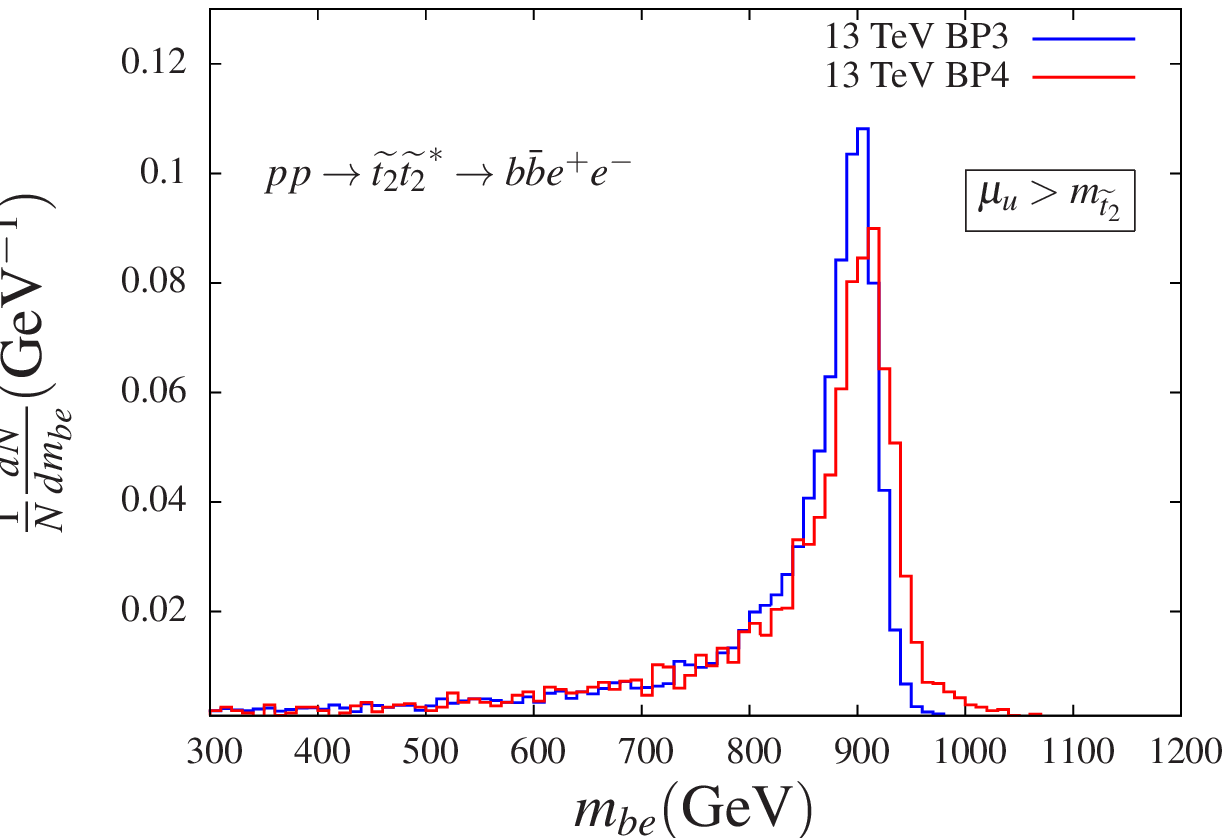}
\end{subfigure}\\
\caption{
Invariant mass distributions for the appropriate pairs of $b$-jet-eletron
      (positron) systems (with low MET characteristic of $R$-parity
      violating decays of $\widetilde{t_2}$ to a bottom quark and an electron (positron))
      for $\mu_u < m_{\tilde{t_2}}$ (left) and  $\mu_u > m_{\tilde{t_2}}$ (right).
      The distributions are obtained by imposing $p_T>$ 200 GeV for the leading
      electron and $\cancel{E}_T<$~50 GeV.
}
\label{fig:invmass}
\end{figure}
%%%%%%%%%%%%%%%%%%%%%%%%%%%%%%%%%%%%%%%%%%%%%%%%%%%%%%%%%%%%%%%%%%%%%%%%%%%%%%%%%%%%

In figure \ref{fig:invmass} we present the invariant mass distributions 
of appropriately chosen pairs of $b$-jet and an electron (positron). Guided
by figures \ref{fig:tot2} and \ref{fig:tot4}, a high $p_T$ threshold of 200 
GeV for the leading electron is demanded along with requiring a $\cancel{E}_T<$ 50
GeV to ensure that we mostly confine ourselves to the signal region. The
left (right) panel of figure \ref{fig:invmass} represent the case with
$\mu_u < m_{\widetilde{t_2}}$ ($\mu_u > m_{\widetilde{t_2}}$). We find that
in both cases clear peaks at $m_{\widetilde{t_2}}$ show up thus raising the 
hope that not only $\widetilde{t_2}$-s could be discovered in this mode but 
also a reliable estimate of its mass would be possible.
\end{itemize}

Before we close this subsection we like to mention that although we have only
discussed two broad scenarios, i.e., $\mu_u < m_{\widetilde{t_2}}$ and 
$\mu_u > m_{\widetilde{t_2}}$, other intermediate situations are all a priori
viable. However, the expectations under those scenarios could be substantiated
in a straightforward manner from the two cases we present. For example,
an increase in value of $\mu_u$ from that in Case 1 would result in
suppression of the branching fractions to Higgsino-like neutralinos and
charginos. With increasing $\mu_u$, at some point, these decay-modes 
(see table 3) would be closed for $\widetilde{t_1}$ and 
BR($\widetilde{t_1} \to t \tilde{\chi}_2^0$) = 1. At the same time,
branching fractions to the Higgsino-like states for $\widetilde{t_2}$ would
also get suppressed before these decay-modes get completely closed as it
happens in Case 2. A detail study of possible correlations
among the event rates in various final states could, in principle, shed light
on the relative value of $\mu_u$ with respect to $m_{\widetilde{t_1}}$ and
$m_{\widetilde{t_2}}$. However, this is beyond the scope of the present work.

%%%%%%%%%%%%%%%%%%%%%%%%%%%%%%%%%%%%%%%%%%%%%%%%%%%%%%%%%%%%%%%%%%%%%%%%%%%%%%%%%%%%
\subsection{Signal significance and the reach}
\label{subsec:reach}
%%%%%%%%%%%%%%%%%%%%%%%%%%%%%%%%%%%%%%%%%%%%%%%%%%%%%%%%%%%%%%%%%%%%%%%%%%%%%%%%%%%%
The signal significance ($\sigma$) is estimated using the expression~\cite{CMS-sig}
%%%%%%%%%%%%%%%%%%%%%%%%%%%%%%%%%
\begin{eqnarray}
\sigma=\sqrt{2\Big[(S+B)\text{ln}\Big(1+\frac{S}{B}\Big)-S\Big]}
\label{sig}
\end{eqnarray}
%%%%%%%%%%%%%%%%%%%%%%%%%%%%%%%%%
which is appropriate for the situation with small number of events (in particular
when the number of background events is less than 50). Equation~(\ref{sig}) is based
on likelihood-ratios and follows from the Poisson distribution.
Here, $S$ and $B$ stand for the numbers of the signal and the background events, 
respectively
after imposition of the set of optimal cuts discussed in section \ref{subsec:selection}.
The $K$-factors for $\widetilde t\widetilde t^*$ are computed using 
{\tt Prospino2 (v2.1)}~\cite{P1,Prospino}. 

We now estimate the required integrated
luminosities for a $5\sigma$ reach of $\widetilde{t}_1$ and $\widetilde{t}_2$
in the four benchmark scenarios we
consider. The final states we focus on are 2 $b$-jets+2 leptons+$\cancel{E}_T$
and $b\bar b e^+ e^-$, which stem from the decay of $\widetilde t_1$ and
$\widetilde t_2$, respectively.
%%%%%%%%%%%%%%%%%%%%%%%%%%%%%%%%%%%%%%%%%%%%%%%%%%%%%%%%%%%%%%%%%%%%%%%%%%%%%%%%%%%%
%%%%%%%%%%%%%%%%%%%%%%%%%%%%%%%%%%%%%%%%%%%%%%%%%%%%%%%%%%%%%%%%%%%%%%%%%%%%%%%%%%%%
\begin{table}[h!]
\centering
 \begin{tabular}{||c|c|c|c|c||} 
 \hline
 $pp\rightarrow\widetilde t_1\widetilde t_1^*\rightarrow 2~b$-$\text{jets}
+2~\text{leptons}+\cancel{E}_T$ & BP1 & BP2 & BP3 & BP4 \\ [0.5ex] 
 \hline\hline
 $\sigma (pp\rightarrow\widetilde t_1\widetilde t_1^*)$ (fb) & 428.9 & 463.4 & 193.0 & 73.6 \\ 
 \hline
Cut acceptance for signal & $1.5\times 10^{-2}$ & 1.6$\times 10^{-2}$ & 
4.4$\times 10^{-3}$ & $4.4\times 10^{-3}$ \\
($\cancel{E}_T>$ 200 GeV for BP1 and BP2) & & & &\\
($M_{T2}^{\ell\ell}>150$ GeV for BP3 and BP4) & & & & \\
\hline 
Required $\mathcal{L}~(\text{fb}^{-1})$ for $5\sigma$ significance  & 256.0 & 316 & 2350.0 & 3000 (3$\sigma$) \\ [1ex]
 \hline 
\end{tabular}
\caption{Required values of integrated luminosities ($\mathcal{L}$) to obtain a $5\sigma$ significance
        in the final state at $\sqrt{s}=13$ TeV. The most important SM background arising
        from $t\bar{t}$ pair production is normalised to a cross section of $\approx 816$ pb
        obtained at the NLO+NNLL level (see section \ref{subsec:smbg}). The cut  acceptance
        for the background is 2.3$\times 10^{-4}$. A flat $b$-tagging
        efficiency of 60\% is used. } 
\label{stop:1}
\end{table}
%%%%%%%%%%%%%%%%%%%%%%%%%%%%%%%%%%%%%%%%%%%%%%%%%%%%%%%%%%%%%%%%%%%%%%%%%%%%%%%%%%%%
For the first case (see table~\ref{stop:1}) the dominant background comes from $t\bar t$ pair 
production which subsequently decays to the 2 $b$-jet+2 lepton+$\cancel{E}_T$ final 
state. An appropriate $K$-factor of $\approx$ 1.6 is used to derive the NLO cross sections
from the LO ones for $\widetilde t$-pair production.

We note that a $5\sigma$ signal significance can 
be achieved for BP-1 and BP-2, with an integrated luminosity around 100 fb$^{-1}$. 
To achieve a similar significance for BP-3 and BP-4, one has to wait for 
a much higher accumulated luminosity, for example, 500 fb$^{-1}$ and 3000 fb$^{-1}$, 
respectively
at $\sqrt{s}=13$ TeV.

%%%%%%%%%%%%%%%%%%%%%%%%%%%%%%%%%%%%%%%%%%%%%%%%%%%%%%%%%%%%%%%%%%%%%%%%%%%%%%%%%%%
\begin{table}[h!]
\centering
 \begin{tabular}{||c|c|c|c|c||} 
 \hline
 $p p\rightarrow\widetilde t_2\widetilde t_2^*\rightarrow 2~b$-$\text{jets}+e^+e^-$ & BP1 & BP2 & BP3 & BP4 \\ [0.5ex] 
 \hline\hline
 $\sigma (pp\rightarrow\widetilde t_2\widetilde t_2^*)$ (fb) & 7.83 & 8.69 & 8.45 & 8.48 \\ 
 \hline
 Cut acceptance for signal & 1.96$\times 10^{-2}$ 
 & 5.26$\times 10^{-2}$ & 1.9$\times 10^{-1}$ & 1.9$\times 10^{-1}$ \\
 ($p_T > 200$ GeV, $\cancel{E}_T<$ 50 GeV) & & & & \\ 
 \hline
 Required $\mathcal{L}~(\text{fb}^{-1})$ for $5\sigma$ significance & 501.23 & 63.92 & 6.85 & 6.61 \\ [1ex]
 \hline 
\end{tabular}
\caption{Required values of integrated luminosities ($\mathcal{L}$) to obtain a 
        $5\sigma$ significance in the 2 $b$-jets + $e^+e^-$ final state at 
$\sqrt{s}=13$ TeV. The SM background
        (see section \ref{subsec:smbg}) at NLO is found to be 9.43 pb (see section
        \ref{subsec:smbg}). The cut acceptance
        for the background is 1.25$\times 10^{-4}$. A flat $b$-tagging
        efficiency of 60\% is used.
}
\label{stop:2}
\end{table}
%%%%%%%%%%%%%%%%%%%%%%%%%%%%%%%%%%%%%%%%%%%%%%%%%%%%%%%%%%%%%%%%%%%%%%%%%%%%%%%%%%%%

In table~\ref{stop:2} we present the required luminosities for a $5\sigma$ 
reach of $\widetilde{t_2}$ for the four benchmark points. The dominant 
SM background comes from $b\bar{b}Z/\gamma^*$ production followed by
$Z/\gamma^*$ giving rise to $e^+e^-$ pairs. This can be efficiently suppressed by
using an on-shell $Z$-veto for the $e^+e^-$ pairs, as discussed in section
\ref{subsec:smbg}.
Thus, as can be seen from this table, a $5\sigma$
significance can be obtained with an integrated luminosity as low as $< 10$ fb$^{-1}$
for the benchmark scenarios BP-3 and BP-4 with $\sqrt{s}=13$ TeV. 
In addition, we also study the $H_T$ distribution, i.e., the scalar
sum of the $p_T$ of the $e^+ e^-$ pair and the reconstructed $b$-jets and
the improvements are marginal.
The wildly
varying integrated luminosities across the benchmark points are the artifact of
varying branching fractions that are instrumental, as has been pointed out in
section \ref{benchmarks}.

%%%%%%%%%%%%%%%%%%%%%%%%%%%%%%%%%%%%%%%%%%%%%%%%%%%%%%%%%%%%%%%%%%%%%%%%%%%%%%%%%%%%
\begin{figure}[t!]
    \centering
    \begin{subfigure}[t]{0.5\textwidth}
        \centering
        \includegraphics[height=2.3in]{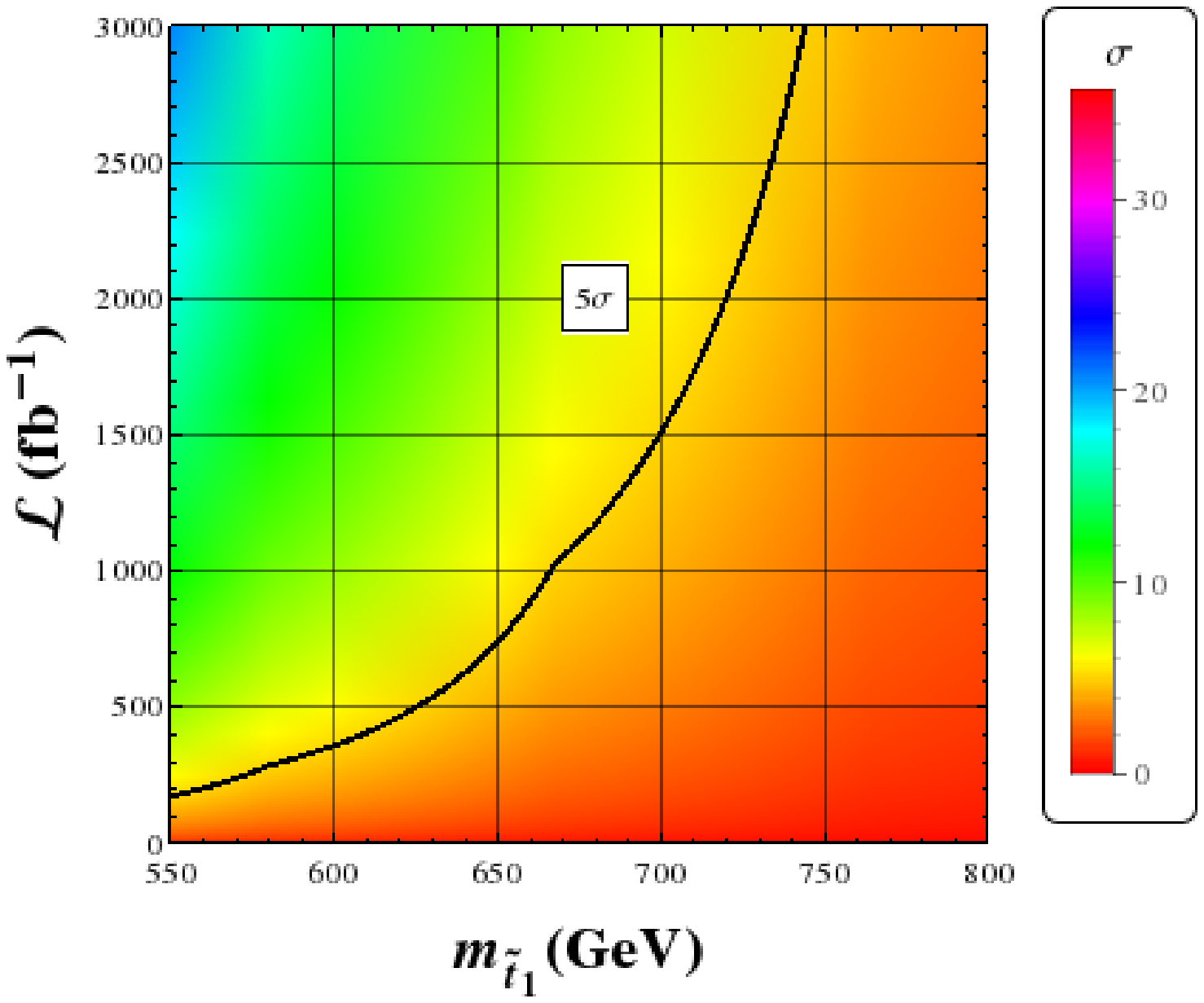}
    \end{subfigure}%
    ~~ 
    \begin{subfigure}[t]{0.5\textwidth}
        \centering
        \includegraphics[height=2.3in]{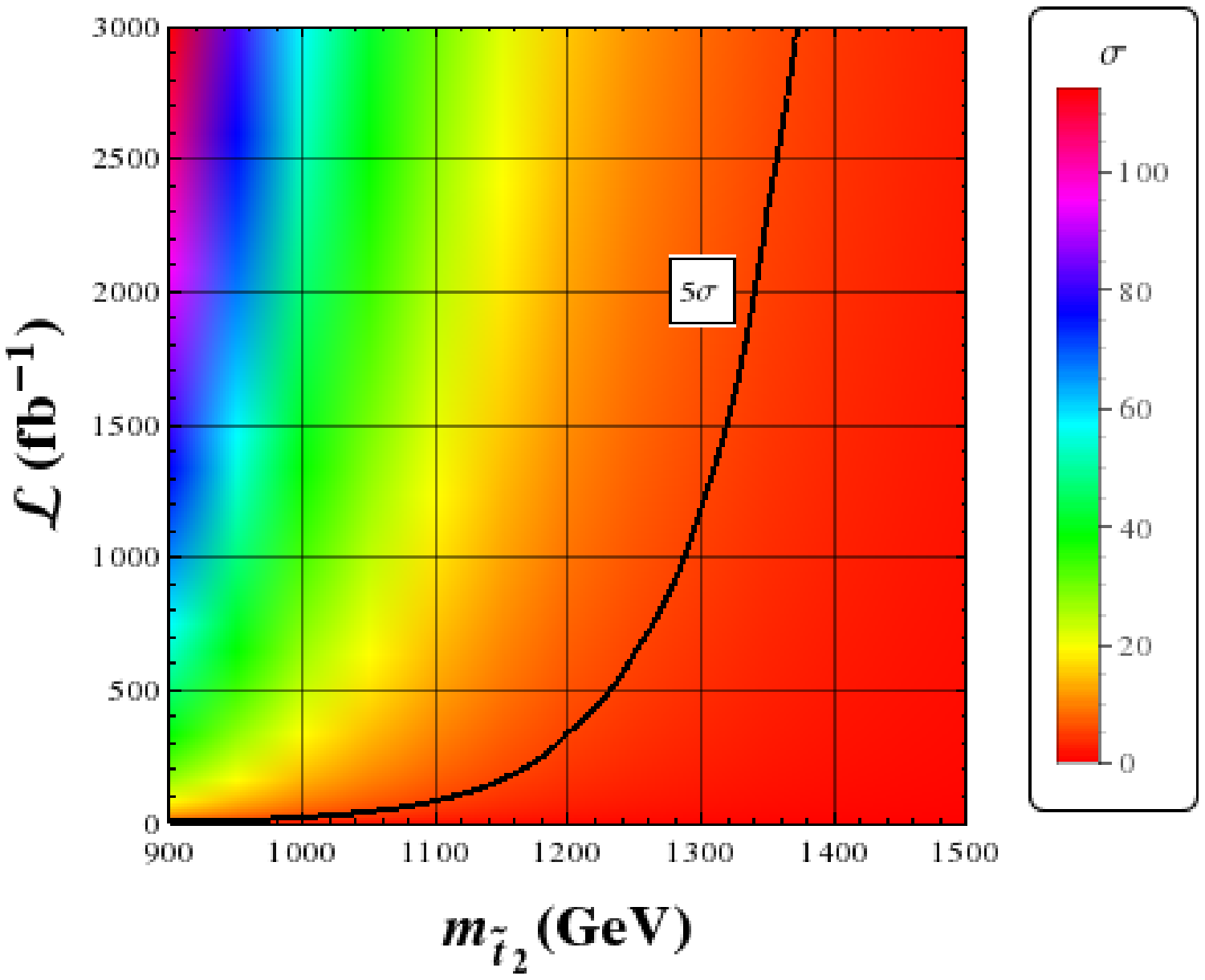}
    \end{subfigure}\\
    \caption{Density plot reflecting the reach for top squark masses via
            2 $b$-jet+2 lepton+$\cancel{E}_T$ final state in BP-$2$ in the
            $m_{\widetilde t_1}$-$\mathcal{L}$ plane (left) and via 2 $b$-jet+
            $e^+e^-$ final state in BP-$4$ in the $m_{\widetilde t_2}$-$\mathcal{L}$
            plane (right). The thick black curves are contours of $5\sigma$
            significance above and on the left of which the masses can be explored
            with $\geq 5\sigma$ significance. The cuts are kept fixed at values
            mentioned in the text for the respective cases.
             }
    \label{fig:density}
\end{figure}
%%%%%%%%%%%%%%%%%%%%%%%%%%%%%%%%%%%%%%%%%%%%%%%%%%%%%%%%%%%%%%%%%%%%%%%%%%%%%%%%%%%%

Figure~\ref{fig:density} summarises the mass-reach for the two top squarks with
varying accumulated integrated luminosities (or, in other words, luminosity
required to probe a certain top squark mass) at the 13 TeV LHC. The left panel
illustrates the case for $\widetilde t_1$ in the final state 2$b$+2 lepton+$\cancel{E}_T$
in BP-2 while the right one does the same for $\widetilde t_2$ via 2$b$+$e^+e^-$
final state in BP-4. As indicated by tables~\ref{stop:1} and \ref{stop:2}, figure
\ref{fig:density} also reveals that $\widetilde t_2$ has a significantly better reach
compared to $\widetilde t_1$ with the final states under consideration. 
This may lead to a tantalising possibility of discovering $\widetilde t_2$
of such a scenario much earlier than $\widetilde t_1$ and the former could guide
us to find the latter. 
We observe that at the 13 TeV LHC and with 
the mass-reaches for $\widetilde t_1$
and $\widetilde t_2$ are around 575 (750) GeV and 1.2 (1.4) TeV respectively, with an integrated
luminosity of 300 (3000) fb$^{-1}$. 

%%%%%%%%%%%%%%%%%%%%%%%%%%%%%%%%%%%%%%%%%%%%%%%%%%%%%%%%%%%%%%%%%%%%%%%%%%%%%%%%%%%%
\section{The `stealth' top  squark scenario}
\label{stealth}
%%%%%%%%%%%%%%%%%%%%%%%%%%%%%%%%%%%%%%%%%%%%%%%%%%%%%%%%%%%%%%%%%%%%%%%%%%%%%%%%%%%%
The SUSY model under consideration, with super-light carriers of MET like
$\widetilde{\chi}_1^0 \equiv \nu_e$ and an MeV neutralino LSP
($\widetilde{\chi}_2^0$) can easily 
conceive a rather low mass top squark lying right in the so-called 
`stealth' window of $197~\text{GeV} \lesssim m_{\widetilde t_1} \lesssim 205~\text{GeV}$
~\cite{Aad:2015pfx}. 
As discussed in section \ref{bounds}, the experimental 
lower bound on $m_{\widetilde t_L}$ is more stringent considering its decay 
modes. Hence we choose $\widetilde t_R$ to be the lightest top squark 
($\widetilde{t_1}$). A benchmark point can be obtained by choosing $(m^2_u)_{33}
=-2.5\times 10^4$ GeV$^2$. This results in $m_{\widetilde t_1}\sim 200$ GeV. 
Such a light top squark cannot provide enough correction to the Higgs mass. 
Hence we choose a relatively large value of $\lambda_S$ (=1.28) so that the 
radiatively generated additional quartic contributions could lift the Higgs 
boson mass to the observed range. All other parameters are fixed at the 
values mentioned in BP-1~(see table~\ref{Tab:Spec1}). Note that the additional
tree level contribution proportional to the neutrino Yukawa coupling `$f$' 
remains small (even for its order one value)
because of large values of $\tan\beta$ that we require.
As a result, $\widetilde{t_1}$ mostly 
decays to $t\widetilde\chi_2^0$ and $\widetilde t_1\rightarrow t\nu_e$ with 
$\sim$ 85\% and $\sim$ 15\% branching fractions, respectively. The possible
final state topologies are exactly the same as those result from top quark 
pair production. 

\begin{figure}[t!]
    \centering
    \begin{subfigure}[t]{0.5\textwidth}
        \centering
        \includegraphics[height=2.3in,width=2.8in]{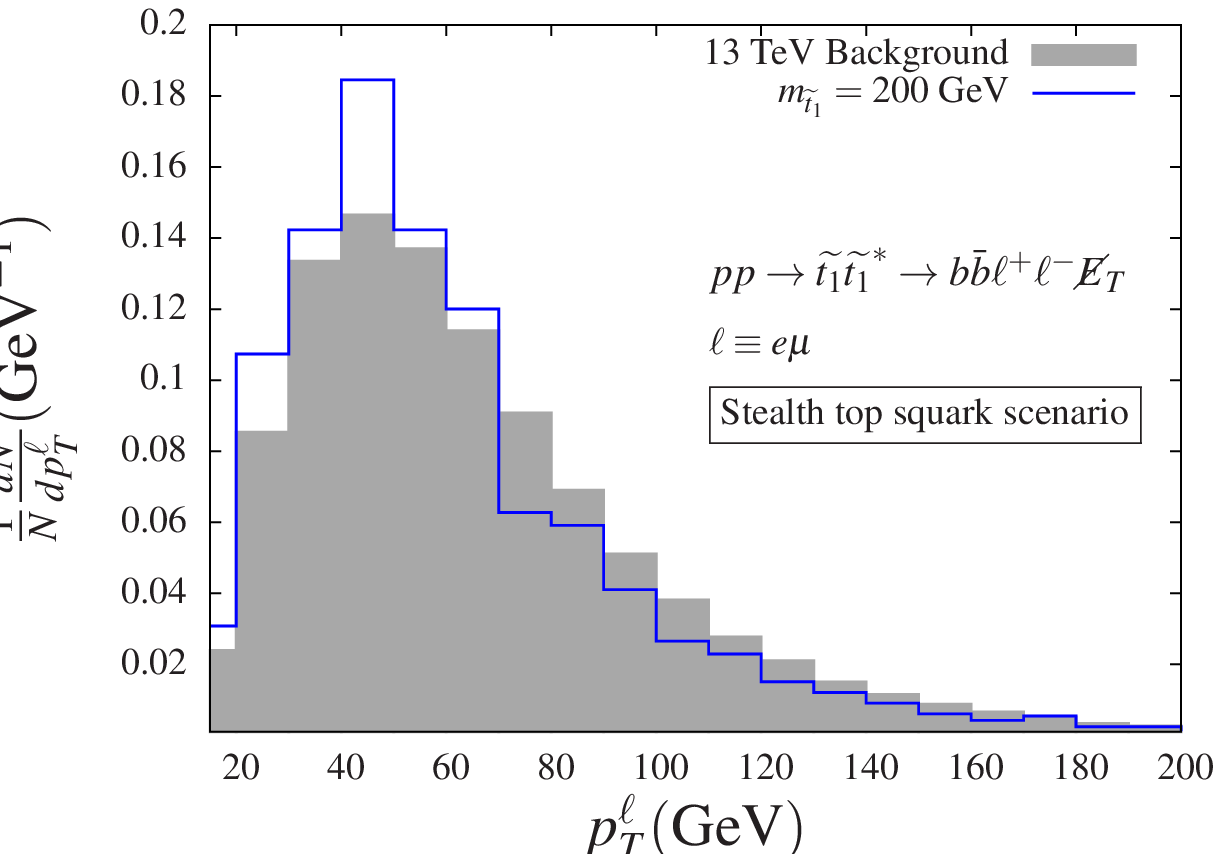}
    \end{subfigure}%
    ~~
        \begin{subfigure}[t]{0.5\textwidth}
        \centering
        \includegraphics[height=2.3in,width=2.8in]{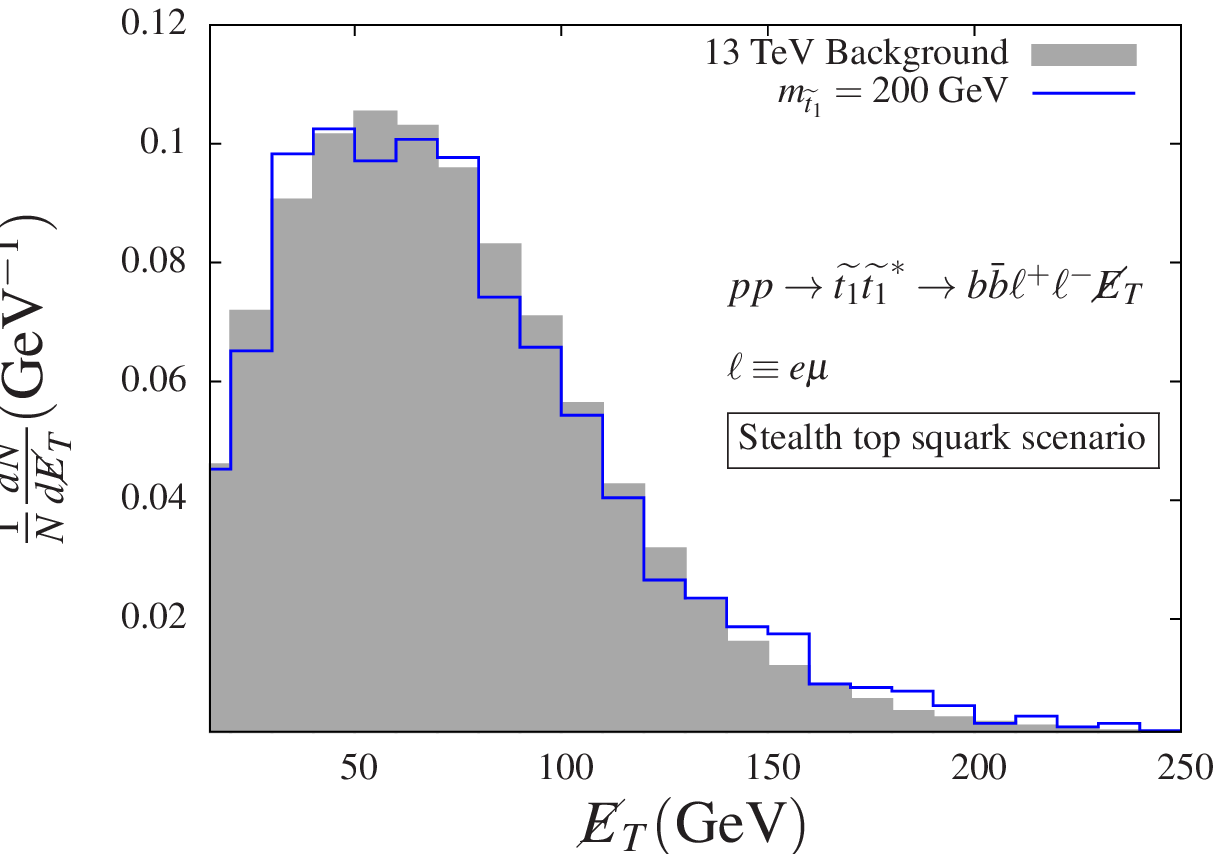}
    \end{subfigure}\\
    \caption{$p_T$ distributions of the harder lepton (left) and the
            $\cancel{E}_T$ distributions (right) for the background and the
            signal in the 2$b$-jet+2 $\text{lepton} +\cancel{E}_T$ final state arising from 
            the decays of a pair of $\widetilde t_1$-s in the `stealth' top squark 
            scenario.}
    \label{ss}
    \end{figure}

We again analyse the final 
state with 2 $b$-jets+ 2 leptons+$\cancel{E}_T$. We checked that the
distributions of various kinematic observables look very similar for the
signal and the $t\bar{t}$ background, which is something literally expected
of a `stealth' top squark and what makes it so elusive. In figure \ref{ss}
we present the $p_T$ distribution of the harder lepton (left panel) and
the $\cancel{E}_T$ ~distribution (right panel) which clearly demonstrate how similar
the behaviors of the SM background and the signal could get. 
In this context, techniques to exploit differences in spin-correlations
inherent to $t\bar{t}$ and $\widetilde t \widetilde t^*$ systems~
\cite{stealth-stop}, use of various transverse mass variables \cite{Cho:2014yma} 
including the one like $m_{T_2}$ in the dileptonic decay 
channel~\cite{Kilic}, incorporating a new variable like `topness' \cite{Graesser:2012qy}
using asymmetric decays of the top squarks have been proposed to study the 
`stealth' top squark regime in search for an
improved sensitivity. Clearly the issue demands dedicated addressal which is beyond
the scope of the present discussion.
%%%%%%%%%%%%%%%%%%%%%%%%%%%%%%%%%%%%%%%%%%%%%%%%%%%%%%%%%%%%%%%%%%%%%%%%%%%%%%%%%%%%
%%%%%%%%%%%%%%%%%%%%%%%%%%%%%%%%%%%%%%%%%%%%%%%%%%%%%%%%%%%%%%%%%%%%%%%%%%%%%%%%%%%%
%%%%%%%%%%%%%%%%%%%%%%%%%%%%%%%%%%%%%%%%%%%%%%%%%%%%%%%%%%%%%%%%%%%%%%%%%%%%%%%%%%%%
\section{Summary and conclusions}
\label{conclusion}
%%%%%%%%%%%%%%%%%%%%%%%%%%%%%%%%%%%%%%%%%%%%%%%%%%%%%%%%%%%%%%%%%%%%%%%%%%%%%%%%%%%%
We study a $U(1)_R-$lepton number model augmented with a right handed neutrino
superfield. The $R$-charges are identified with the lepton numbers in such a 
way that the sneutrinos acquire large $vev$-s. Such large $vev$-s for sneutrinos are 
not prohibited since, in such a scenario, the same are not constrained by the Majorana
masses of the neutrinos. In this paper, we choose to work in a basis in which
only the electron type sneutrino acquires a nonzero $vev$ whereas the $vev$-s of the other
two sneutrinos are rotated away. This simple extension with a right handed neutrino 
superfield is rather interesting in the sense that the Higgs boson mass gets a tree 
level contribution which can be substantial in the low $\tan\beta$ regime and for 
order one neutrino Yukawa coupling `$f$'. Also present are the large one loop 
contribution to the Higgs boson mass arising from new couplings in the theory. Thus, 
one can easily accommodate a scenario where both top squarks are light. In addition, 
a very light bino-like neutralino comes out naturally in this scenario
along with an active neutrino endowed with an appropriately small Majorana mass.
Therefore, rich and interesting collider signatures are expected in such a scenario.
The signature of a top squark decaying to a top quark and a neutralino and/or a
neutrino is similar to the top quark pair production in the SM. Under favorable circumstances, 
top squark could also decay to a bottom quark and a chargino leading to a similar final 
state containing 2 $b$-jets and 2 leptons along with MET. In this mode, top squark 
mass of around 575 (750) GeV can be probed with 300 (3000) $\textrm{fb}^{-1}$ of 
integrated luminosity. 

Furthermore, in an $R$-parity violating scenario such as ours, the charginos mix with the 
electron. The decay width of the top squark to
a bottom quark and an electron (positron) is enhanced because of the enhanced coupling 
$\lambda^{\prime}_{133}$ as well as an unsuppressed phase space. Hence we study in detail 
the final state with 2 $b$-jets accompanied by an $e^+e^-$ pair arising from such a
dominant decay. We show that 
even when the top squark is heavy ($m_{\widetilde t_2}\lsim 1.2 (1.4)$ TeV), this particular 
channel could deliver a large signal significance with 300 (3000)~$\textrm{fb}^{-1}$ of integrated 
luminosity. 

In the model discussed in this work, only ${\widetilde t_2} (\approx {\widetilde t_L})$ 
decays to $b e^+$. This is an artifact of no $L$-$R$ mixing in the top squark sector. 
Also, note that the scenario prohibits $\widetilde t_2$ decaying to $b\mu^+$ or $b\tau^+$.
The final state arising from both top squarks decaying to $be^+ (\bar b e^-)$ mode could carry MET 
which can only be of spurious origin (mis-measurements of various visible momenta, 
defects in the detector, etc.) and hence is characteristically small. This feature 
can be used to establish such a model and differentiate it from other competing ones. 
Due to a relatively clean final state and hence, a possibility to reconstruct the 
heavier top squark mass reasonably efficiently, such a state could be within an easier 
reach of the current LHC run when compared to its lighter peer. Such a scenario thus, 
gives rise to an interesting possibility that $\widetilde t_2$ can be found much earlier 
than $\widetilde t_1$ at the LHC and could carry a reliable hint as to where exactly 
to look for the latter. The signal region for $\widetilde{t_1}$ is attributed with a 
much larger MET as is usual in searches for new heavy states in scenarios with a stable 
charge- and color-neutral particle(s). This is in sharp contrast with the case of 
$\widetilde t_2$ in such a scenario.

Although the analyses in this work are presented in terms of two broad scenarios, viz., 
$\mu_u < m_{\widetilde{t_2}}$ and $\mu_u > m_{\widetilde{t_2}}$, it is pointed out that 
the signatures discussed are robust under intermediate situations except for some 
obvious quantitative issues getting in. Simultaneous searches in various channels 
described in this work are expected to shed light on the detailed aspect of the spectrum 
and the involved new couplings of such a scenario.

Finally, we have demonstrated how the `stealth' top squark can appear in our model
naturally. However, probing such a window, 
197 GeV $\lesssim m_{\widetilde t} \lesssim$ 205 GeV~\cite{Aad:2015pfx} 
needs dedicated analysis which is an active area of research on its own merit. 
Overall, characteristic signatures for these light top squark states
at the LHC even have the potential to discriminate between competing scenarios that
may give rise to such a light pair of top squarks. In addition, such issues
and projections are not readily available for 13/14 TeV run. Therefore, it is 
important to study all these issues at the dawn of 13 TeV run of the LHC.

%%%%%%%%%%%%%%%%%%%%%%%%%%%%%%%%%%%%%%%%%%%%%%%%%%%%%%%%%%%%%%%%%%%%%%%%%%%%%%%%%%%%
\acknowledgments
SC would like to thank the Council of Scientific and Industrial Research, 
Government of India for the financial support received as a Senior Research 
Fellow. It is a pleasure to thank Florian Staub for many helpful discussions
regarding {\tt SARAH}. SC would also like to thank Subhadeep Mondal, Arghya Choudhury 
and Amit Chakraborty for many
helpful discussions. AD would like to thank the Department of Theoretical Physics,
IACS for hospitality during the course of this work.
SR acknowledges the hospitality of the University 
of Helsinki and Helsinki Institute of Physics where this work was initiated.  
KH and HW acknowledge support from the Academy of Finland (project no 137960).

%%%%%%%%%%%%%%%%%%%%%%%%%%%%%%%%%%%%%%%%%%%%%%%%%%%%%%%%%%%%%%%%%%%%%%%%%%%%%%%%%%%%

%%%%%%%%%%%%%%%%%%%%%%%%%%%%%%%%%%%%%%%%%%%%%%%%%%%%%%%%%%%%%%%%%%%%%%%%%%%%%%%%%%%%

\begin{thebibliography}{99}
%%%%%%%%%%%%%%%%%%%%%%%%%%%%%%%%%%%%%%%%%%%%%%%%%%%%%%%%%%%%%%%%%%%%%%%%%%%%%%%%%%%%
\bibitem{ATLAS-higgs}
  G.~Aad {\it et al.}  [ATLAS Collaboration],
%  ``Observation of a new particle in the search for the Standard Model Higgs 
%   boson with the ATLAS detector at the LHC,''
  Phys.\ Lett.\ B {\bf 716} (2012) 1
  [arXiv:1207.7214 [hep-ex]].

%%%%%%%%%%
\bibitem{CMS-higgs}
  S.~Chatrchyan {\it et al.}  [CMS Collaboration],
%  ``Observation of a new boson at a mass of 125 GeV with the CMS experiment at the LHC,''
  Phys.\ Lett.\ B {\bf 716} (2012) 30
  [arXiv:1207.7235 [hep-ex]].

\bibitem{CMS-bound}
  S.~Chatrchyan {\it et al.}  [CMS Collaboration],
%  ``Search for supersymmetry in hadronic final states with missing transverse 
%   energy using the variables $\alpha_T$ and b-quark multiplicity in pp collisions 
%  at $\sqrt s=8$ TeV,''
  Eur.\ Phys.\ J.\ C {\bf 73} (2013) 9,  2568
  [arXiv:1303.2985 [hep-ex]].

\bibitem{ATLAS-bound}
  G.~Aad {\it et al.}  [ATLAS Collaboration],
%  ``Search for new phenomena in final states with large jet multiplicities and 
%    missing transverse momentum at $\sqrt{s}$=8 TeV proton-proton collisions using 
%    the ATLAS experiment,''
  JHEP {\bf 1310} (2013) 130
   [Erratum-ibid.\  {\bf 1401} (2014) 109]
  [arXiv:1308.1841 [hep-ex]].

\bibitem{ATLAS-bound-1}
  G.~Aad {\it et al.}  [ATLAS Collaboration],
%  ``Search for squarks and gluinos with the ATLAS detector in final states with 
%  jets and missing transverse momentum using $\sqrt{s}=8$ TeV proton--proton collision data,''
  JHEP {\bf 1409} (2014) 176
  [arXiv:1405.7875 [hep-ex]].
%%%%%%%%%%
\bibitem{Fayet}
  P.~Fayet,
%  ``Supersymmetry and Weak, Electromagnetic and Strong Interactions,''
  Phys.\ Lett.\ B {\bf 64} (1976) 159.

\bibitem{Polchinski}
J.~Polchinski and L.~Susskind,
%``Breaking Of Supersymmetry At Intermediate-Energy,''
Phys.\ Rev.\  D {\bf 26}, 3661 (1982).

\bibitem{Hall-1}
  L.~J.~Hall,
%  ``Alternative Low-energy Supersymmetry,''
  Mod.\ Phys.\ Lett.\ A {\bf 5} (1990) 467.

\bibitem{Hall-2}
  L.~J.~Hall and L.~Randall,
%  ``U(1)-R symmetric supersymmetry,''
  Nucl.\ Phys.\ B {\bf 352} (1991) 289.

\bibitem{Jack}
  I.~Jack and D.~R.~T.~Jones,
%  ``Nonstandard soft supersymmetry breaking,''
  Phys.\ Lett.\ B {\bf 457} (1999) 101
  [hep-ph/9903365].

\bibitem{Fox}
  P.~J.~Fox, A.~E.~Nelson and N.~Weiner,
%  ``Dirac gaugino masses and supersoft supersymmetry breaking,''
  JHEP {\bf 0208} (2002) 035
  [hep-ph/0206096].

\bibitem{Nelson}
  A.~E.~Nelson, N.~Rius, V.~Sanz and M.~Unsal,
%  ``The Minimal supersymmetric model without a mu term,''
  JHEP {\bf 0208} (2002) 039
  [hep-ph/0206102].
 
\bibitem{Chacko}
  Z.~Chacko, P.~J.~Fox and H.~Murayama,
%  ``Localized supersoft supersymmetry breaking,''
  Nucl.\ Phys.\ B {\bf 706} (2005) 53
  [hep-ph/0406142].

\bibitem{Antoniadis}
I.~Antoniadis,  K.~Benakli, A.~Delgado, M.~Quiros and M.~Tuckmantel,
%``Splitting extended supersymmetry,''
Phys.\ Lett.\  B {\bf 634}, 302 (2006)
[arXiv:hep-ph/0507192];
\bibitem{Antoniadis1}
I.~Antoniadis, K.~Benakli, A.~Delgado, M.~Quiros and M.~Tuckmantel,
%``Split extended supersymmetry from intersecting branes,''
Nucl.\ Phys.\  B {\bf 744}, 156 (2006)
[arXiv:hep-th/0601003].

\bibitem{Antoniadis-1}
I.~Antoniadis, K.~Benakli, A.~Delgado and M.~Quiros,
%``A new gauge mediation theory,''
Adv.\ Stud.\ Theor.\ Phys.\  {\bf 2}, 645 (2008)
[arXiv:hep-ph/0610265].
  
\bibitem{kribs}
  G.~D.~Kribs, E.~Poppitz and N.~Weiner,
%  ``Flavor in supersymmetry with an extended R-symmetry,''
  Phys.\ Rev.\ D {\bf 78} (2008) 055010
  [arXiv:0712.2039 [hep-ph]].

\bibitem{Choi}
  S.~Y.~Choi, M.~Drees, A.~Freitas and P.~M.~Zerwas,
%  ``Testing the Majorana Nature of Gluinos and Neutralinos,''
  Phys.\ Rev.\ D {\bf 78} (2008) 095007
  [arXiv:0808.2410 [hep-ph]].

\bibitem{Amigo}
S.~D.~L.~Amigo, A.~E.~Blechman, P.~J.~Fox and E.~Poppitz,
%``R-symmetric gauge mediation,''
JHEP {\bf 0901}, 018 (2009)
[arXiv:0809.1112 [hep-ph]];

\bibitem{Benakli}
  K.~Benakli and M.~D.~Goodsell,
%  ``Dirac Gauginos in General Gauge Mediation,''
  Nucl.\ Phys.\ B {\bf 816} (2009) 185
  [arXiv:0811.4409 [hep-ph]].

\bibitem{Blechman}
 A.~E.~Blechman,
% ``R-symmetric Gauge Mediation and the MRSSM,''
 Mod.\ Phys.\ Lett.\  A {\bf 24} (2009) 633
 [arXiv:0903.2822 [hep-ph]].

\bibitem{Belanger}
  G.~Belanger, K.~Benakli, M.~Goodsell, C.~Moura and A.~Pukhov,
%  ``Dark Matter with Dirac and Majorana Gaugino Masses,''
  JCAP {\bf 0908} (2009) 027
  [arXiv:0905.1043 [hep-ph]].

\bibitem{Benakli-m}
  K.~Benakli and M.~D.~Goodsell,
%  ``Dirac Gauginos and Kinetic Mixing,''
  Nucl.\ Phys.\ B {\bf 830} (2010) 315
  [arXiv:0909.0017 [hep-ph]].

\bibitem{Kumar}
  A.~Kumar, D.~Tucker-Smith and N.~Weiner,
%  ``Neutrino Mass, Sneutrino Dark Matter and Signals of 
  Lepton Flavor Violation in the MRSSM,''
  JHEP {\bf 1009} (2010) 111
  [arXiv:0910.2475 [hep-ph]].

\bibitem{Fox-1}
  B.~A.~Dobrescu and P.~J.~Fox,
%  ``Uplifted supersymmetric Higgs region,''
  Eur.\ Phys.\ J.\ C {\bf 70} (2010) 263
  [arXiv:1001.3147 [hep-ph]].

\bibitem{Benakli-m2}
  K.~Benakli and M.~D.~Goodsell,
%  ``Dirac Gauginos, Gauge Mediation and Unification,''
  Nucl.\ Phys.\ B {\bf 840} (2010) 1
  [arXiv:1003.4957 [hep-ph]].
  %%CITATION = ARXIV:1003.4957;%%

\bibitem{Choi-1}
  S.~Y.~Choi, D.~Choudhury, A.~Freitas, J.~Kalinowski, J.~M.~Kim and P.~M.~Zerwas,
%  ``Dirac Neutralinos and Electroweak Scalar Bosons of N=1/N=2 Hybrid 
   Supersymmetry at Colliders,''
  JHEP {\bf 1008} (2010) 025
  [arXiv:1005.0818 [hep-ph]].
  %%CITATION = ARXIV:1005.0818;%%

\bibitem{Carpenter}
  L.~M.~Carpenter,
%  ``Dirac Gauginos, Negative Supertraces and Gauge Mediation,''
  JHEP {\bf 1209} (2012) 102
  [arXiv:1007.0017 [hep-th]].
  %%CITATION = ARXIV:1007.0017;%%

\bibitem{Kribs-1}
  G.~D.~Kribs, T.~Okui and T.~S.~Roy,
%  ``Viable Gravity-Mediated Supersymmetry Breaking,''
  Phys.\ Rev.\  D {\bf 82} (2010) 115010
  [arXiv:1008.1798 [hep-ph]].
  %%CITATION = PHRVA,D82,115010;%%

\bibitem{Abel}
  S.~Abel and M.~Goodsell,
%  ``Easy Dirac Gauginos,''
  JHEP {\bf 1106} (2011) 064
  [arXiv:1102.0014 [hep-th]].
  %%CITATION = ARXIV:1102.0014;%%

\bibitem{Benakli-2}
  K.~Benakli, M.~D.~Goodsell and A.~-K.~Maier,
%  ``Generating mu and Bmu in models with Dirac Gauginos,''
  Nucl.\ Phys.\ B {\bf 851} (2011) 445
  [arXiv:1104.2695 [hep-ph]].

\bibitem{Kalinowski}
  J.~Kalinowski,
%  ``Phenomenology of R-symmetric supersymmetry,''
  Acta Phys.\ Polon.\ B {\bf 42} (2011) 2425.
  %%CITATION = APPOA,B42,2425;%%

\bibitem{Benakli-1}
  K.~Benakli,
%  ``Dirac Gauginos: A User Manual,''
  Fortsch.\ Phys.\  {\bf 59} (2011) 1079
  [arXiv:1106.1649 [hep-ph]].

\bibitem{Gregoire}
  C.~Frugiuele and T.~Gregoire,
%  ``Making the Sneutrino a Higgs with a $U(1)_R$ Lepton Number,''
  Phys.\ Rev.\ D {\bf 85} (2012) 015016
  [arXiv:1107.4634 [hep-ph]].

\bibitem{ItoyamaMaru}
  H.~Itoyama and N.~Maru,
%  ``D-term Dynamical Supersymmetry Breaking Generating Split N=2 
  Gaugino Masses of Mixed Majorana-Dirac Type,''
  Int.\ J.\ Mod.\ Phys.\ A {\bf 27} (2012) 1250159
  [arXiv:1109.2276 [hep-ph]].
 
\bibitem{Katz}
  C.~Brust, A.~Katz, S.~Lawrence and R.~Sundrum,
%  ``SUSY, the Third Generation and the LHC,''
  JHEP {\bf 1203} (2012) 103
  [arXiv:1110.6670 [hep-ph]].

\bibitem{Rehermann}
  K.~Rehermann and C.~M.~Wells,
%  ``Weak Scale Leptogenesis, R-symmetry, and a Displaced Higgs,''
  arXiv:1111.0008 [hep-ph].
  %%CITATION = ARXIV:1111.0008;%%

\bibitem{Davies}
  R.~Davies and M.~McCullough,
%  ``Small neutrino masses due to R-symmetry breaking for a small cosmological constant,''
  Phys.\ Rev.\ D {\bf 86} (2012) 025014
  [arXiv:1111.2361 [hep-ph]].

\bibitem{Bertuzzo}
  E.~Bertuzzo and C.~Frugiuele,
%  ``Fitting Neutrino Physics with a $U(1)_R$ Lepton Number,''
  JHEP {\bf 1205} (2012) 100
  [arXiv:1203.5340 [hep-ph]].

\bibitem{Davies-1}
  R.~Davies,
%  ``Dirac gauginos and unification in F-theory,''
  JHEP {\bf 1210} (2012) 010
  [arXiv:1205.1942 [hep-th]].
  %%CITATION = ARXIV:1205.1942;%%

\bibitem{Argurio}
  R.~Argurio, M.~Bertolini, L.~Di Pietro, F.~Porri and D.~Redigolo,
%  ``Holographic Correlators for General Gauge Mediation,''
  JHEP {\bf 1208} (2012) 086
  [arXiv:1205.4709 [hep-th]].
  %%CITATION = ARXIV:1205.4709;%%

\bibitem{Goodsell:2012fm}
  M.~D.~Goodsell,
%  ``Two-loop RGEs with Dirac gaugino masses,''
  JHEP {\bf 1301} (2013) 066
  [arXiv:1206.6697 [hep-ph]].

\bibitem{Fok}
  R.~Fok, G.~D.~Kribs, A.~Martin and Y.~Tsai,
%  ``Electroweak Baryogenesis in R-symmetric Supersymmetry,''
  Phys.\ Rev.\ D {\bf 87} (2013) 5,  055018
  [arXiv:1208.2784 [hep-ph]].
  %%CITATION = ARXIV:1208.2784;%%

\bibitem{Argurio-1}
  R.~Argurio, M.~Bertolini, L.~Di Pietro, F.~Porri and D.~Redigolo,
%  ``Exploring Holographic General Gauge Mediation,''
  JHEP {\bf 1210} (2012) 179
  [arXiv:1208.3615 [hep-th]].
  %%CITATION = ARXIV:1208.3615;%%

\bibitem{Kumar-1}
  C.~Frugiuele, T.~Gregoire, P.~Kumar and E.~Ponton,
%  ``'L=R' - $U(1)_R$ as the Origin of Leptonic 'RPV',''
  JHEP {\bf 1303} (2013) 156
  [arXiv:1210.0541 [hep-ph]].

\bibitem{Claudia}
  C.~Frugiuele, T.~Gregoire, P.~Kumar and E.~Ponton,
%  ``'L=R' -- $U(1)_R$ Lepton Number at the LHC,''
  JHEP {\bf 1305} (2013) 012
  [arXiv:1210.5257 [hep-ph]].

\bibitem{Goodsell}
  K.~Benakli, M.~D.~Goodsell and F.~Staub,
%  ``Dirac Gauginos and the 125 GeV Higgs,''
  JHEP {\bf 1306} (2013) 073
  [arXiv:1211.0552 [hep-ph]].

\bibitem{Riva}
  F.~Riva, C.~Biggio and A.~Pomarol,
%  ``Is the 125 GeV Higgs the superpartner of a neutrino?,''
  JHEP {\bf 1302} (2013) 081
  [arXiv:1211.4526 [hep-ph]].

\bibitem{IM1} 
  H.~Itoyama and N.~Maru,
%  ``D-term Triggered Dynamical Supersymmetry Breaking,''
  Phys.\ Rev.\ D {\bf 88} (2013) 025012
  [arXiv:1301.7548 [hep-ph], arXiv:1301.7548 [hep-ph]].
 
\bibitem{IM2} 
  H.~Itoyama and N.~Maru,
%  ``126 GeV Higgs Boson Associated with D-term Triggered Dynamical Supersymmetry Breaking,''
  arXiv:1312.4157 [hep-ph].

\bibitem{Agrawal}
  P.~Agrawal and C.~Frugiuele,
%  ``Mixing stops at the LHC,''
  JHEP {\bf 1401} (2014) 115
  [arXiv:1304.3068 [hep-ph], arXiv:1304.3068].

\bibitem{SC-1}
  S.~Chakraborty and S.~Roy,
%  ``Higgs boson mass, neutrino masses and mixing and keV dark 
  matter in an $U(1)_R-$ lepton number model,''
  JHEP {\bf 1401} (2014) 101
  [arXiv:1309.6538 [hep-ph]].


\bibitem{Csaki}
  C.~Csaki, J.~Goodman, R.~Pavesi and Y.~Shirman,
%  ``The $m_D-b_M$ problem of Dirac gauginos and its solutions,''
  Phys.\ Rev.\ D {\bf 89} (2014) 5,  055005
  [arXiv:1310.4504 [hep-ph]].


\bibitem{Dudas}
  E.~Dudas, M.~Goodsell, L.~Heurtier and P.~Tziveloglou,
%  ``Flavour models with Dirac and fake gluinos,''
  Nucl.\ Phys.\ B {\bf 884} (2014) 632
  [arXiv:1312.2011 [hep-ph]].
%%%%%%%%%%%%%%%

\bibitem{Beauchesne}
  H.~Beauchesne and T.~Gregoire,
%  ``Electroweak precision measurements in supersymmetric models 
%  with a U(1)$_R$ lepton number,''
  JHEP {\bf 1405} (2014) 051
  [arXiv:1402.5403 [hep-ph]].

\bibitem{Bertuzzo:2014bwa}
  E.~Bertuzzo, C.~Frugiuele, T.~Gregoire and E.~Ponton,
%  ``Dirac gauginos, R symmetry and the 125 GeV Higgs,''
  JHEP {\bf 1504} (2015) 089
  [arXiv:1402.5432 [hep-ph]].


\bibitem{Benakli:2014cia} 
  K.~Benakli, M.~Goodsell, F.~Staub and W.~Porod,
%  ``The Constrained Minimal Dirac Gaugino Supersymmetric Standard Model,''
  Phys.\ Rev.\ D {\bf 90}, 045017 (2014)
  [arXiv:1403.5122 [hep-ph]].

\bibitem{SC-2}
  S.~Chakraborty, D.~K.~Ghosh and S.~Roy,
%  ``7 keV Sterile neutrino dark matter in $U(1)_R-$ lepton number model,''
  JHEP {\bf 1410} (2014) 146
  [arXiv:1405.6967 [hep-ph]].

\bibitem{Goodsell:2014dia}
  M.~D.~Goodsell and P.~Tziveloglou,
%  ``Dirac Gauginos in Low Scale Supersymmetry Breaking,''
  Nucl.\ Phys.\ B {\bf 889} (2014) 650
  [arXiv:1407.5076 [hep-ph]].
\bibitem{Ipek:2014moa}
  S.~Ipek, D.~McKeen and A.~E.~Nelson,
%  ``CP Violation in Pseudo-Dirac Fermion Oscillations,''
  Phys.\ Rev.\ D {\bf 90} (2014) 7,  076005
  [arXiv:1407.8193 [hep-ph]].

\bibitem{Busbridge:2014sha} 
  D.~Busbridge,
%  ``Constrained Dirac gluino mediation,''
  arXiv:1408.4605 [hep-ph].
\bibitem{Diessner:2014ksa}
  P.~Die\ss{}ner, J.~Kalinowski, W.~Kotlarski and D.~St\"{o}ckinger,
%  ``Higgs boson mass and electroweak observables in the MRSSM,''
  JHEP {\bf 1412} (2014) 124
  [arXiv:1410.4791 [hep-ph]].

\bibitem{SC-3}
  S.~Chakraborty, A.~Datta and S.~Roy,
%  ``$h\rightarrow\gamma\gamma$ in U(1)$_{R}$ -lepton number model with a right-handed neutrino,''
  JHEP {\bf 1502} (2015) 124
  [arXiv:1411.1525 [hep-ph]], Erratum to be published.
%\cite{Nelson:2015cea}
\bibitem{Tuhin}
  A.~E.~Nelson and T.~S.~Roy,
%  ``New Supersoft Supersymmetry Breaking Operators and a Solution to the $\mu$ Problem,''
  Phys.\ Rev.\ Lett.\  {\bf 114} (2015) 201802
  [arXiv:1501.03251 [hep-ph]].

%\cite{Martin:2015eca}
\bibitem{Martin:2015eca}
  S.~P.~Martin,
%  ``Non-standard supersymmetry breaking and Dirac gaugino masses without supersoftness,''
  arXiv:1506.02105 [hep-ph].

%\cite{Berger:2015qra}
\bibitem{Berger:2015qra}
  J.~Berger, J.~A.~Dror and W.~H.~Ng,
%  ``Sneutrino Higgs models explain lepton non-universality in CMS excesses,''
  arXiv:1506.08213 [hep-ph].

%\cite{Goodsell:2015ura}
\bibitem{Goodsell:2015ura}
  M.~D.~Goodsell, M.~E.~Krauss, T.~M\"{u}ller, W.~Porod and F.~Staub,
%  ``Dark matter scenarios in a UV-complete model with Dirac gauginos,''
  arXiv:1507.01010 [hep-ph].
%%%%%%%%%%%%%%%%%%%%%%%%%%%%%%%%%%%%%%%%%%%%%%%%%%%%%%%%
%%%%%%%%%%
\bibitem{Nu-1}
  T.~Schwetz, M.~Tortola and J.~W.~F.~Valle,
%  ``Global neutrino data and recent reactor fluxes: status of three-flavour oscillation parameters,''
  New J.\ Phys.\  {\bf 13} (2011) 063004
  [arXiv:1103.0734 [hep-ph]].

\bibitem{Nu-2}
  T.~Schwetz, M.~Tortola and J.~W.~F.~Valle,
%  ``Where we are on $\theta_{13}$: addendum to `Global neutrino data and recent reactor fluxes: status of three-flavour oscillation parameters',''
  New J.\ Phys.\  {\bf 13} (2011) 109401
  [arXiv:1108.1376 [hep-ph]].

\bibitem{Nu-3}
  D.~V.~Forero, M.~Tortola and J.~W.~F.~Valle,
%  ``Global status of neutrino oscillation parameters after Neutrino-2012,''
  Phys.\ Rev.\ D {\bf 86} (2012) 073012
  [arXiv:1205.4018 [hep-ph]].

\bibitem{Nu-4}
  M.~C.~Gonzalez-Garcia, M.~Maltoni, J.~Salvado and T.~Schwetz,
%  ``Global fit to three neutrino mixing: critical look at present precision,''
  JHEP {\bf 1212} (2012) 123
  [arXiv:1209.3023 [hep-ph]].
%%%%%%%%%%%%%%%%%%%%%%%%%%%%%%%%%%%%%%%%%%%%%%%%%%%%%%%%
\bibitem{Romao}
  W.~Porod, M.~Hirsch, J.~Romao and J.~W.~F.~Valle,
%  ``Testing neutrino mixing at future collider experiments,''
  Phys.\ Rev.\ D {\bf 63} (2001) 115004
  [hep-ph/0011248].
%%%%%%%%%%%%%%%%%%%%%%%%%%%%%%%%%%%%%%%%%%%%%%%%%%%%%%%%
\bibitem{Staub}  
  F.~Staub,
%  ``Sarah,''
  arXiv:0806.0538 [hep-ph].
\bibitem{Staub1}
  F.~Staub,
%  ``SARAH 3.2: Dirac Gauginos, UFO output, and more,''
  Comput.\ Phys.\ Commun.\  {\bf 184}, pp. 1792 (2013)
  [Comput.\ Phys.\ Commun.\  {\bf 184}, 1792 (2013)]
  [arXiv:1207.0906 [hep-ph]].
\bibitem{Staub2}
  F.~Staub,
%  ``Exploring new models in all detail with SARAH,''
  arXiv:1503.04200 [hep-ph].
%%%%%%%%%%%%%%%%%%%%%%%%%%%%%%%%%%%%%%%%%%%%%%%%%%%%%%%%
\bibitem{Curtin}
  D.~Curtin, P.~Meade and P.~J.~Tien,
%  ``Natural SUSY in Plain Sight,''
  Phys.\ Rev.\ D {\bf 90} (2014) 11,  115012
  [arXiv:1406.0848 [hep-ph]].


\bibitem{Beuria}
  J.~Beuria, A.~Chatterjee, A.~Datta and S.~K.~Rai,
%  ``Two Light Stops in the NMSSM and the LHC,''
  arXiv:1505.00604 [hep-ph].
\bibitem{Aad:2015pfx}
  G.~Aad {\it et al.} [ATLAS Collaboration],
%  ``ATLAS Run 1 searches for direct pair production of 
%third-generation squarks at the Large Hadron Collider,''
  arXiv:1506.08616 [hep-ex].
%%%%%%%%%%%%%%%%%%%%%%%%%%%%%%%%%%%%%%%%%%%%%%%%%%%%%%%%

\bibitem{Fan1}
  J.~Fan, M.~Reece and J.~T.~Ruderman,
%  ``Stealth Supersymmetry,''
  JHEP {\bf 1111} (2011) 012
  [arXiv:1105.5135 [hep-ph]].

\bibitem{Csaki1}
  C.~Csaki, L.~Randall and J.~Terning,
%  ``Light Stops from Seiberg Duality,''
  Phys.\ Rev.\ D {\bf 86} (2012) 075009
  [arXiv:1201.1293 [hep-ph]].

\bibitem{Fan}
  J.~Fan, M.~Reece and J.~T.~Ruderman,
%  ``A Stealth Supersymmetry Sampler,''
  JHEP {\bf 1207} (2012) 196
  [arXiv:1201.4875 [hep-ph]].

\bibitem{stealth-stop}
  Z.~Han and A.~Katz,
%  ``Stealth Stops and Spin Correlation: A Snowmass White Paper,''
  arXiv:1310.0356 [hep-ph].

\bibitem{Weiler}
  M.~Czakon, A.~Mitov, M.~Papucci, J.~T.~Ruderman and A.~Weiler,
%  ``Closing the stop gap,''
  Phys.\ Rev.\ Lett.\  {\bf 113} (2014) 20,  201803
  [arXiv:1407.1043 [hep-ph]].

\bibitem{Till}
  T.~Eifert and B.~Nachman,
%  ``Sneaky light stop,''
  Phys.\ Lett.\ B {\bf 743} (2015) 218
  [arXiv:1410.7025 [hep-ph]].
\bibitem{Future}
S.~Chakraborty, AseshKrishna Datta, Katri Huitu, Sourov Roy, Harri Waltari,
Work in progress.

\bibitem{SSHD}
  H.~K.~Dreiner, M.~Hanussek, J.~S.~Kim and S.~Sarkar,
%  ``Gravitino cosmology with a very light neutralino,''
  Phys.\ Rev.\ D {\bf 85} (2012) 065027
  [arXiv:1111.5715 [hep-ph]].

%%%%%%%%%%%%%%%%%%%%%%%%%%%%%%%%%%%%%%%%%%%%%%%%%%%%%%%%
\bibitem{LHC-stop}
  [ATLAS Collaboration],
%  ``Search for direct top squark pair production in final states with one isolated 
%  lepton, jets, and missing transverse momentum in $\sqrt{s}=8,$TeV $pp$ collisions 
%  using 21 fb$^{-1}$ of ATLAS data,''
  ATLAS-CONF-2013-037, ATLAS-COM-CONF-2013-038.\\
\bibitem{LHC-s1}[CMS Collaboration],
%  ``Search for direct top squark pair production in events with a single isolated 
%  lepton, jets and missing transverse energy at $\sqrt{s}$ = 8 TeV,''
  CMS-PAS-SUS-12-023.\\
\bibitem{LHC-s2} The ATLAS collaboration,
%  ``Search for strong production of supersymmetric particles in final states with 
%  missing transverse momentum and at least three b-jets using 20.1 $\text{fb}^{-1}$ of pp 
%  collisions at $\sqrt{s}$ = 8 TeV with the ATLAS Detector.,''
  ATLAS-CONF-2013-061, ATLAS-COM-CONF-2013-071.\\
\bibitem{LHC-s3}  
  S.~Chatrchyan {\it et al.}  [CMS Collaboration],
%  ``Search for top-squark pair production in the single-lepton final state in pp 
%  collisions at $\sqrt{s}$ = 8 TeV,''
  Eur.\ Phys.\ J.\ C {\bf 73} (2013) 12,  2677
  [arXiv:1308.1586 [hep-ex]].

\bibitem{ATLAS-RPV}
  The ATLAS collaboration,
%  ``A search for $B-L$ $R$-Parity violating scalar top decays in $\sqrt{s} = 8$ 
% TeV $pp$ collisions with the ATLAS experiment,''
  ATLAS-CONF-2015-015, ATLAS-COM-CONF-2015-017.
\bibitem{CMS-RPV}
  V.~Khachatryan {\it et al.} [CMS Collaboration],
%  ``Search for pair production of third-generation scalar leptoquarks and top squarks in proton-proton collisions at sqrt(s) = 8 TeV,''
  Phys.\ Lett.\ B {\bf 739} (2014) 229
  [arXiv:1408.0806 [hep-ex]].
%%%%%%%%%%%%%%%%%%%%%%%%%%%%%%%%%%%%%%%%%%%%%%%%%%%%%%%%
\bibitem{Barbier}
  R.~Barbier, C.~Berat, M.~Besancon, M.~Chemtob, A.~Deandrea, E.~Dudas, P.~Fayet and S.~Lavignac {\it et al.},
%  ``R-parity violating supersymmetry,''
  Phys.\ Rept.\  {\bf 420} (2005) 1
  [hep-ph/0406039].

\bibitem{Ambrosanio}
  S.~Ambrosanio, G.~L.~Kane, G.~D.~Kribs, S.~P.~Martin and S.~Mrenna,
%  ``Search for supersymmetry with a light gravitino at the Fermilab Tevatron and CERN LEP colliders,''
  Phys.\ Rev.\ D {\bf 54} (1996) 5395
  [hep-ph/9605398].
%%%%%%%%%%%%%%%%%%%%%%%%%%%%%%%%%%%%%%%%%%%%%%%%%%%%%%%%
\bibitem{Gunion}
  J.~F.~Gunion and H.~E.~Haber,
%  ``Higgs Bosons in Supersymmetric Models. 1.,''
  Nucl.\ Phys.\ B {\bf 272} (1986) 1
   [Erratum-ibid.\ B {\bf 402} (1993) 567].

\bibitem{Low}
  I.~Low,
%  ``Polarized charginos (and top quarks) in scalar top quark decays,''
  Phys.\ Rev.\ D {\bf 88} (2013) 9,  095018
  [arXiv:1304.0491 [hep-ph]].

%%%%%%%%%%%%%%%%%%%%%%%%%%%%%%%%%%%%%%%%%%%%%%%%%%%%%%%%
\bibitem{ATLAS-191}
  G.~Aad {\it et al.}  [ATLAS Collaboration],
%  ``Measurement of Spin Correlation in Top-Antitop Quark Events and Search 
%  for Top Squark Pair Production in pp Collisions at $\sqrt{s}=8$ TeV Using the ATLAS Detector,''
  Phys.\ Rev.\ Lett.\  {\bf 114} (2015) 14,  142001
  [arXiv:1412.4742 [hep-ex]].

\bibitem{ATLAS-470}
  G.~Aad {\it et al.}  [ATLAS Collaboration],
%  ``Search for direct top-squark pair production in final states with 
%two leptons in pp collisions at $\sqrt{s} =$ 8TeV with the ATLAS detector,''
  JHEP {\bf 1406} (2014) 124
  [arXiv:1403.4853 [hep-ex]].

\bibitem{CMS-stop}
  V.~Khachatryan {\it et al.}  [CMS Collaboration],
%  ``Search for pair-produced resonances decaying to jet pairs in proton-proton collisions at $\sqrt{s}$ = 8 TeV,''
  Phys.\ Lett.\ B {\bf 739} (2014) 229
  [arXiv:1408.0806 [hep-ex]].

\bibitem{DF}
  H.~K.~Dreiner, K.~Nickel, F.~Staub and A.~Vicente,
%  ``New bounds on trilinear R-parity violation from lepton flavor violating observables,''
  Phys.\ Rev.\ D {\bf 86} (2012) 015003
  [arXiv:1204.5925 [hep-ph]].
%%%%%%%%%%%%%%%%%%%%%%%%%%%%%%%%%%%%%%%%%%%%%%%%%%%%%%%%
\bibitem{Datta}
  A.~Datta and B.~Mukhopadhyaya,
%  ``Are messages of R-parity violating supersymmetry hidden within top quark signals?,''
  Phys.\ Rev.\ Lett.\  {\bf 85} (2000) 248
  [hep-ph/0003174].

\bibitem{Sujoy}
  S.~P.~Das, A.~Datta and S.~Poddar,
%  ``Top squark and neutralino decays in a R-parity violating model constrained by neutrino oscillation data,''
  Phys.\ Rev.\ D {\bf 73} (2006) 075014
  [hep-ph/0509171].

\bibitem{Datta:2006ak}
  A.~Datta and S.~Poddar,
%  ``New signals of a R-parity violating model of neutrino mass at the Tevatron,''
  Phys.\ Rev.\ D {\bf 75} (2007) 075013
  [hep-ph/0611074].

\bibitem{Datta:2009dc}
  A.~Datta and S.~Poddar,
%  ``Probing R-parity violating models of neutrino mass at the LHC via top squark decays,''
  Phys.\ Rev.\ D {\bf 79} (2009) 075021
  [arXiv:0901.1619 [hep-ph]].

\bibitem{Marshall}
  Z.~Marshall, B.~A.~Ovrut, A.~Purves and S.~Spinner,
%  ``Spontaneous $R$-Parity Breaking, Stop LSP Decays and the Neutrino Mass Hierarchy,''
  Phys.\ Lett.\ B {\bf 732} (2014) 325
  [arXiv:1401.7989 [hep-ph]].

\bibitem{Bose:2014vea}
  R.~Bose, A.~Datta, A.~Kundu and S.~Poddar,
%  ``LHC signatures of neutrino mass generation through R-parity violation,''
  Phys.\ Rev.\ D {\bf 90} (2014) 3,  035007
  [arXiv:1405.1282 [hep-ph]].

\bibitem{Chun}
  E.~J.~Chun, S.~Jung, H.~M.~Lee and S.~C.~Park,
%  ``Stop and Sbottom LSP with R-parity Violation,''
  Phys.\ Rev.\ D {\bf 90} (2014) 11,  115023
  [arXiv:1408.4508 [hep-ph]].

%%%%%%%%%%%%%%%%%%%%%%%%%%%%%%%%%%%%%%%%%%%%%%%%%%%%%%%%
\bibitem{Bedo}
  A.~Choudhury and A.~Datta,
%  ``New limits on top squark NLSP from LHC 4.7 $fb^{-1}$ data,''
  Mod.\ Phys.\ Lett.\ A {\bf 27} (2012) 1250188
  [arXiv:1207.1846 [hep-ph]].


\bibitem{Grober}
  R.~Grober, M.~Muhlleitner, E.~Popenda and A.~Wlotzka,
%  ``Light Stop Decays: Implications for LHC Searches,''
  arXiv:1408.4662 [hep-ph].


\bibitem{Ferretti}
  G.~Ferretti, R.~Franceschini, C.~Petersson and R.~Torre,
%  ``Spot the stop with a b-tag,''
  Phys.\ Rev.\ Lett.\  {\bf 114} (2015) 201801
  [arXiv:1502.01721 [hep-ph]].

\bibitem{Belangert}
  G.~Belanger, D.~Ghosh, R.~Godbole and S.~Kulkarni,
%  ``Light stop in the MSSM after LHC Run 1,''
  arXiv:1506.00665 [hep-ph].
%%%%%%%%%%%%%%%%%%%%%%%%%%%%%%%%%%%%%%%%%%%%%%%%%%%%%%%%

\bibitem{CMS-FV}
  S.~Chatrchyan {\it et al.}  [CMS Collaboration],
%  ``Search for top squarks in $R$-parity-violating supersymmetry using three or more leptons and b-tagged jets,''
  Phys.\ Rev.\ Lett.\  {\bf 111} (2013) 22,  221801
  [arXiv:1306.6643 [hep-ex]].

\bibitem{ATLAS-FV}
  G.~Aad {\it et al.}  [ATLAS Collaboration],
%  ``Search for pair-produced third-generation squarks decaying via charm quarks or in compressed supersymmetric scenarios in $pp$ collisions at $\sqrt{s}=8~$TeV with the ATLAS detector,''
  Phys.\ Rev.\ D {\bf 90} (2014) 5,  052008
  [arXiv:1407.0608 [hep-ex]].

%%%%%%%%%%%%%%%%%%%%%%%%%%%%%%%%%%%%%%%%%%%%%%%%%%%%%%%%
\bibitem{Amit1}
  A.~Chakraborty, D.~K.~Ghosh, D.~Ghosh and D.~Sengupta,
%  ``Stop and sbottom search using dileptonic $M_{T2}$ variable and boosted top technique at the LHC,''
  JHEP {\bf 1310} (2013) 122
  [arXiv:1303.5776 [hep-ph]].

\bibitem{Eckel}
  J.~Eckel, S.~Su and H.~Zhang,
%  ``Exotic Stop Decay at the LHC,''
  arXiv:1411.1061 [hep-ph].

\bibitem{Amit2}
  A.~Chakraborty, D.~K.~Ghosh, S.~Mondal, S.~Poddar and D.~Sengupta,
%  ``Probing the NMSSM via Higgs boson signatures from stop cascade decays at the LHC,''
  Phys.\ Rev.\ D {\bf 91} (2015) 115018
  [arXiv:1503.07592 [hep-ph]].

%%%%%%%%%%%%%%%%%%%%%%%%%%%%%%%%%%%%%%%%%%%%%%%%%%%%%%%%
\bibitem{Kim}
  J.~S.~Kim, K.~Rolbiecki, K.~Sakurai and J.~Tattersall,
%  `Stop' that ambulance! New physics at the LHC?,
  JHEP {\bf 1412} (2014) 010
  [arXiv:1406.0858 [hep-ph]].

\bibitem{Abe}
  H.~Abe, J.~Kawamura and Y.~Omura,
%  ``LHC phenomenology of natural MSSM with non-universal gaugino masses at the unification scale,''
  arXiv:1505.03729 [hep-ph].

\bibitem{Han:2015tua}
  T.~Han, S.~Su, Y.~Wu, B.~Zhang and H.~Zhang,
%  ``To the Bottom of the Sbottom,''
  arXiv:1507.04006 [hep-ph].

%%%%%%%%%%%%%%%%%%%%%%%%%%%%%%%%%%%%%%%%%%%%%%%%%%%%%%%%
\bibitem{CMS-PAS}
  CMS Collaboration [CMS Collaboration],
% ``Search for direct production of bottom squark pairs,''
  CMS-PAS-SUS-13-018. 
%%%%%%%%%%%%%%%%%%%%%%%%%%%%%%%%%%%%%%%%%%%%%%%%%%%%%%%%
\bibitem{Porod}
  W.~Porod,
%  ``SPheno, a program for calculating supersymmetric spectra, SUSY particle decays and SUSY particle production at e+ e- colliders,''
  Comput.\ Phys.\ Commun.\  {\bf 153} (2003) 275
  [hep-ph/0301101].

\bibitem{Porod1}
  W.~Porod and F.~Staub,
%  ``SPheno 3.1: Extensions including flavour, CP-phases and models beyond the MSSM,''
  Comput.\ Phys.\ Commun.\  {\bf 183} (2012) 2458
  [arXiv:1104.1573 [hep-ph]].
%%%%%%%%%%%%%%%%%%%%%%%%%%%%%%%%%%%%%%%%%%%%%%%%%%%%%%%%
\bibitem{Vicente}
  W.~Porod, F.~Staub and A.~Vicente,
%  ``A Flavor Kit for BSM models,''
  Eur.\ Phys.\ J.\ C {\bf 74} (2014) 8,  2992
  [arXiv:1405.1434 [hep-ph]].
%%%%%%%%%%%%%%%%%%%%%%%%%%%%%%%%%%%%%%%%%%%%%%%%%%%%%%%%
\bibitem{Higgsbounds}
  P.~Bechtle, O.~Brein, S.~Heinemeyer, G.~Weiglein and K.~E.~Williams,
%  ``HiggsBounds: Confronting Arbitrary Higgs Sectors with Exclusion Bounds 
%   from LEP and the Tevatron,''
  Comput.\ Phys.\ Commun.\  {\bf 181} (2010) 138
  [arXiv:0811.4169 [hep-ph]].\\
\bibitem{Higgsbounds1}
  P.~Bechtle, O.~Brein, S.~Heinemeyer, G.~Weiglein and K.~E.~Williams,
%  ``HiggsBounds 2.0.0: Confronting Neutral and Charged Higgs Sector Predictions 
%  with Exclusion Bounds from LEP and the Tevatron,''
  Comput.\ Phys.\ Commun.\  {\bf 182} (2011) 2605
  [arXiv:1102.1898 [hep-ph]].\\
\bibitem{Higgsbounds2}
  P.~Bechtle, O.~Brein, S.~Heinemeyer, O.~Stal, T.~Stefaniak, G.~Weiglein and 
  K.~Williams,
%  ``Recent Developments in HiggsBounds and a Preview of HiggsSignals,''
  PoS CHARGED {\bf 2012} (2012) 024
  [arXiv:1301.2345 [hep-ph]].\\
\bibitem{Higgsbounds3}
  P.~Bechtle, O.~Brein, S.~Heinemeyer, O.~Stal, T.~Stefaniak, G.~Weiglein and 
  K.~E.~Williams,
%  ``$\mathsf{HiggsBounds}-4$: Improved Tests of Extended Higgs Sectors against 
%  Exclusion Bounds from LEP, the Tevatron and the LHC,''
  Eur.\ Phys.\ J.\ C {\bf 74} (2014) 3,  2693
  [arXiv:1311.0055 [hep-ph]].
%%%%%%%%%%%%%%%%%%%%%%%%%%%%%%%%%%%%%%%%%%%%%%%%%%%%%%%%
\bibitem{HiggsSignals1}
  P.~Bechtle, S.~Heinemeyer, O.~Stål, T.~Stefaniak and G.~Weiglein,
%  ``$HiggsSignals$: Confronting arbitrary Higgs sectors with measurements at the Tevatron and the LHC,''
  Eur.\ Phys.\ J.\ C {\bf 74} (2014) 2,  2711
  [arXiv:1305.1933 [hep-ph]].

\bibitem{HiggsSignals2}
  P.~Bechtle, S.~Heinemeyer, O.~Stål, T.~Stefaniak and G.~Weiglein,
%  ``Probing the Standard Model with Higgs signal rates from the Tevatron, the LHC and a future ILC,''
  JHEP {\bf 1411} (2014) 039
  [arXiv:1403.1582 [hep-ph]].

%%%%%%%%%%%%%%%%%%%%%%%%%%%%%%%%%%%%%%%%%%%%%%%%%%%%%%%%
\bibitem{Madgraph}
  J.~Alwall, R.~Frederix, S.~Frixione, V.~Hirschi, F.~Maltoni, O.~Mattelaer, 
  H.-S.~Shao and T.~Stelzer {\it et al.},
%  ``The automated computation of tree-level and next-to-leading order differential 
%   cross sections, and their matching to parton shower simulations,''
  JHEP {\bf 1407} (2014) 079
  [arXiv:1405.0301 [hep-ph]].

\bibitem{Pumplin:2002vw}
  J.~Pumplin, D.~R.~Stump, J.~Huston, H.~L.~Lai, P.~M.~Nadolsky and W.~K.~Tung,
%  ``New generation of parton distributions with uncertainties from global QCD analysis,''
  JHEP {\bf 0207} (2002) 012
  [hep-ph/0201195].
\bibitem{P1}
  W.~Beenakker, M.~Kramer, T.~Plehn, M.~Spira and P.~M.~Zerwas,
%  ``Stop production at hadron colliders,''
  Nucl.\ Phys.\ B {\bf 515} (1998) 3
  [hep-ph/9710451].


\bibitem{Prospino}
  W.~Beenakker, R.~Hopker and M.~Spira,
%  ``PROSPINO: A Program for the production of supersymmetric particles in next-to-leading order QCD,''
  hep-ph/9611232.

%%%%%%%%%%%%%%%%%%%%%%%%%%%%%%%%%%%%%%%%%%%%%%%%%%%%%%%%
\bibitem{Pythia}
  T.~Sjostrand, S.~Mrenna and P.~Z.~Skands,
%  ``PYTHIA 6.4 Physics and Manual,''
  JHEP {\bf 0605} (2006) 026
  [hep-ph/0603175].

\bibitem{ttsm} 
{\footnotesize \url{https://twiki.cern.ch/twiki/bin/view/LHCPhysics/TtbarNNLO#Top_quark_pair_cross_sections_at}}

\bibitem{Campbell}
  J.~M.~Campbell and R.~K.~Ellis,
%  ``Radiative corrections to Z b anti-b production,''
  Phys.\ Rev.\ D {\bf 62} (2000) 114012
  [hep-ph/0006304].

\bibitem{Dawson:2003zu}
  S.~Dawson, C.~Jackson, L.~H.~Orr, L.~Reina and D.~Wackeroth,
%  ``Associated Higgs production with top quarks at the large hadron collider: NLO QCD corrections,''
  Phys.\ Rev.\ D {\bf 68} (2003) 034022
  [hep-ph/0305087].
  %%CITATION = HEP-PH/0305087;%%

%\cite{FebresCordero:2008ci}
\bibitem{FebresCordero:2008ci}
  F.~Febres Cordero, L.~Reina and D.~Wackeroth,
%  ``NLO QCD corrections to $Z b \bar{b}$ production with massive bottom quarks at the Fermilab Tevatron,''
  Phys.\ Rev.\ D {\bf 78} (2008) 074014
  [arXiv:0806.0808 [hep-ph]].
  %%CITATION = ARXIV:0806.0808;%%

\bibitem{Frederix}
  R.~Frederix, S.~Frixione, V.~Hirschi, F.~Maltoni, R.~Pittau and P.~Torrielli,
%  ``W and $Z/\gamma*$ boson production in association with a bottom-antibottom pair,''
  JHEP {\bf 1109} (2011) 061
  [arXiv:1106.6019 [hep-ph]].

\bibitem{Ox}
\url{http://www.hep.phy.cam.ac.uk/~lester/mt2/}

%\cite{ATLAS:2015jla}
\bibitem{ATLAS:2015jla}
The ATLAS collaboration [ATLAS Collaboration],
%``A search for $B-L$ $R$-Parity violating scalar top decays in  $\sqrt{s} = 8$ TeV $pp$ collisions with the ATLAS experiment,''
ATLAS-CONF-2015-015.
%%CITATION = ATLAS-CONF-2015-015;%%

\bibitem{CMS-sig}
  D.~G.~d'Enterria {\it et al.}  [CMS Collaboration],
%  ``CMS physics technical design report: Addendum on high density QCD with heavy ions,''
  J.\ Phys.\ G {\bf 34} (2007) 2307.

\bibitem{Cho:2014yma}
  W.~S.~Cho, J.~S.~Gainer, D.~Kim, K.~T.~Matchev, F.~Moortgat, L.~Pape and M.~Park,
%  ``Improving the sensitivity of stop searches with on-shell constrained invariant mass variables,''
  JHEP {\bf 1505} (2015) 040
  [arXiv:1411.0664 [hep-ph]].
  %%CITATION = ARXIV:1411.0664;%%

\bibitem{Kilic}
  C.~Kilic and B.~Tweedie,
%  ``Cornering Light Stops with Dileptonic mT2,''
  JHEP {\bf 1304} (2013) 110
  [arXiv:1211.6106 [hep-ph]].

\bibitem{Graesser:2012qy}
  M.~L.~Graesser and J.~Shelton,
%  ``Hunting Mixed Top Squark Decays,''
  Phys.\ Rev.\ Lett.\  {\bf 111} (2013) 12,  121802
  [arXiv:1212.4495 [hep-ph]].
  %%CITATION = ARXIV:1212.4495;%%




\end{thebibliography}
\end{document}